\documentclass{resonancearXiv}

\usepackage{babel}

\usepackage{wrapfig,framed}

\usepackage{textcmds}
\usepackage{bibunits}
\usepackage{graphicx}
\usepackage{dcolumn}
\usepackage{bm}
\usepackage[breaklinks,colorlinks = true,linkcolor = red,urlcolor=blue,citecolor=red]{hyperref}
\usepackage{multirow}
\usepackage{array}
\usepackage{booktabs}
\usepackage{upgreek}
\usepackage{epsfig}
\usepackage{mathrsfs}
\usepackage{amssymb}
\usepackage{amsbsy}

\usepackage{color}
\usepackage{cancel}
\usepackage{pifont}
\usepackage{marginnote}
\usepackage{float}
\usepackage{verbatim}
\usepackage{bbm, dsfont}
\usepackage{lineno}
\usepackage{amsmath}
\usepackage{caption}
\usepackage{subcaption}

\usepackage{chemfig}

\usepackage{mdframed}

\usepackage{lipsum}

\definecolor{skincolor}{rgb}{0.98, 0.92, 0.84}
\mdfdefinestyle{mystyle}{innertopmargin=0.5cm,innerbottommargin=0.5cm,roundcorner=10pt,backgroundcolor= skincolor,linewidth=0pt,splittopskip=0.5cm}

\newcommand{\highlight}[1]{%
  \colorbox{red!50}{$\displaystyle#1$}}

\newcommand{\bcen}{\begin{center}}
\newcommand{\ecen}{\end{center}}
\newcommand{\btab}{\begin{tabular}}
\newcommand{\etab}{\end{tabular}}
\newcommand{\bdes}{\begin{description}}
\newcommand{\edes}{\end{description}}

\newcommand{\beq}{\begin{equation}}
\newcommand{\eeq}{\end{equation}}
\newcommand{\bea}{\begin{eqnarray}}
\newcommand{\eea}{\end{eqnarray}}

\newcommand{\bary}{\begin{array}}
	\newcommand{\eary}{\end{array}}
\newcommand{\benum}{\begin{enumerate}}
	\newcommand{\eenum}{\end{enumerate}}
\newcommand{\bitem}{\begin{itemize}}
	\newcommand{\eitem}{\end{itemize}}

\newcommand{\bk} { \bm{k} }
\newcommand{\br} { \boldsymbol{r}}

\newcommand{\bA} { \mbox{\boldmath $A$}}

\newcommand{\eqn}[1] {eqn.~(\ref{#1})}
\newcommand{\Eqn}[1] {Eqn.~(\ref{#1})}

\newcommand{\sect}[1] {Section~\ref{#1}}

\newcommand{\fig}[1]{fig.~\ref{#1}}
\newcommand{\Fig}[1]{Fig.~\ref{#1}}

\makeatletter

\newcommand{\Rmnum}[1]{\expandafter\@slowromancap\romannumeral #1@}
\makeatother

\mathchardef\mhyphen="2D
\usepackage[utf8]{inputenc}

\begin{document}

\title{Exploring ideas in topological quantum phenomena}
\secondTitle{A journey through the SSH model}

\author{Anantha Hegde, Adarsh Kumar, Adhip Agarwala and Bhaskaran Muralidharan

\footnote{hegdeanantha@gmail.com,kumar.adarsh@iitb.ac.in,adhip@pks.mpg.de,bm@ee.iitb.ac.in }

}

\maketitle

\authorIntro{
Anantha and Adarsh just graduated from IIT Bombay (India) with a B.Tech and an M.Tech in Electrical Engineering. Adhip is a postdoctoral fellow at Max Planck Institute for the Physics of Complex Systems, Dresden (Germany). Bhaskaran is a Professor in the Department of Electrical Engineering at IIT Bombay (India).
}

\begin{sloppypar}

\begin{abstract}
Geared as an invitation for undergraduates, beginning graduate students, we present a pedagogical introduction to one-dimensional topological phases -- in particular the Su-Schrieffer-Heeger model. In the process, we delve upon ideas of entanglement using the correlator method and the von-Neumann density-matrix method, geometric phase, polarization, transport signatures and the role of electron-electron interactions. Through hands-on numerical experiments, whose \href{https://github.com/hnoend/SSH_codes}{codes} are shared, we try to drive home the message why a program of simulating quantum electronics with topological toy models is the store house for discovering fantastic physics ideas.
\end{abstract}

\monthyear{July 2021}
\artNature{GENERAL  ARTICLE}

\section{Phases of matter that surround us}

The understanding of the various phases of matter which surround us is often characterised by the way {\it we} react to them. For instance, it would hardly be the best idea to touch a live wire which is attached to the electric socket. That a material is a metal is often defined by its ability to conduct electric current. Therefore while we move around, we intuitively classify phases around us, things which are metallic (copper, the steel dining plates, the spoon) and things which are insulating (that abhor current) such as a piece of wood, glass etc. Things which stick to each other (magnets) and things that would rather burn than to conduct heat (plastics or polymers, say over a metal). However intuitive and obvious while these may look, the microscopic mechanisms behind their behavior is rather intricate and has a rich display of quantum mechanics which we often fail to appreciate. The following pages will introduce you, {\it the reader}, to some of these fascinating stories which are hidden behind seemingly plain phenomena. The write-up will assume you are an undergraduate student who has had a first course in quantum mechanics at the level of the introductory book by D.J. Griffiths and some exposure to solid-state physics. The latter is not an essential pre-requisite though.

\section{Taking the first shot}
While every material has a unique set of atoms, bond-strengths, ionization energies, stiffness etc.~and therefore \qq{solving} any material using the rule of quantum mechanics is essentially impossible. However using the most simplest of assumptions and keeping the minimal but essential ingredients one can still understand a variety of phenomena. For instance a class of materials which we understand in a relatively straightforward manner are metals and insulators. 

\leftHighlight{
    \includegraphics[width=1.0\linewidth]{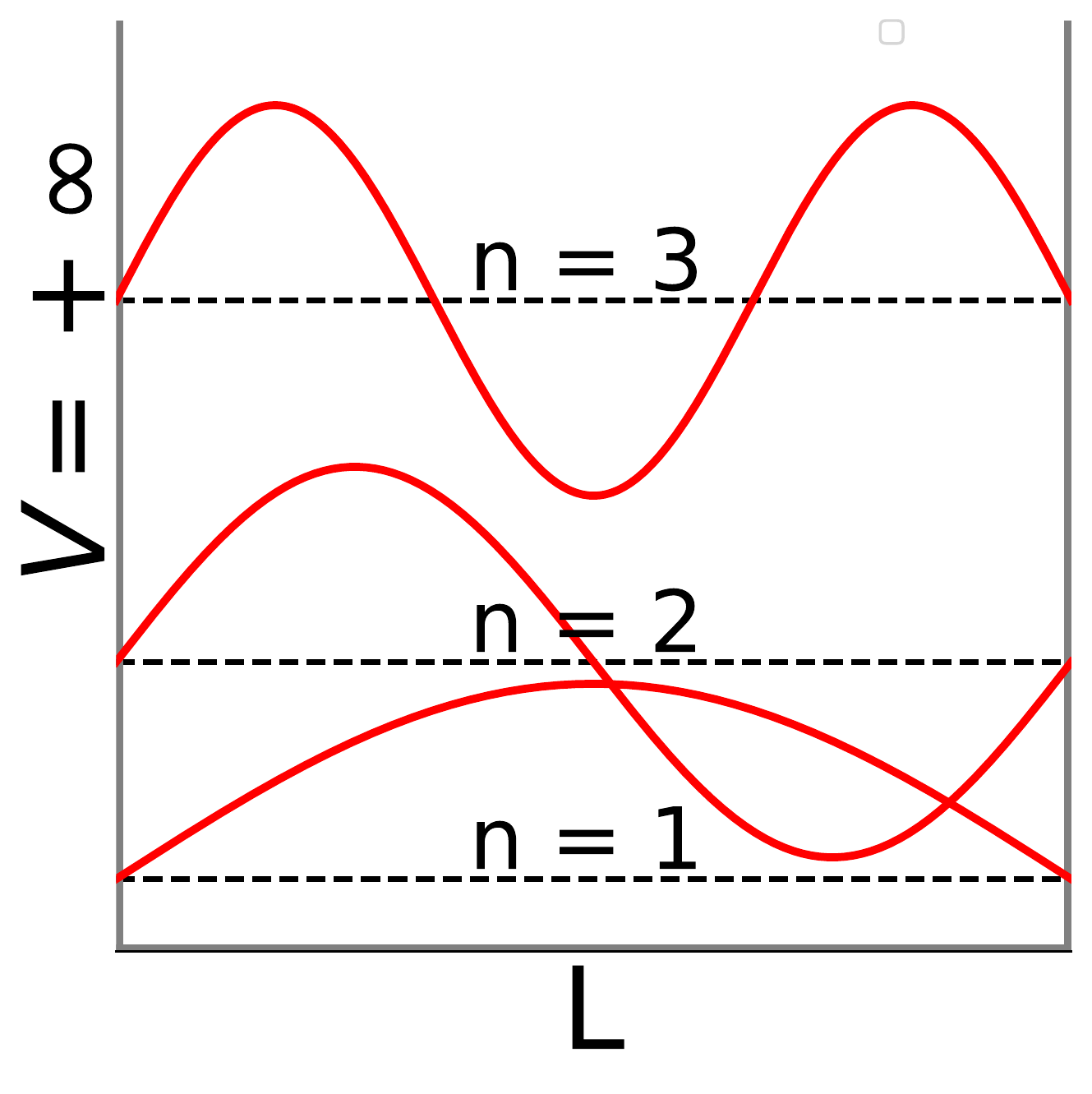}
    First three momentum states for a particle in a one dimensional box
}

In quantum mechanics 101, we have solved for an electron in a potential box. There, the laws of quantum mechanics dictate that each electron, instead of being thought of as a particle should rather be thought of as a wave characterized by a wavelength $\lambda$ (and the momentum vector $\vec{k} = \frac{2\pi}{\lambda} \hat{\br}$). When we wish to fill in many electrons we invoke Pauli exclusion principle --  and demand that each of these states (labelled by a distinct $\bk = \frac{\pi n}{L}$ ($n=1,2,\ldots$)) should be occupied by just one electron. If one wishes to add more electrons one needs to add them in a different momentum state. If our universe was just a box, and all we had were non-interacting electrons -- our universe could be exactly solvable but also boring. Thankfully, even in a usual metal -- that is not the case. Electrons as they move in a material are constantly pulled and pushed by the ions which surround them. In reality electrons also push and pull each other -- because they have the same electronic charge and repel. Often their interactions cannot be ignored, one of the reasons many of us are still doing research. 

But, as is true for many of us when we appear for exams, we answer what we know best -- and see how good or bad it really is. 

So let us assume electrons don't interact -- what does that mean? We can solve for just \q{one} electron in a potential landscape which gives us all the one-particle states it can occupy -- and then when we have more electrons to worry about, we just fill them one by one keeping in mind the Pauli exclusion principle. If electrons {\it did} interact -- addition of another electron could have disturbed the existing ones and therefore we would have had to solve for them again.

\subsection{Modelling a solid}

To model a solid, what potential landscape should one then use? Schrodinger's equation is a differential equation in space and time. And again solving this is not very easy -- specially when potentials are complicated. Now given a material has a host of atoms each with different number of protons, one could imagine that the electrostatic potential they generate for an electron can be really weird. One could of course try to solve it numerically and obtain stationary solutions, by discretizing the Schrodinger equation as follows:
\bea
E \psi(\br) &=& \Big(-\frac{\partial^2}{\partial^2 \br} + V(\br) \Big) \psi (\br) \\
E \psi(\br_i) &=&  -\frac{ \psi(\br_{i+1}) + \psi(\br_{i-1}) -2\psi(\br_{i})}{\br_{i+1}-\br_{i}}  +   V(\br_i) \psi(\br_{i})
\label{discrete_hamiltonian}
\eea
This leads to a matrix form of this equation which can be numerically diagonalized to find the eigenvalues.

Even while the above method may just seem like a numerical convenience -- diagonalizing a matrix form of a Hamiltonian is a very standard practice in quantum condensed matter and has very concrete physical motivations to do so. This framework of solving Hamiltonians is called the {\it  tight-binding framework}. It introduces some assumptions in order to make our job of analysis simpler for e.g.~that the atoms which are placed in real space are sufficiently far such that their atomic orbital wavefunctions don't overlap. 

More concretely, consider two atoms $A,B$ governed with a Hamiltonian $H$ with their orbital wavefunctions $|\phi_A\rangle$ and $|\phi_B\rangle$. Even while we assume $\langle \phi_A|\phi_B\rangle = 0$, this does not imply that $\langle \phi_A|H|\phi_B\rangle $ is {\it zero}. In fact, the term $\langle \phi_A|H|\phi_B\rangle $ is crucially non-zero and is called the hopping amplitude for an electron to move between site $A$ and site $B$. This reflects the idea that electrons can freely move in a lattice potential allowing for such hopping processes. The above way of defining our system already assumes that a general solution for the wavefunction for this system is a linear combination of the orbital wavefunctions on the two atoms separately. Are such assumptions physically meaningful in a a realistic setting? In order to make sense of this assumptions we consider two finite potential wells separated by a barrier and analyze what happens to the ground state and the first excited state as the width of the barrier is tuned (see \hyperlink{example1}{Example 1}).

{\leftHighlight{Example 1}}
\begin{mdframed}[style = mystyle, align=center,userdefinedwidth=1.0\columnwidth]

\hypertarget{example1}{{\it \textbf{Example:}}}
For an electron being shared between two atoms, is it fair to assume (i) that the atomic orbitals of the atoms don't overlap? and (ii) that the complete solution is just a linear combination of individual wavefunctions on the two atoms?

Consider a one dimensional (D) system of two finite potential wells (of depth $V_0$) each of width $a$ and separated by a distance $b$ (see \Fig{fig:double_well_schematic}). We will call these atoms $A$ and $B$.

A single potential well of width $a$ centered at $x=x_o$ with a depth of $V_0$ has a lowest energy state $\phi(x)$ which can be numerically obtained.
Now our orbital wavefunctions on atoms $A$ and $B$ would be shifted versions of $\phi(x)$. We'd have $\phi_A(x) = \phi(x+(a+b)/2)$ and $\phi_B(x) = \phi(x-(a+b)/2)$. Since $\phi(x)$ would have an exponentially decreasing tail, the overlap $\langle \phi_A|\phi_B \rangle$ would decrease as $b$ increases. What about the ground state for the double well system? According to our assumption, a trial solution is

\beq
\psi_{trial} = \frac{1}{\sqrt{2}} \Big( \phi_A(x) + \phi_B(x)\Big)
\eeq

Let's try and calculate the actual solution.
What will the ground state $\psi_0$ (with energy $E$) of such a system be? 
\begin{figure}[H]
    \centering
    \includegraphics[width = 0.8\linewidth]{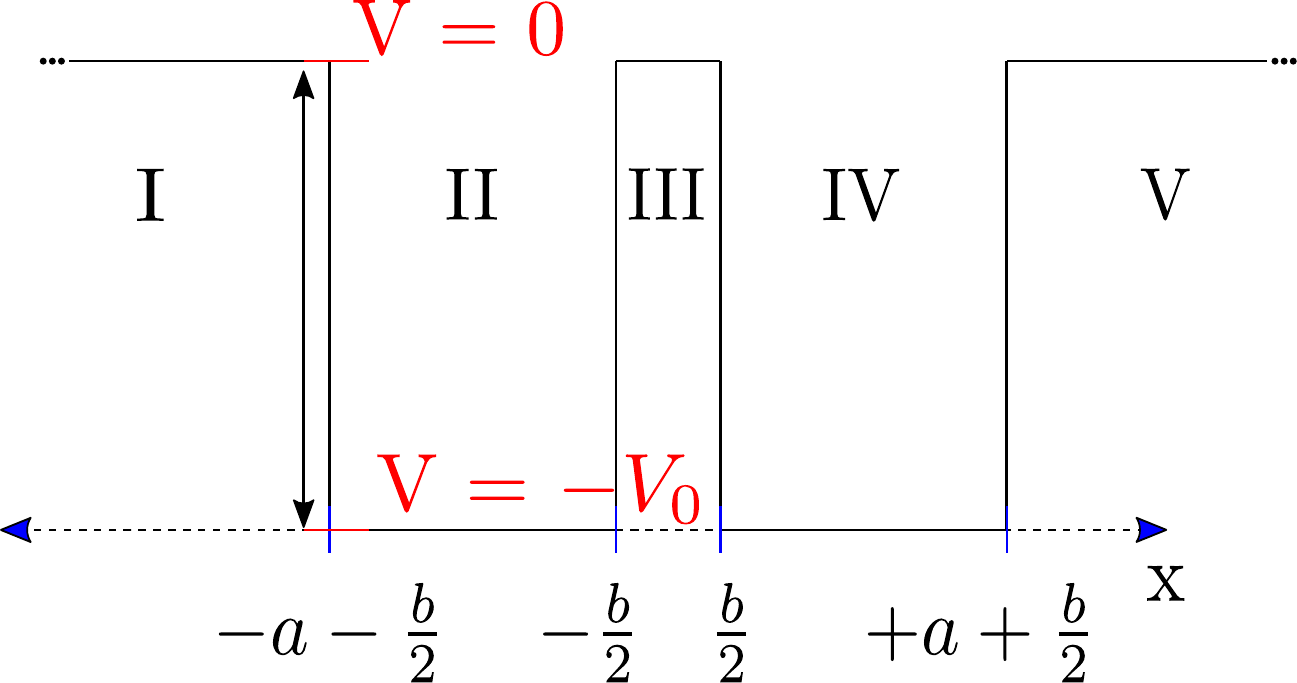}
    \caption{A double potential well of depth $V_{0}$ in one dimension. Width of each well is $a$ and they are separated by a distance $b$. The system is centered at $x=0$. }
    \label{fig:double_well_schematic}
\end{figure}
The potential landscape divides the 1D space into five distinct regions. Looking for solutions with $E<0$ such that in regions of potential $V=-V_o$ we have a standing wave, while for regions with $V=0$ we have a decaying solution we assume the following form of the wavefunction 
\begin{align*}
Ae^{-kx}           &  \ \ \ in\ region\ I\\
B\cos(px) - C\sin(px)     &  \ \ \ in\ region\ II\\
D(e^{kx}+e^{-kx})   &  \ \ \ in\ region\ III\\
B\cos(px) + C\sin(px)     &  \ \ \ in\ region\ IV\\
Ae^{-kx}   &  \ \ \ in\ region\ V
\end{align*}
where $k=\sqrt{\frac{-2mE}{\hbar^2}}$ and $p=\sqrt{\frac{2m(V_0+E)}{\hbar^2}}$. Matching the wavefunction and its derivative at the boundaries of the various regions, we arrive at the following quantization condition
\beq
k^2(1-e^{-kb})\sin(pa) - p^2(1+e^{-kb})\sin(pa) + 2kp\cos(pa) = 0
\label{2well_cond}
\eeq
Looks daunting, but let's see what happens when $b\gg a$. In that case we can drop the $e^{-kb}$ term as it would be much smaller compared to 1. Thus, we get 
\[k^2\sin(pa) - p^2\sin(pa) + 2kp\cos(pa) = 0 \]
Trying to solve this as a quadratic equation in $k$, we get two possible conditions as 
\[ k = p\tan(pa/2) \ \ or \ \ k = -p\cot(pa/2) \]
which are the quantization conditions for the even and odd solutions of a single potential well! One can numerically solve \eqn{2well_cond} for the ground state $\Psi_0$ and plot the wavefunctions for various values of $b$ to see what is happening.

\begin{figure}[H]
    \centering
    \includegraphics[width = 0.8\linewidth]{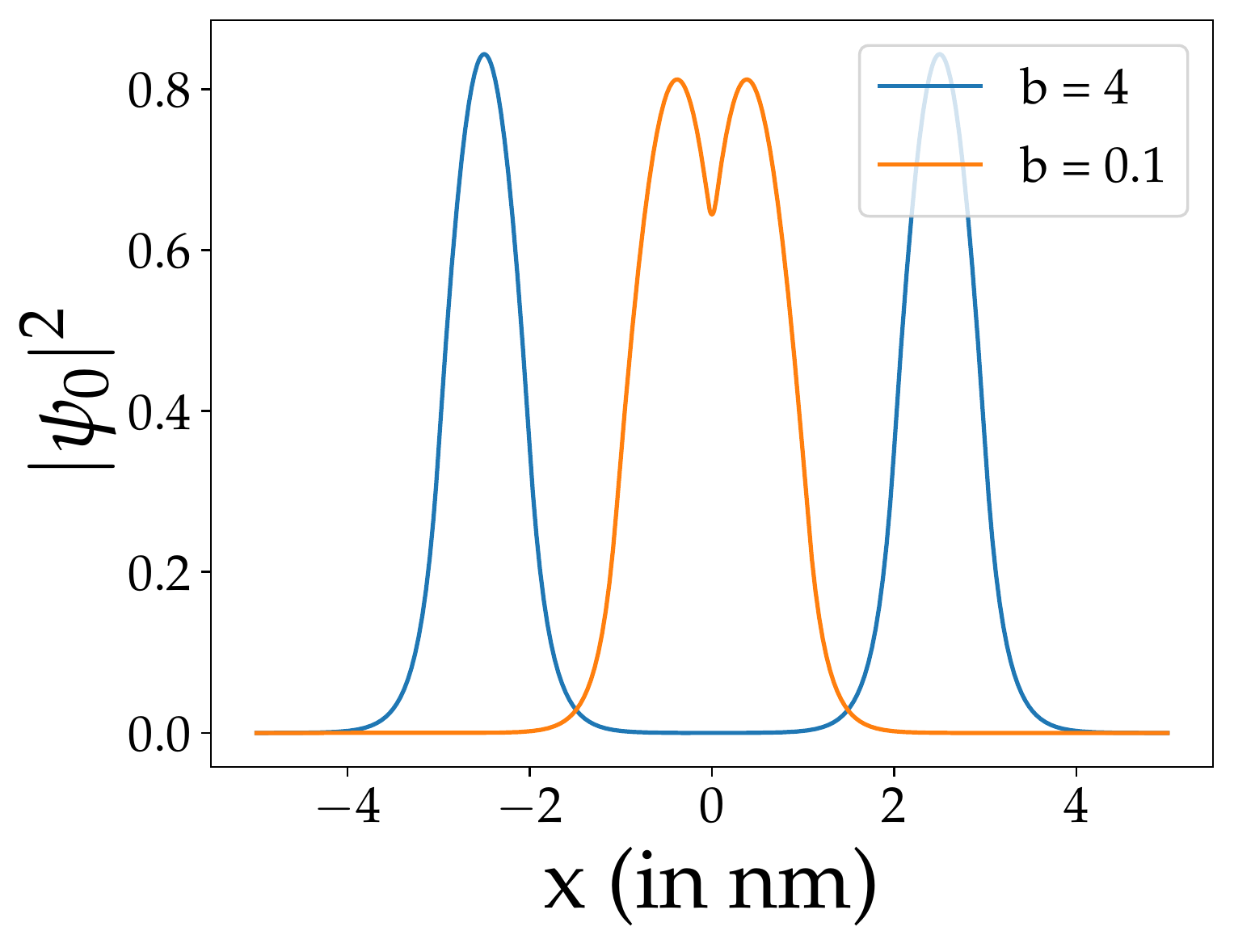}
    \caption{The ground state $\Psi_0$ for $b = 0.1$nm and $b = 4$nm. Width of each well is $a=1$nm and $V_0 = 1.14$eV. }
    \label{fig:double_well_schematic2}
\end{figure}

The \href{https://github.com/hnoend/SSH_codes/blob/main/Figure2_double_quantum_well.m}{code} for \fig{fig:double_well_schematic2} shows the numerical calculations. An interesting exercise is to check what happens to the overlap $\langle \psi_{trial}|\psi_0\rangle$ as a function of $b/a$. We find that for $b \gg a$ the wavefunction for a double potential well problem can be very approximated by a linear combination of two separate potential wells when the distance between these 'atoms' is larger than the size of the atoms.

\end{mdframed}

The above example shows that instead of solving for a differential equation in all of space we can just worry about two coefficients in the tight-binding limit and the Hamiltonian just becomes a $2 \times 2$ matrix. In general, for a system with $N$ sites, each having  $m$ atomic orbitals, the Hamiltonian becomes a $Nm \times Nm$ matrix. This is now a matrix problem which can be diagonalized to get the energy eigenvalues for the electrons and the corresponding wave functions. The hopping parameters which enter this Hamiltonian matrix are often motivated from the chemistry of the material or theoreticians often  choose them to model a particular phenomena. In the following examples (\hyperlink{example2}{Example 2} and \hyperlink{example3}{3}) we will visit this framework for some toy problems. 

\leftHighlight{Example 2}
\begin{mdframed}[style = mystyle, align=center,userdefinedwidth=1.0\columnwidth]
\hypertarget{example2}{{\it \textbf{Example:}}} Sweet spot for the electron \\
Let's model an electron on a molecule of four sites forming a letter $T$. 
\begin{center}
\includegraphics[width=0.2\columnwidth]{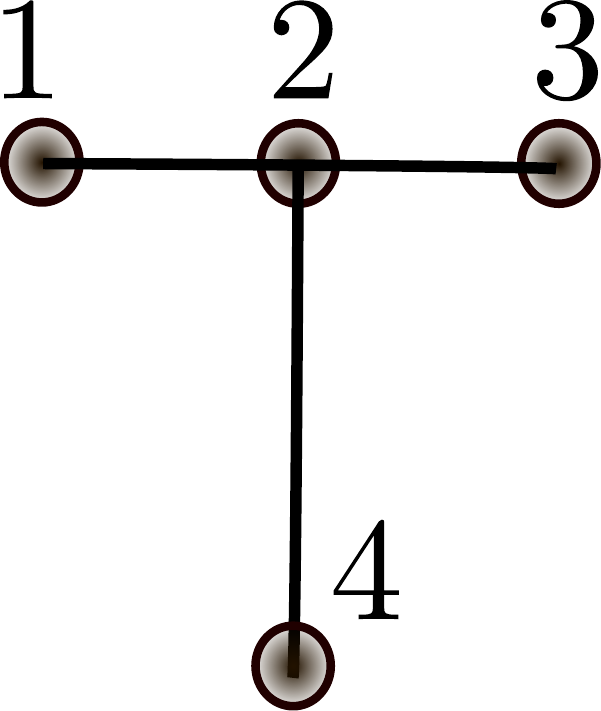}
 \end{center}
Consider fermions hopping on four sites labelled by $1,2,\ldots,4$ and let the hopping amplitude be $t$. 
The Hamiltonian then is determined by the hopping process on the three bonds
\beq
H = t \Big( |1\rangle \langle 2| + |2\rangle \langle  1| \Big) + t \Big( |2\rangle \langle 3| + |3\rangle \langle  2| \Big) + t \Big( |2\rangle \langle 4| + |4\rangle \langle  2| \Big)
 \eeq
The Hamiltonian matrix is then be given by: \\
\begin{center}
\begin{tabular}{|c||c|c|c|c|c|}
\hline
	 & $|1\rangle$ & $|2\rangle$  & $|3\rangle$ & $|4\rangle$ \\ 
	\hline \hline
 $|1\rangle$ & 0 & t & 0  & 0  \\ 
	\hline 
$|2\rangle$ & t	& 0 & t  & t  \\ 
	\hline 
$|3\rangle$ & 0	& t & 0 & 0 \\ 
	\hline
$|4\rangle$ & 0	& t  & 0 & 0  \\ 
	\hline 
\end{tabular} 
 \end{center}
which can be straightforwardly diagonalized. {\it Before} you diagonalize the matrix, where do you think the electron will like to reside the most?
\end{mdframed}

\leftHighlight{Example 3}
\begin{mdframed}[style = mystyle, align=center,userdefinedwidth=1.0\columnwidth]

{
\hypertarget{example3}{{\it \textbf{Example:}}} 
Electron in a 'one-dimensional' solid. 
\begin{center}
   \includegraphics[width=0.6\columnwidth]{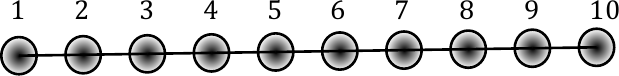}
\end{center}
Consider a lattice of 10 sites labelled by $1,2,\ldots,10$ with a distance of 1 unit between them. We shall consider a hopping term $t$ only between neighbouring sites. In order to make our life simpler, we will assume a periodic boundary condition (PBC) and also consider a hopping term between site 1 and site 10. The reasoning behind this assumption is model an infinite chain where any electron will never encounter a boundary. The Hamiltonian then becomes
\beq
H = t \Big( \sum_{j=1}^9   |j\rangle \langle j+1| + |j+1\rangle \langle j|  \Big) + t\Big( |1\rangle \langle 10| + |10\rangle \langle 1| \Big)
\label{tightH}
\eeq
The Hamiltonian matrix is then given by: 
\begin{center}
\begin{tabular}{|c||c|c|c|c|c|c|c|c|c|c|c|}
\hline
	 & $|1\rangle$ & $|2\rangle$  & $|3\rangle$ & $|4\rangle$ & $|5\rangle$ & $|6\rangle$  & $|7\rangle$ & $|8\rangle$ & $|9\rangle$ & $|10\rangle$ \\ 
	\hline \hline
$|1\rangle$ & 0 & t & 0 & 0 & 0 & 0 & 0 & 0 & 0 & t  \\ 
	\hline
$|2\rangle$ & t & 0 & t & 0 & 0 & 0 & 0 & 0 & 0 & 0  \\ 
	\hline
$|3\rangle$ & 0 & t & 0 & t & 0 & 0 & 0 & 0 & 0 & 0  \\ 
	\hline
$|4\rangle$ & 0 & 0 & t & 0 & t & 0 & 0 & 0 & 0 & 0  \\ 
	\hline
$|5\rangle$ & 0 & 0 & 0 & t & 0 & t & 0 & 0 & 0 & 0  \\ 
	\hline
$|6\rangle$ & 0 & 0 & 0 & 0 & t & 0 & t & 0 & 0 & 0  \\ 
	\hline
$|7\rangle$ & 0 & 0 & 0 & 0 & 0 & t & 0 & t & 0 & 0  \\ 
	\hline
$|8\rangle$ & 0 & 0 & 0 & 0 & 0 & 0 & t & 0 & t & 0  \\ 
	\hline
$|9\rangle$ & 0 & 0 & 0 & 0 & 0 & 0 & 0 & t & 0 & t  \\ 
	\hline
$|10\rangle$ & t & 0 & 0 & 0 & 0 & 0 & 0 & 0 & t & 0  \\ 
	\hline
\end{tabular} 
 \end{center}
How would we diagonalize this? Borrowing from math (the $n_{th}$ roots of unity), you can check that one can write the solution as 
\beq
|\psi\rangle = \sum_{j=1}^{10} \psi_j |j\rangle =  \frac{1}{\sqrt{10}}\sum_{j=1}^{10} e^{ikj} |j\rangle
\label{tightsi}
\eeq
where $i=\sqrt{-1}$, and $k$ is a parameter as in the particle in a box case. Now we need to find a condition on $k$ that would lead to a consistent solution. In the eigenvalue equation, observe that for rows 2 to 9 of the matrix, we get the following equation
\[
te^{ik(j-1)} +te^{ik(j+1)} = E e^{ikj}
\]
which suggests $E = 2t\cos(k)$. However for the first row,
\[
te^{ik(10)} +te^{ik(2)} = E e^{ik(1)}
\]
If we just had $e^{i10k} = 1$ it would solve our problem. Check that this takes care of the last row as well. This leads to quantization of $k$ where $k$ can be set to $\frac{2\pi m}{10}$ where $m$ ranges from 1 to 10, and each $k$ gives a state with energy $2t\cos(k)$ . An interesting observation to make is that, all the coefficients have the same magnitude and hence the electron is equally likely to reside at any of these 10 sites. Since, we were initially looking for 10 eigenvalues and we have found 10 values of $k$, we are done! Usually we often define $k$ within the range of $-\pi$ to $\pi$ also referred to as the momentum space or the Brillouin zone.
}
\end{mdframed}

If you have worked out the examples, they may seem too simple to be of any practical relevance -- but there are good reasons to discuss them.

Let's move to a bit more complicated compound - copper or silicon. But before, we need a small help with notation. We are going to introduce a second quantization notation \cite{2ndq, altland_simons_2006}.  where each site ($i$)  of the system can contribute two states -- (i) vacuum $|0\rangle$ and (ii) one with an electron $|1\rangle$. Operators called creation operators $c^\dagger_i$ creates an electron on a vacuum state and annihilation operators ($c_i$) kill the same.
\bea
c^\dagger_i |0\rangle \rightarrow |1\rangle \\
c_i |1\rangle \rightarrow |0\rangle
\eea
Thus another way of writing the Hamiltonian is where every row and column of the matrix can be associated with creation and annihilation operators on a particular position $i$. When an electron moves from position $i$ to $j$ one could write this as a term.
\beq
\text{electron moves from site $i$ to $j$} \rightarrow c^\dagger_{j} c_i
\eeq
So the Hamiltonians we discussed just now in \hyperlink{example2}{Examples 2} and \hyperlink{example3}{3} can also be written in second quantized form as

\rightHighlight{Second quantized form}
\begin{mdframed}[style = mystyle, align=center,userdefinedwidth=1.0\columnwidth]

{\it for \hyperlink{example2}{Example 2:}} 
\beq
 H = t (c^\dagger_{1} c_2 + c^\dagger_{2} c_1 ) + t (c^\dagger_{2} c_3 + c^\dagger_{3} c_2 ) + t (c^\dagger_{2} c_4 + c^\dagger_{4} c_2 )
\eeq
{\it for \hyperlink{example3}{Example 3}:}
\beq
 H = t \Big( \sum_{j=1}^9 c^\dagger_{j} c_{j+1} + c^\dagger_{j+1} c_{j} \Big) + t (c^\dagger_{1} c_{10} + c^\dagger_{10} c_{1} )
\eeq
\end{mdframed}

Let us move on to some real life materials. Consider copper, usually the material that forms the wires of our home, or salt which we eat. They are crystals and electrons move around in them.

\rightHighlight{Question: Is brass, or steel a crystal? If not, what are they? How does one draw band diagrams for them?}

\begin{figure}[h!]
\begin{centering}
\begin{framed}

    \begin{subfigure}{\linewidth}
    \centering
        \includegraphics[width=1.0\textwidth]{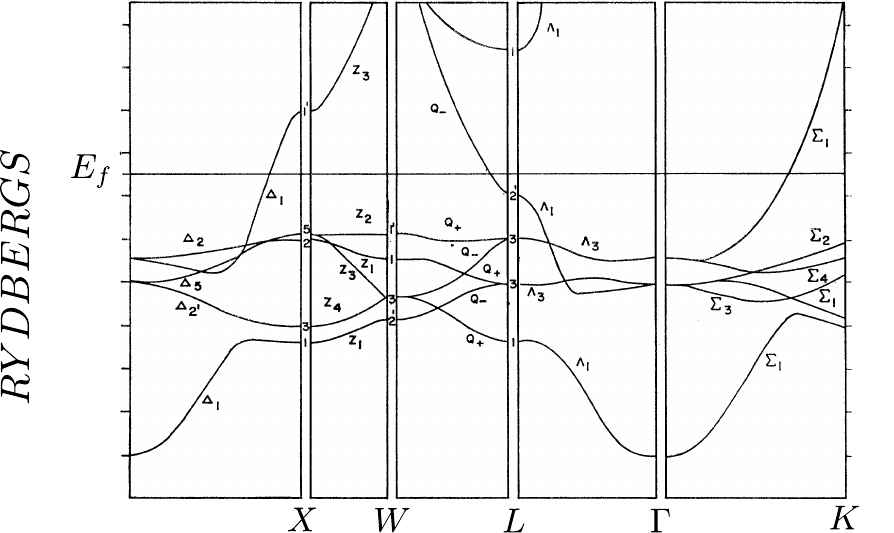}
        \caption*{(a)}
    \end{subfigure} %
    \begin{subfigure}{\linewidth}
    \centering
        \includegraphics[width=1.0\textwidth]{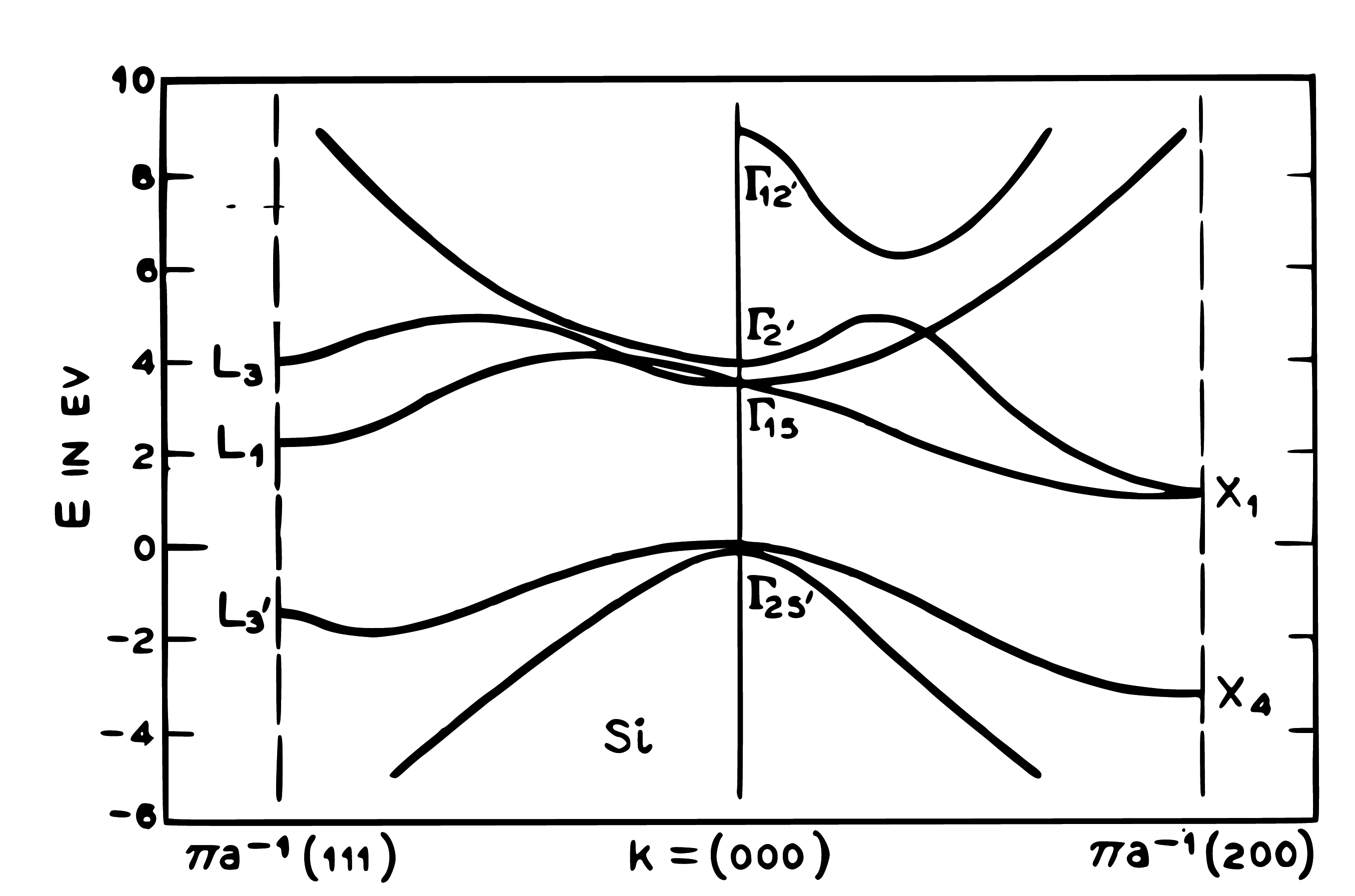}
        \caption*{(b)}
    \end{subfigure} %
 \caption{(a) Band structure of copper \cite{Burdick_PR_1963} which is a metal commonly used in electrical wires. Energies are measured in Rydbergs ($=13.6eV$). Fermi energy (see $E_f$) crosses a band signifying a metal. (b) Band structure of silicon \cite{silicon_bs} which is a semiconductor used extensively to form the substrate of electronic chips. Fermi energy sits at $E=zero$ which lies in the band gap.}
 \label{fig:BandD}
 \end{framed}
\end{centering}
\end{figure}

Then, should just the lattice and saying electrons move from one site to another -- tell us everything about them? The answer is clearly no. How many atoms we have,  how different are they, what are the different hopping parameters -- all forms the first basic skeletal structure to help us start making the first guesses. Real life systems have extremely complicated band structures, which can be found to a great extent of accuracy using modern computational tools where the chemistry of the atoms are taken into account to a great deal. Since the momentum space is in three dimensions -- now we have three momentum points $(k_x, k_y, k_z)$ which are associated with various labels such as $\Gamma, X,W$ etc. A typical band diagram is shown in 
\Fig{fig:BandD}. 
Notice the ways various momentum eigenstates move around in energies as we move around in momentum space. Interestingly at zero energy -- notice in copper we have a line with cuts through, saying that energy required to shift an electron to a new state is $zero$ while for silicon there is a `gap'! Which means one needs a finite amount of energy to add a particle to the next band (see \Fig{fig:BandD}). This is the crucial feature which distinguishes a metal from an insulator.

Let's therefore ask which is the simplest way of modeling a metal or an insulator in one dimension?

\begin{mdframed}[style=mystyle]
{
\leftHighlight{Example 4}
\hypertarget{example4}{{\it \textbf{Example:}}} On-site energies makes things interesting! \\
So far, we have considered the onsite energy to be $zero$. That is if you check \hyperlink{example2}{Example 2}, $\langle j|H|j\rangle$ is taken to be $zero$ for every site $j$. It could have been some constant for every site, but that would just be equivalent to adding a scaled identity matrix to the Hamiltonian. This would not change the eigenvectors and the eigenvalues would just have a constant shift. But now, let's consider the case where we have a onsite energy $\epsilon_0$ for the odd numbered sites and different onsite energy $\epsilon_1$ for the even numbered sites.  The Hamiltonian then becomes
\begin{align}
H = & t \Big( \sum_{j=1}^9   |j\rangle \langle j+1| + |j+1\rangle \langle j|  \Big) + t\Big( |1\rangle \langle 10| + |10\rangle \langle 1| \Big) \nonumber \\ + & \Big( \sum_{j=1}^5 \epsilon_0 |2j-1\rangle \langle 2j-1| + \epsilon_1 |2j\rangle \langle 2j| \Big) 
\end{align}

The Hamiltonian matrix is then given by:
\begin{center}
\begin{tabular}{|c||c|c|c|c|c|c|c|c|c|c|c|}
\hline
	 & $|1\rangle$ & $|2\rangle$  & $|3\rangle$ & $|4\rangle$ & $|5\rangle$ & $|6\rangle$  & $|7\rangle$ & $|8\rangle$ & $|9\rangle$ & $|10\rangle$ \\ 
	\hline \hline
$|1\rangle$ & $\epsilon_0$ & t & 0 & 0 & 0 & 0 & 0 & 0 & 0 & t  \\ 
	\hline
$|2\rangle$ & t & $\epsilon_1$ & t & 0 & 0 & 0 & 0 & 0 & 0 & 0  \\ 
	\hline
$|3\rangle$ & 0 & t & $\epsilon_0$ & t & 0 & 0 & 0 & 0 & 0 & 0  \\ 
	\hline
$|4\rangle$ & 0 & 0 & t & $\epsilon_1$ & t & 0 & 0 & 0 & 0 & 0  \\ 
	\hline
$|5\rangle$ & 0 & 0 & 0 & t & $\epsilon_0$ & t & 0 & 0 & 0 & 0  \\ 
	\hline
$|6\rangle$ & 0 & 0 & 0 & 0 & t & $\epsilon_1$ & t & 0 & 0 & 0  \\ 
	\hline
$|7\rangle$ & 0 & 0 & 0 & 0 & 0 & t & $\epsilon_0$ & t & 0 & 0  \\ 
	\hline
$|8\rangle$ & 0 & 0 & 0 & 0 & 0 & 0 & t & $\epsilon_1$ & t & 0  \\ 
	\hline
$|9\rangle$ & 0 & 0 & 0 & 0 & 0 & 0 & 0 & t & $\epsilon_0$ & t  \\ 
	\hline
$|10\rangle$ & t & 0 & 0 & 0 & 0 & 0 & 0 & 0 & t & $\epsilon_1$  \\ 
	\hline
\end{tabular} 
 \end{center}
 How do we diagonalize this? We need to be able to exploit some periodic pattern. Consider the following solution
 \beq
|\psi\rangle = \frac{1}{\sqrt{10}}\sum_{j=1}^{5} e^{ikj} \Big( a_k |2j-1\rangle +b_k |2j\rangle \Big)
\label{tightsi2}
\eeq
From our previous experience let us also put a constraint on $e^{i5k}$ to be 1. Now, if we observe the eigenvalue equation in the $3^{rd}$ and $4^{th}$ row, we get
\beq
\begin{bmatrix} \epsilon_{0} & t(1+e^{-ik}) \\ t(1+e^{ik})  & \epsilon_{1} \end{bmatrix} \begin{bmatrix} a_k \\ b_k \end{bmatrix}
= E_k \begin{bmatrix} a_k \\ b_k \end{bmatrix}
\label{2b2te}
\eeq
You can check that the same equation is obtained for other pairs of rows as well. For a given $k$ the energy eigenvalues are obtained by solving \eqn{2b2te} . It is a quadratic equation and yields two values of energy. From the condition of $e^{i5k} = 1$, $k$ is $\frac{2\pi m}{5}$ where $m$ ranges from -2 to 2 (keeping $k$ between $-\pi$ and $\pi$). For each value of $k$ we have two values of energy giving 10 eigenvalues as is expected. What's interesting to note now is that $a_k$ and $b_k$ need not have the same magnitude and the electron can now have a preference for one of the sites based on whether it is odd or even numbered! 

Another takeaway is that if we plot energy as a function of $k$, we get 2 bands as opposed to one band in the previous example. If $\epsilon_0$ and $\epsilon_1$ are unequal, then there exists a \q{gap} between these bands.

Now if the system is half-filled i.e.,~the Fermi energy lies in the middle of the two bands (Fermi energy is the energy below which all states are filled), then the gapped system is an insulator while the gapless one is a metal! 

For $\epsilon_0 = \epsilon_1$ there is no gap and for $\epsilon_0 = \epsilon_1 = 0$ this problem is identical to the one in \hyperlink{example3}{Example 3}. But we have two bands here and we only had one band previously. Work out the wavefunctions for both cases and convince yourself that the solutions actually turn out to be the same. Labelling the parameter space in different ways is just an effective tool, to interpret the eigenvalues and eigenvectors we obtain.

}
\end{mdframed}

While the simplistic examples may miss everything about what copper or silicon does, we find that just these two parameters allow us {\it to model} one essential phenomena which we saw in the real material -- of having a gapless and gapped energy spectra.

\section{The curious case of a polymer}

Polymers are compounds where a `large molecule' repeats itself periodically. For instance in the usual polythene which we often use at home to carry fruits and vegetables is a chain made of a molecular unit which looks like this

\begin{center}
\chemfig{-C(-[2]H)(-[6]H)-C(-[2]H)(-[6]H)-}
\end{center}

Here each carbon with a valency of four is strongly covalently bonded with two other carbons and two hydrogen atoms. 
It is an insulating system with quite a large band gap and low melting point.  Let's focus on another polymer now which looks like this:

\begin{center}
\chemfig{-C(-[6]H)=C(-[2]H)-C(-[6]H)=C(-[2]H)-C(-[6]H)=C(-[2]H)-C(-[6]H)=C(-[2]H)-}
\end{center}

Unlike the previous example, here every carbon just has one associated hydrogen atom -- and carbons now share three electrons. Notice that even bonds share two electrons each! But the fact we drew the double bonds on even numbered bonds is just a choice. We could have instead put them on the odd ones. For instance this, 
\begin{center}
\chemfig{=C(-[6]H)-C(-[2]H)=C(-[6]H)-C(-[2]H)=C(-[6]H)-C(-[2]H)=C(-[6]H)-C(-[2]H)=}
\end{center}
In the actual molecules the bond angles are not $90^\circ$ but we are going to ignore this specificity in our discussion here. Both these configurations have the same energies, and therefore in a system they would be equally likely. Let's call them version PA-$\alpha$, and PA-$\beta$. Now, polymers are long chains where such units just attach to each other. So it may seem that we can have two possible ways of attaching them:
\beq
\text{Junction 1:}\ \ \boxed{PA-\alpha}--\boxed{PA-\alpha} \ or \ \boxed{PA-\beta}--\boxed{PA-\beta}
\label{junction_type1}
\eeq
\beq
\text{Junction 2:}\ \
\boxed{PA-\alpha}--\boxed{PA-\beta} \ or \ \boxed{PA-\beta}--\boxed{PA-\alpha} 
\label{junction_type2}
\eeq

{\it Are these two ways really the same?} \\

History of this polymer is quite amazing. The physics of attaching these various chains and that it was far from trivial is tied to some remarkable work which showed an anomalous signal in a magnetic susceptibility experiment.

\rightHighlight{Benzene too has a similar bond structure with even and odd bonds, just that it has 6 atoms of carbon. Do they have two different versions of the compound -- like polyacetelyne?
If yes, {\it why?} If not, {\it why not?}
Answer: It doesn't. Now think about why not.}
W.P. Su who was then a graduate student in UPenn, John Schrieffer, who had received the Nobel prize a few years earlier for his work on superconductivity and was a theorist in the same department, and A.J. Heeger -- who was an expert on experiments on polymers, again working there -- came together to understand this and wrote a series of papers on just trying to understand this simple looking polymer that had nothing but carbon and hydrogens. While these set of papers founded a new sub-field and ideas that stemmed topological physics -- Heeger continued his investigations on polymers. He eventually won the Nobel prize in Chemistry in 2000 for his work on 
polymers, {\it in particular} polyacetelyne -- the polymer you see above. In case, you haven't appreciated -- just sit back and think for a minute the impact these polymers have had on physics, chemistry, and our whole planet! And all of it had people scribbling funny bond diagrams.

\section{Welcome to the SSH chain}

In the last few sections we looked at a few tight binding Hamiltonians. We also noticed on how to think about metals and insulators.

{\it Let's recap.} Basically given a Hamiltonian we look at its energy spectrum. This is often done assuming periodic boundary conditions and invoking the Bloch wave functions. The energy levels which represent allowed states, when closely spaced, can form bands. 

\leftHighlight{Consider polythene -- whose band gap was $\sim 8eV$ -- how much voltage would one need to apply for the electron to conduct?}

Often when the number of electrons in a lattice is half of the number of atoms --  we have a system which we call {\it half filling}. The energy upto which the electrons gets filled is called Fermi energy or the chemical potential. The lower~(filled) bands are called the valence bands, and the upper bands~(empty) are termed the conduction bands. For {\it insulators} the chemical potential or the Fermi energy lies in the band gap and hence need sufficient (often humongous amounts of) energy to excite electrons and get the material to conduct. 
For metals such as copper, there is no gap, these bands criss-cross each other and therefore any little energy allows for electrons to move. 

So what about polyacetelyne? 

\begin{figure}[t]
    \centering
    \begin{framed}

    \includegraphics[width=1.0\linewidth]{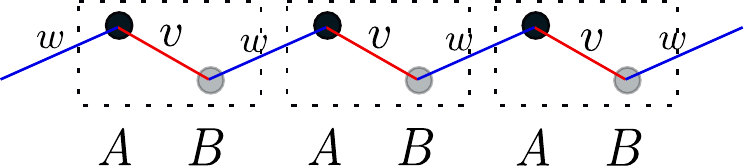}\\
  \caption{Schematic for the SSH Model. A and B within the same unit cell are coupled with the intracellular hopping parameter $v$, and adjacent unit cells are coupled through the intercellular hopping parameter $w$.}
  \label{SSH_schematic}
      \end{framed}
\end{figure}

Su-Schrieffer-Heeger (SSH) modelled the chain saying we have two kinds of atoms -- A and B (see \Fig{SSH_schematic}). Now even though both atoms are really carbon, their bond-strengths aren't the same. We know the system likes to either have stronger even or odd bonds and can have two configurations. So consider the Hamiltonian has parameters $w$ and $v$ which models the bond strengths.

The Hamiltonian is then given by
\beq
 H = v\sum_{j = 1}^{N/2}|2j-1\rangle\langle 2j| +  w\sum_{j = 1}^{N/2}|2j+1\rangle\langle 2j| +  h.c.
 \label{ham}
 \eeq
 
 In second quantized form this becomes 
 
 \beq
 H = v\sum_{j = 1}^{N/2} c_{2j-1}^{\dagger}c_{2j} +  w\sum_{j = 1}^{N/2}c_{2j+1}^{\dagger}c_{2j} +  h.c.
 \label{2qham}
 \eeq
 
 Given two distinct atoms it is helpful to consider a unit cell comprising of two atoms. The first atom is then referred to as the $A$ site of the unit cell, and the second atom is referred to as the $B$ site of the unit cell. The Hamiltonian in \eqn{ham}, is written in such a way that the odd numbered sites correspond to the $A$ sites and the even numbered sites correspond to the $B$ sites in \Fig{SSH_schematic}. The Hamiltonian after relabelling this numbering to consider unit cells with their $A$ and $B$ sites is 
 \beq
 H = v\sum_{j = 1}^{N/2}|j,A\rangle\langle j,B| +  w\sum_{j = 1}^{N/2}|j+1,A\rangle\langle j,B| +  h.c.
 \label{hamAB}
 \eeq
 
  \beq
 H = v\sum_{j = 1}^{L} c_{j,A}^{\dagger}c_{j,B} +  w\sum_{j = 1}^{L}c_{j+1,A}^{\dagger}c_{j,B} +  h.c.
 \label{2qhamAB}
 \eeq
 
 where $L$ is the number of unit cells. The number of atoms are $N$ and this gives us the relation $N=2L$ and $j$ labels the unit cells.

 \subsection{Band Diagram}

Given the Hamiltonian for the SSH chain,  we can Fourier transform and diagonalize the Hamiltonian. 
Solving in a similar fashion as in \hyperlink{example4}{Example 4} we get the following $2\times2$ matrix eigenvalue problem

\beq
\begin{bmatrix} 0 & (v+ w e^{-ik}) \\ (v + we^{ik})  & 0 \end{bmatrix} \begin{bmatrix} a_k \\ b_k \end{bmatrix}
= E_k \begin{bmatrix} a_k \\ b_k \end{bmatrix}
\label{ssh_2mat}
\eeq

The energy eigenvalues turn out to be
\beq
E_k = \pm \sqrt{v^2 + w^2 + 2v w \cos(k) \ }
\eeq

\begin{figure}[t]
    \centering
    \begin{framed}

    \includegraphics[width = 0.8\linewidth]{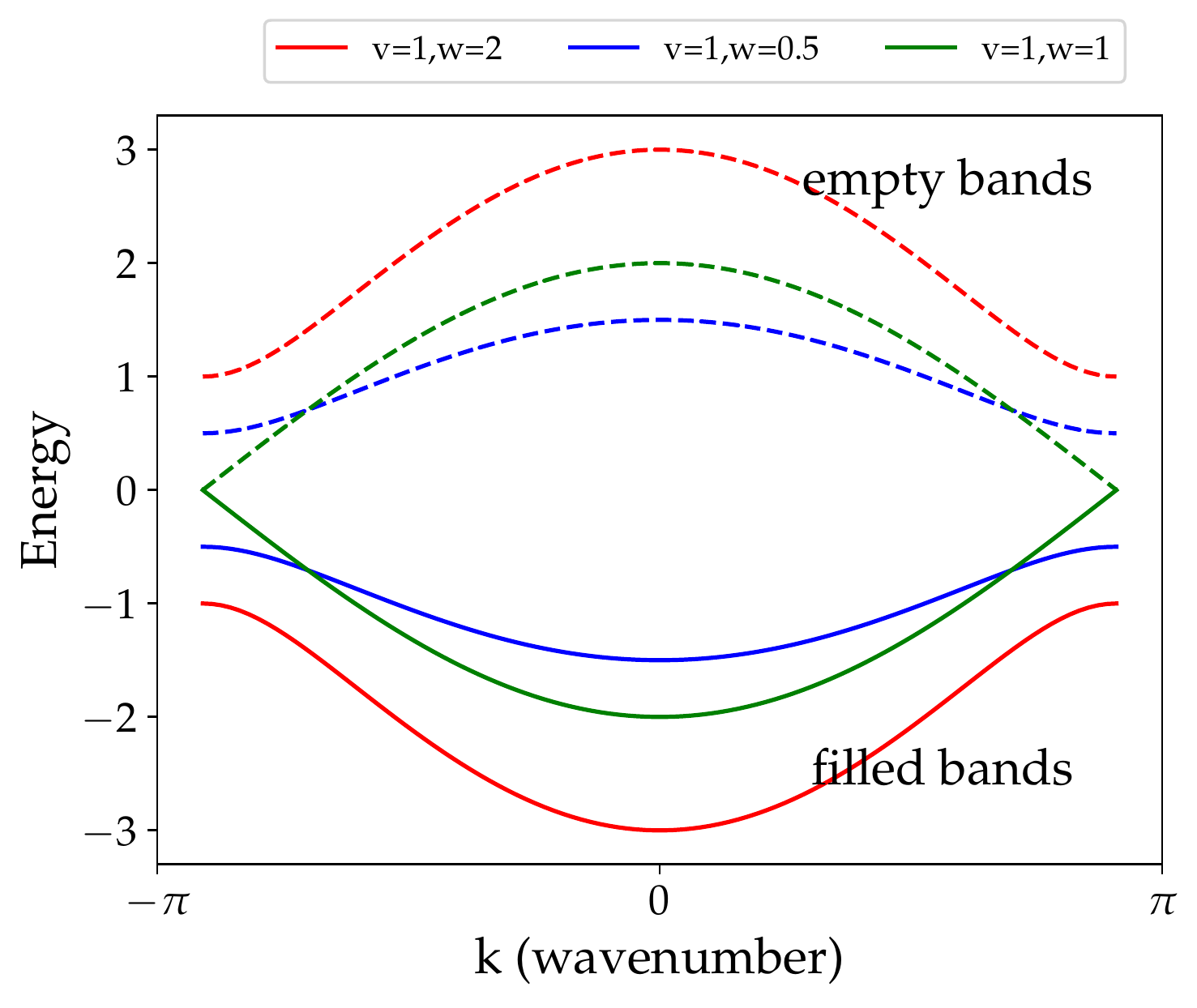}
  \caption{Band diagram of the SSH chain for various parameter values}
  \label{SSH_band_diagram}
      
    \end{framed}
\end{figure}

One thing that might strike as odd at this place is that the energy expression seems to be symmetric with respect to $v$ and $w$ implying that PA-$\alpha$ and PA-$\beta$ are the same when viewed from this lens.
The band diagram for some values of $v$ and $w$ is illustrated in \Fig{SSH_band_diagram}

We now track the band diagram as $w$ and $v$ is altered. We keep in mind that we have a system which is half-filled which means the Fermi energy is pinned to $zero$. We find the system is always an insulator when $v \neq w$ but becomes a metal when $v$ equals $w$! 
 
\subsection{Modelling the junction of polyacetelyne} 
 
 Now for polyacetelyne -- we had two types of phases -- the one with even and odd bonds stronger and then two ways of making junctions out of them. We are now at the stage of asking whether all of these junctions behave similarly. 
 
 \begin{figure}[!htb]
    \centering
    \begin{framed}
    \begin{subfigure}{\linewidth}
    \centering
  \includegraphics[width=0.9\linewidth]{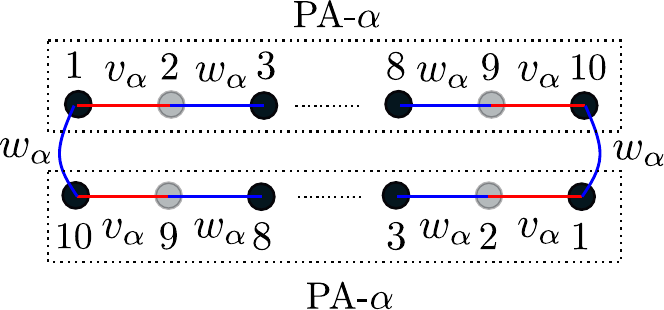}
  \caption{}
  \label{PA_junction1}
    \end{subfigure}
    \begin{subfigure}{\linewidth}
    \centering
  \includegraphics[width=0.9\linewidth]{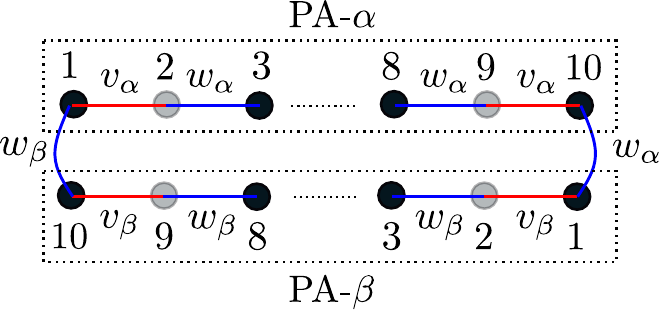}
  \caption{}
  \label{PA_junction2}
    \end{subfigure}
    \caption{Modeling the polyactelyne chain junctions (a) Connecting the last atom of a $PA\mhyphen\alpha$ chain to the first atom of another $PA\mhyphen\alpha$ chain. (b) Connecting the last atom of a $PA\mhyphen\alpha$ chain to the first atom of a $PA\mhyphen\beta$ chain.}
    \label{PA_AB}
        \end{framed}
\end{figure}
 
\begin{figure}[t]
    \centering
    \begin{framed}
    \includegraphics[width=1.0\columnwidth]{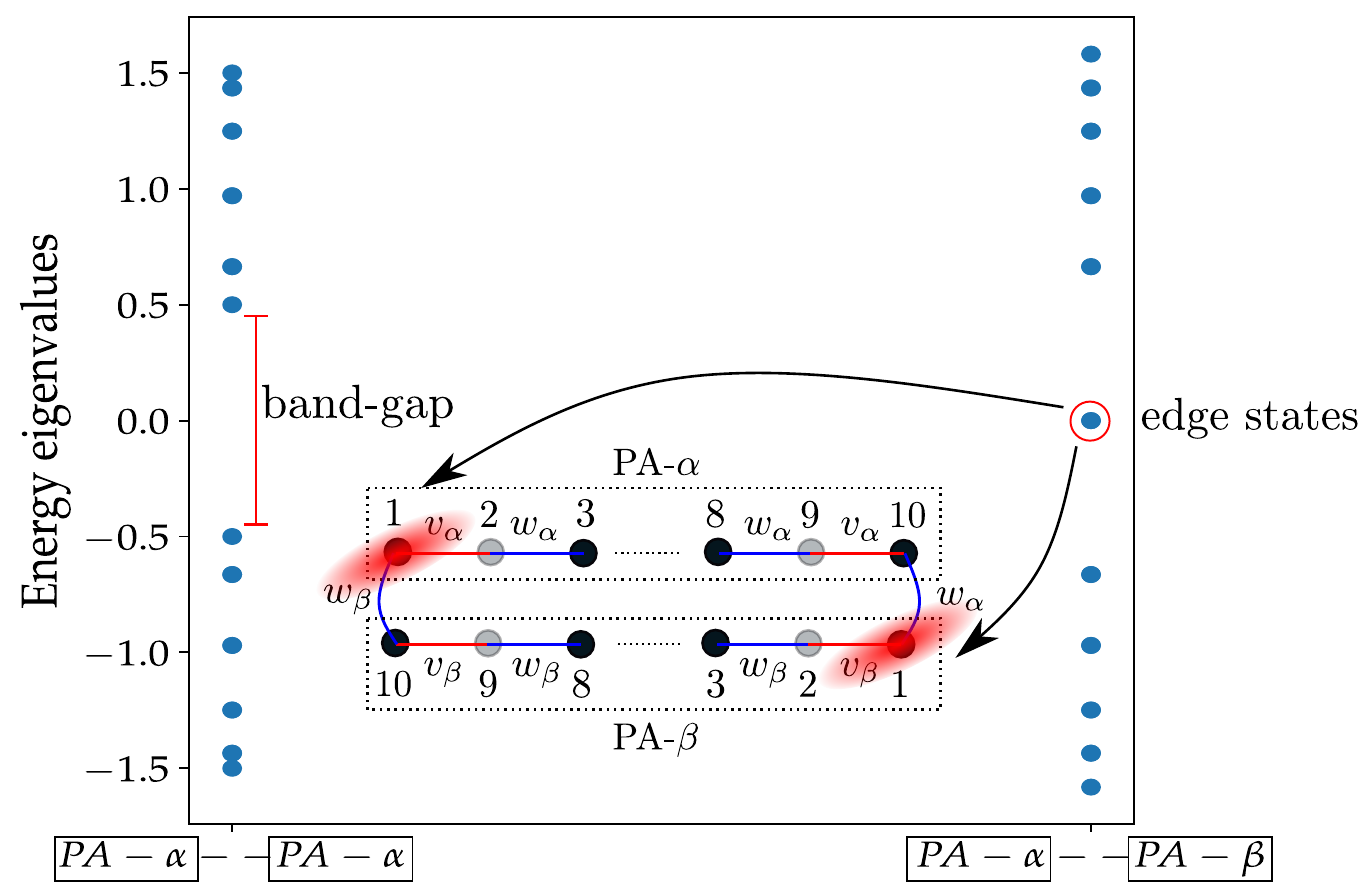}
        \caption{Energy eigenvalue spectrum for the two setups formed in \Fig{PA_AB} has been shown here. Edge states occur when $PA-\alpha$ and $PA-\beta$ are joined end-to-end.}
  \label{fig:junction_eigenvalues}
\end{framed}

\end{figure}

 Let's model an odd and even numbered bonds between carbon atoms via parameters $v$ and $w$ respectively. Let's have $PA\mhyphen\alpha$ with $v_\alpha=1, w_\alpha=0.5$ and $PA\mhyphen\beta$ with $v_\beta=0.5, w_\beta=1.0$ and each having 10 carbon atoms. As seen in \fig{PA_AB} trying to connect them with periodic boundary conditions, we have two possible cases. Connecting two $PA\mhyphen\alpha$s or two $PA\mhyphen\beta$s would be equivalent just giving us a  regular chain with the first junction type from \eqn{junction_type1}.  Connecting two different versions of PA as seen in \Fig{PA_junction2} would give us the second junction type from \eqn{junction_type2}. We now have two consecutive double bonds at one end and two consecutive single bonds at the other end.
 The energy eigenvalues for the two cases have been plotted in \Fig{fig:junction_eigenvalues}.

 Notice in \fig{fig:junction_eigenvalues} for the second type of chain the eigenvalues show some energy states close to zero energy. In fact if we look at the probability density of the electron it resides on one of these junctions. So these electrons just like to reside on the junction and not worry about other things! 
 
 This is a fantastic result -- if for a large polymer which is just made of one version of the PA - there are no such localized electrons, but say when such junctions are present at a finite density -- that means a bunch of electrons just sit on the junctions. It is these electrons which gives an anomalous signal in the experiments and was the discovery SSH gang made.

 \subsection{A single chain with open boundaries}
 
 While the polyacetelyne gave us the motivation to model the junctions -- to think why bond strengths can be different. It really gave us a toy model to play with. We can now forget about the compound, and start finding out what does a single chain tells us. What happens if we now think about $w$ and $v$ as some parameters which can be tuned in a lab.

\begin{figure}[!htb]
    \centering
    \begin{framed}
    \begin{subfigure}{\linewidth}
    \centering
  \includegraphics[width=0.9\linewidth]{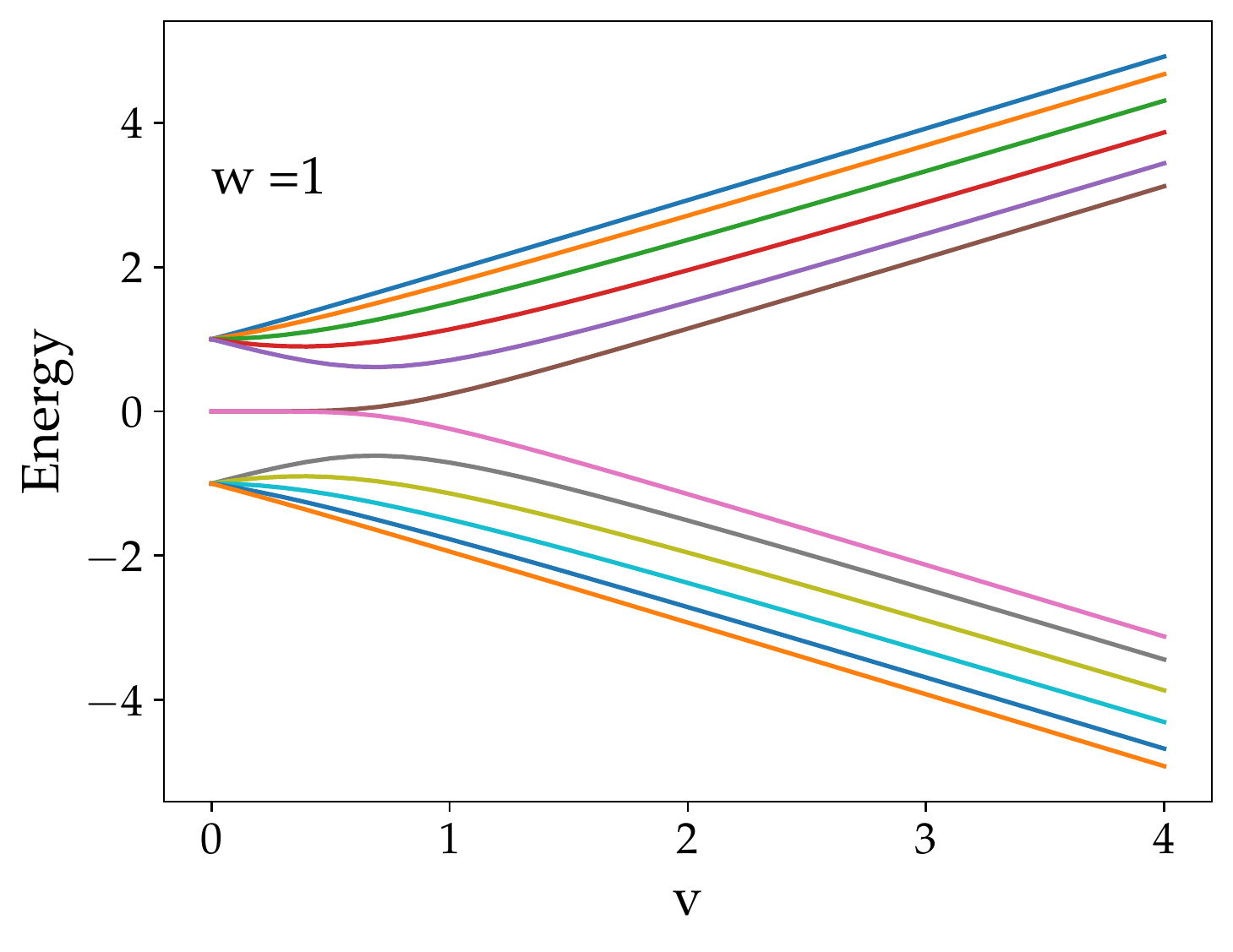}
  \caption{}
  \label{fig:energy_spectrum}
    \end{subfigure}
    \begin{subfigure}{\linewidth}
    \centering
  \includegraphics[width=0.9\linewidth]{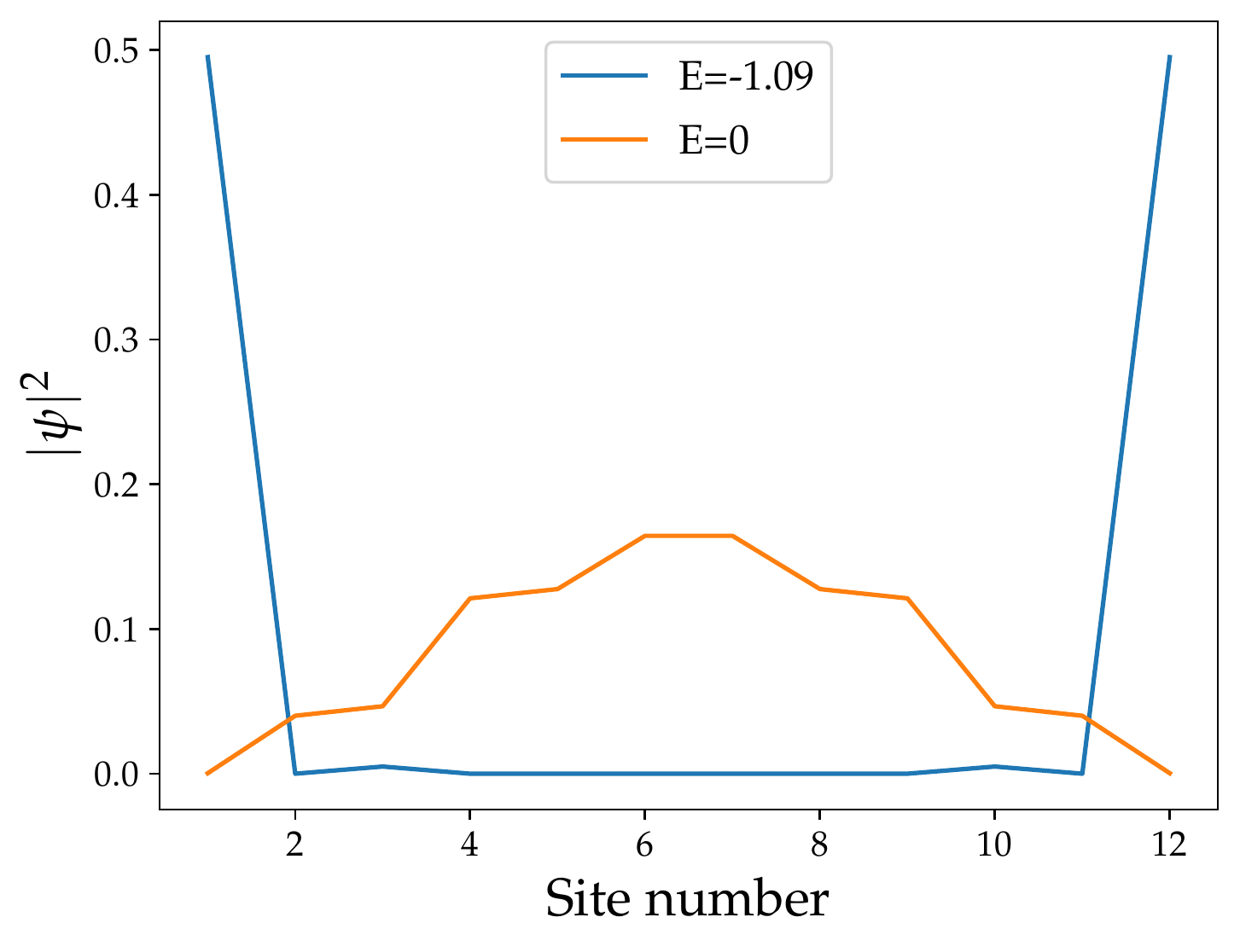}
  \caption{}
  \label{psi2_ssh}
    \end{subfigure}
    \caption{(a) Energy spectrum of a SSH chain with 12 atoms and open boundary conditions. We note that the zero energy states exist in the topological regime and (b) $|\Psi|^2$ for two states, in the topological regime ($v=0.1$, $w=1$). The $E=0$ state is highly localized on the edge sites.}
    \label{fig:my_label_sshloc}
        \end{framed}
\end{figure}

Let us look at the energy spectrum, for $w$ = 1 and $v$ varying from 1 to 4 in \Fig{fig:energy_spectrum}. If we focus on small values of $v$, we see that the spectrum resembles an insulator except for two states with energy very close to \textit{zero}. If we keep the Fermi energy close to \textit{zero}, we can excite these two edge states. But what do these states correspond to? Looking at the probability density of these states, we notice that it is highly localized on the edge sites as shown in \Fig{psi2_ssh}! So although the material is insulating in the bulk, it has some anomalous states at the edges.

To obtain analytic expressions for the wave functions in a arbitrary open system is in general hard. Let's see if a \textit{zero} energy wavefunction exists in this system \cite{short_topo} --

Consider the following form for the wavefunction
\[
|\psi\rangle = \sum_{j=1}^{L} \Big( a_j |j,A\rangle +b_j |j,B\rangle \Big)
\]

For a \textit{zero} energy solution, we need to find coefficients $a_j$ and $b_j$ such that $H|\psi \rangle = 0$, where $H$ is the Hamiltonian for the open SSH chain (see \eqn{2qham}). Looking at each row, we get the following set of equations
\[
va_j + wa_{j+1} = 0 \ \ \ \ \ for \ j = 1\ldots L-1
\]
\[
wb_j + vb_{j+1} = 0 \ \ \ \ \ for \ \ j = 1\ldots L-1
\]
\[
va_{L} = 0  \ \ \ \ \ vb_1 = 0
\]

This would give us the following recurrence relations and a condition
\bea
 a_{j} = \frac{-w}{v}a_{j+1} \ \ \ \ \ for \ j = 1\ldots L-1 \\
 b_{j+1} = \frac{-w}{v}b_{j} \ \ \ \ \ for \ j = 1\ldots L-1 \\
 a_{L} = b_1 = 0
 \label{anb10}
\eea

Looking at these equations one might think that having $b_1 = 0$ would force all other $b_js$ to 0 (similarly for $a_js$) and that no such solution having energy to be exactly \textit{zero} could exist! 

But for large $L$ we could try and cook up exponentially decaying coefficients  to get an approximate \textit{zero} energy solution. Let us define a parameter $\xi$ as follows
\[
\xi = \frac{1}{\log{|w|}-\log{|v|}}
\]

From the recurrence relations we now get 
\bea
|a_{N/2}| = e^{-\frac{L-1}{\xi}}|a_1| \\
|b_{1}| = e^{-\frac{L-1}{\xi}}|b_{L}| 
\label{open_ssh_exp}
\eea

If $\xi$ were negative, as $L \to \infty$ we could satisfy \eqn{anb10}! And $\xi$ is indeed negative if $v<w$! 

Great, so now we can have two possible solutions, one by setting $a_1$ to 0 and $b_{L}$ as non-zero and vice-versa. What do these solutions correspond to? Having $a_1$ as non-zero corresponds to the electron having the highest probability of residing in site $|j=1,A\rangle$ and then decaying exponentially as we move to the right. While having $b_{L}$ as non-zero corresponds to the electron having the highest probability of residing in site $|j=L,B\rangle$ and then decaying exponentially as we move the left. For this reason $\xi$ is also termed as the \q{localization length}. The fun part is that both these solutions reside at the \q{edges}!

\section{It's not just a phase}

In the above analysis we noticed something quite strange. The two Hamiltonians for the polyacetalyne versions seemed quite the same. The eigenvalues in the periodic system were in fact exactly the same -- both describing insulators at half-filling. Yet, in an open system one of them hosts boundary modes and the other doesn't. Are these phases of electrons really the same? Or are these insulators different?

Turns out they {\it are} very subtly different. And this is what we discuss next.

\subsection{Polarization}

Insulators, while electrically may not conduct -- they may still be different. This is because, insulators polarize in presence of electric fields -- and the way they polarize can be different. This is the electrostatic response to an electric field, and to understand its effect we first need to understand how to think about polarization within a tight-binding framework.

Notice we have a charge neutral system so the number of electrons and number of positive charges should be the same. This means for the SSH system, we should effectively consider that every two atoms has {\it one} positive charge to balance one shared electron. This of course assumes we have ignored all other (protons') charges and deeper electron orbitals.

\begin{figure}[t]
    \centering
    \begin{framed}
    \includegraphics[width=0.8\columnwidth]{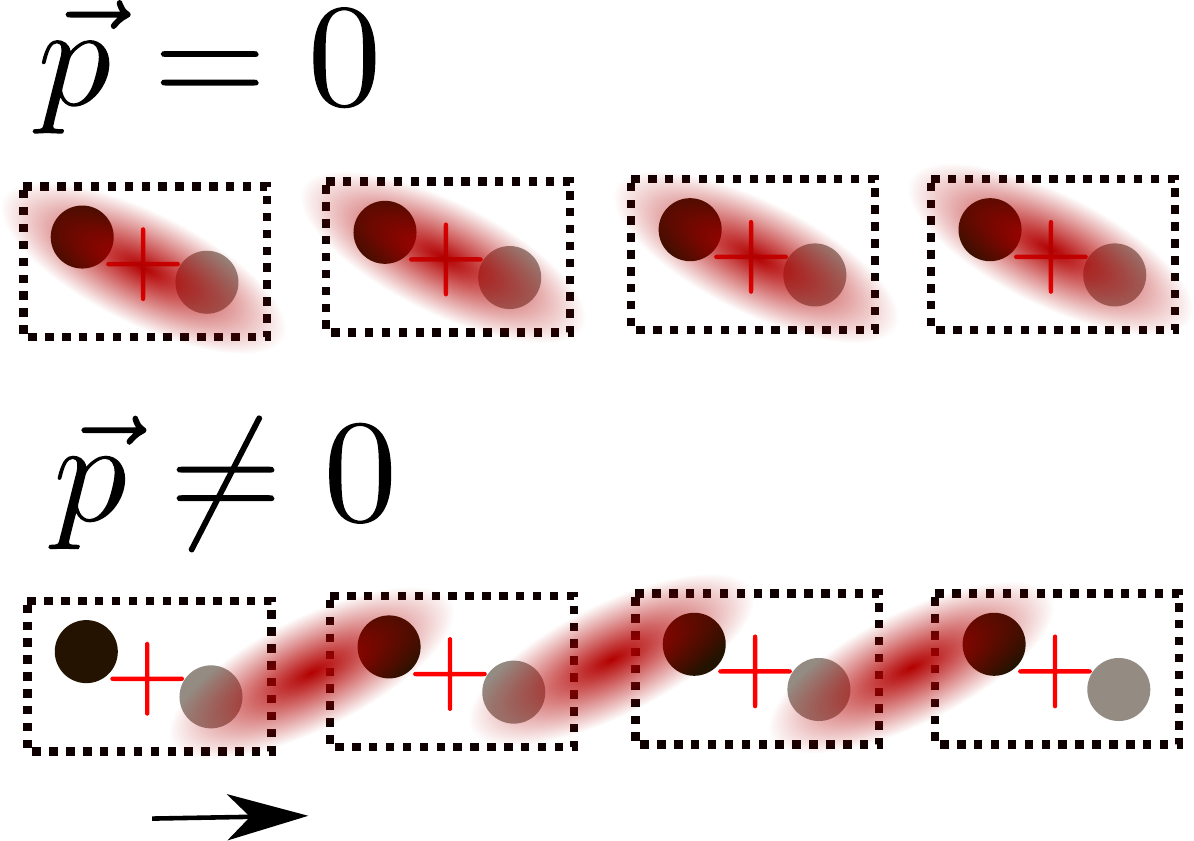}
    \caption{Two insulators, but with different dipole moments.}
    \label{fig:dipole}
        \end{framed}
\end{figure}

Consider two systems (see \Fig{fig:dipole}) where electrons (represented by a shaded orange ellipse) and positive charges shown in squared rectangles have the same centre of mass. Notice this has no dipole moment, however in the other case where the electron cloud is slight shifted -- here the system has a dipole moment.

Both these systems are insulators, given electrons are localized. And their wavefunctions must have an observable that distinguishes them. Clearly, energies to excitations etc.~cannot. One thing which potentially can is the centre of mass of the electrons. Consider operator $n_{jA}$ which measures the electron density on the $A^{th}$ site of the $j^{th}$ unit cell (see \Fig{SSH_schematic}) - the centre of mass is measured by 

\beq
\hat{M} = \sum_{j=1,\ldots,L} j (n_{jA}+n_{jB})
\eeq

However we run into trouble, how can one think about this in a periodic system? After all, the site we label $1$ or $L$ is just a choice we made. And physically meaningful quantities, such as observables shouldn't depend on such choices. We therefore place the chain on a ring -- and calculate the center of mass of the electrons on a chain using this operator instead --

\beq
\hat{O} = \exp(i \frac{2\pi\hat{M}}{L})= \exp \Big(i \frac{2\pi}{L} \sum_{j=1,\ldots,L} j (n_{jA}+n_{jB}) \Big)
\eeq

\begin{figure}[t]
    \centering
    \begin{framed}
    \includegraphics[width=0.8\columnwidth]{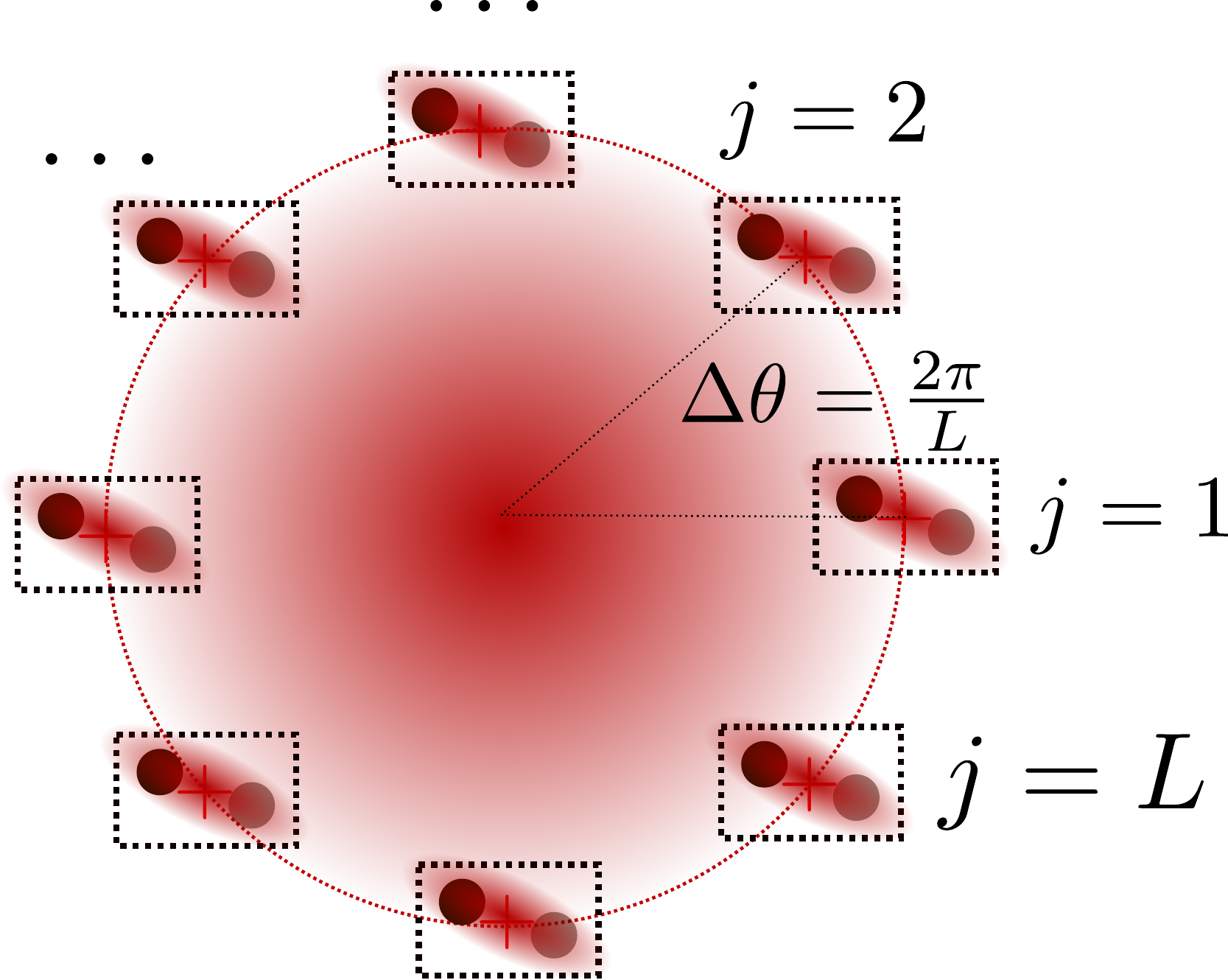}
    \caption{Placing the chain on a ring}
    \label{fig:dipole3}
    \end{framed}
\end{figure}

On a circle, this operator is well defined and doesn't change its eigenvalues for a half-filled system. This operator, as we will see below will play a crucial role in discovering the physics of the SSH chain. However before we calculate these explicitly let us develop the understanding of this operator a bit more.

\subsection{Applying a small electric field}

In the last section we looked at $\hat{O}$ as a centre of mass for electrons, but there is another way of thinking about this. Imagine we wish to  apply a small electric field to an electronic system, in general we could do this by adding a potential of the form to a discrete Hamiltonian
\beq
V =  v \sum_j  j n_j
\eeq
where the electric field in dimensionless units is just $v$. But then we run into the same trouble -- where we do not know how to keep periodicity intact. So we remind ourselves that electric fields can be applied in another way -- by threading a magnetic flux since 
\beq
E = -\frac{\partial A}{\partial t}
\eeq
We need to therefore understand in a tight-binding framework how should one electron sense a magnetic field. As electron moves from a site $i$ to $j$ it should sense a phase $\int_{\br_i}^{\br_j} \bA.d\br$. This often goes by the name of Peierl's substitution \cite{Sakurai} \cite{peierl}
\beq
t_{i j}:=t_{i j} e^{-i \frac{e}{\hbar} \int_{i}^{j} \mathbf{A} \cdot d \mathbf{l}}
\label{Peierl}
\eeq

For instance consider a tight-binding chain with a one of kind of atom as we saw in \hyperlink{example4}{Example 4} whose Hamiltonian is given by

\beq
\hat{H}=\sum_{i=1}^{L}-\left[t c_{i+1}^{\dagger} c_{i}+t c_{i}^{\dagger} c_{i+1}\right]
\eeq

An operator given by 
\beq
\hat{O} = \exp \Big(i \frac{\phi}{L} \sum_{j=1,\ldots,L} j n_{j} \Big)
\eeq

changes
 \beq
c_{l}^{\dagger} \rightarrow e^{i \frac{\phi}{\Phi_{0}}  l}c_{l}^{\dagger} \quad \quad c_{l+1}^{\dagger} c_{l} \rightarrow e^{i(\phi / L) }
\eeq
implementing a Ahranov-Bohm phase \cite{Griffiths2004Introduction} arising from a flux $\phi$ in a ring. Therefore, for $\phi=2\pi$ this really becomes the operator for dipole moment that we were looking for. This implies that as an electron moves through the ring it gains a phase of $2\pi$.

\begin{mdframed}[style=mystyle]
\rightHighlight{Example 5}
\hypertarget{example5}{{\it\textbf{Example:}}} Let us consider a 4 atom square as shown in \Fig{fig:AB_square}.
\begin{figure}[H]
    \centering
    \includegraphics[width=0.5\columnwidth]{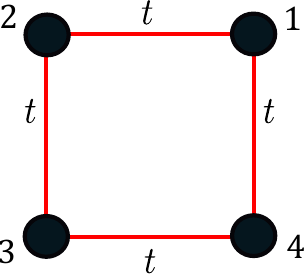}
    \caption{4 atom periodic chain in a magnetic flux}
    \label{fig:AB_square}
\end{figure}
The Hamiltonian of this setup would be:
\beq
H = \begin{bmatrix} 
	 0 & t & 0 & t^\ast \\
	 t^\ast & 0 & t & 0 \\
	 0 & t^\ast & 0 & t \\
	 t & 0 & t^\ast & 0 \\
	 \end{bmatrix}
	\label{eqn:Hamiltonian_4_square}
\eeq
Here, $t$ is the hopping parameter between nearest neighbours of the chain which can be set to 1 without loss of generality. $t^\ast$ denotes the complex conjugate of $t$. Now if a magnetic flux $\phi$ is passed through this setup then the hopping parameters would change as given by Peierl's substitution. Consider $\phi$ = $ \phi_{0}$ where $\phi_{0}$ is a constant. If the length of the entire chain is $L$ (i.e., distance between adjacent sites is $L/4$), then the magnetic vector potential is $A$ = $\phi_{0}/ L$. Now, using \eqn{Peierl}, we get that the hopping parameter modifies to:
\beq
t:= te^{i \frac{\phi_{0}}{4}}
\eeq

Let us now plot the energy eigenvalues of the Hamiltonian under consideration as $\phi_{0}$ changes from 0 to $2\pi$.

\begin{figure}[H]
    \centering
    \includegraphics[width=\linewidth]{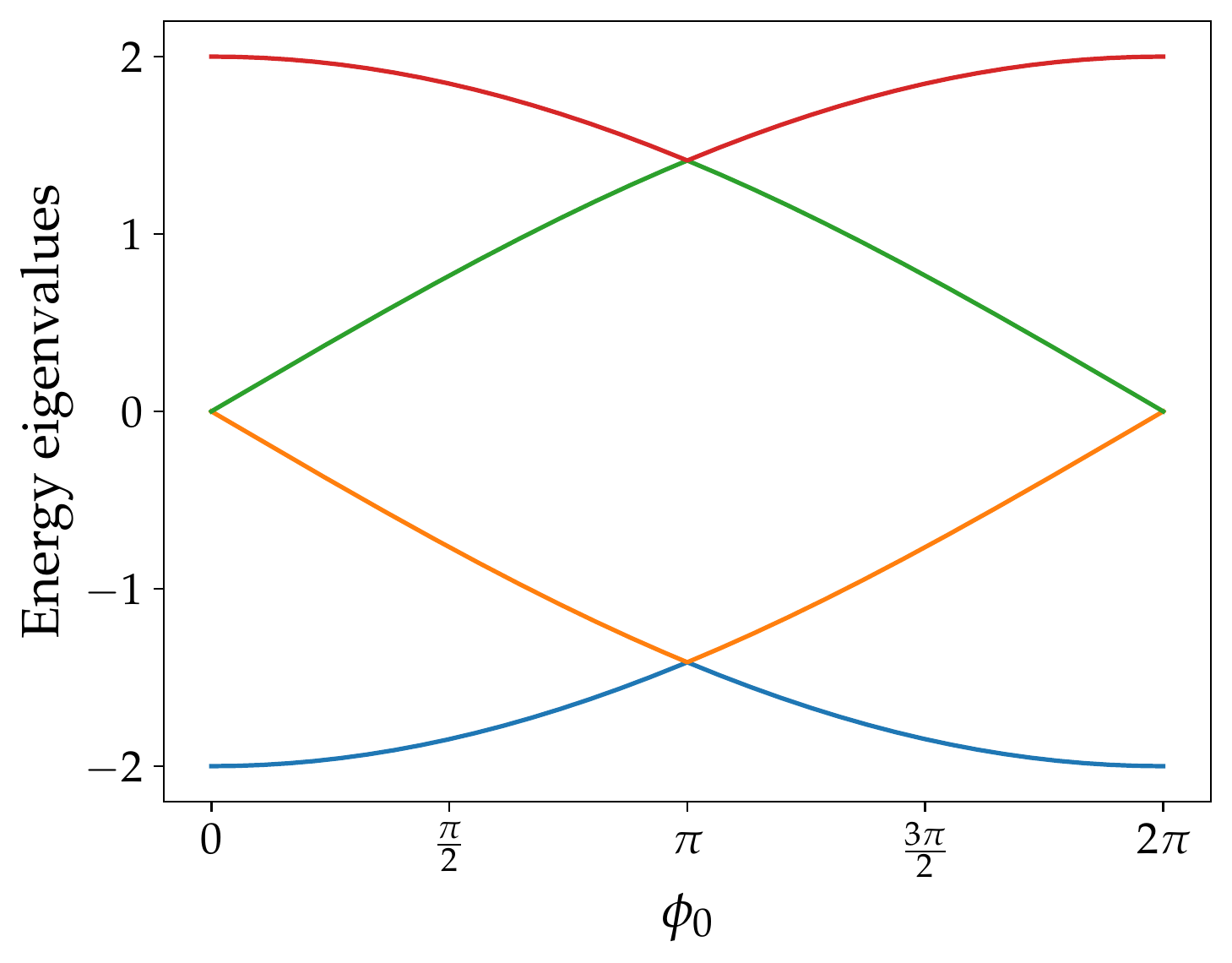}
    \caption{Energy eigenvalues of the 4-atom Hamiltonian given in \eqn{eqn:Hamiltonian_4_square} as a function of magnetic field $\phi_o$.}
    \label{fig:dipole4}
\end{figure}

\end{mdframed}

Therefore for a state $|\Psi\rangle$ which comprises of single particle states given by say the momenta labels $k_1, k_2, k_3 .. k_{N_p}$ where $N_p$ is the number of particles

\beq
|\Psi\rangle = c^\dagger_{k_1}c^\dagger_{k_2}\ldots c^\dagger_{k_{N_p}} |\Omega\rangle
\eeq
and $|\Omega\rangle $ is the vacuum state. We could interpret $\langle \Psi | \hat{O} | \Psi \rangle$ as an expectation of the dipole moment (when placed on a ring) or, how the state has {\it changed} after an action of a small electric field (or a flux of $2\pi$). This has a particularly neat interpretation for what we know of metals and insulators. For instance, let us ask how a single $k$ state behaves under $\hat{O}$
\beq
O c_{k}^{\dagger} O^{\dagger}=c_{k+2 \pi / L}^{\dagger}
\eeq

\begin{figure}[t]
    \centering
    \begin{framed}

    \includegraphics[width = \linewidth]{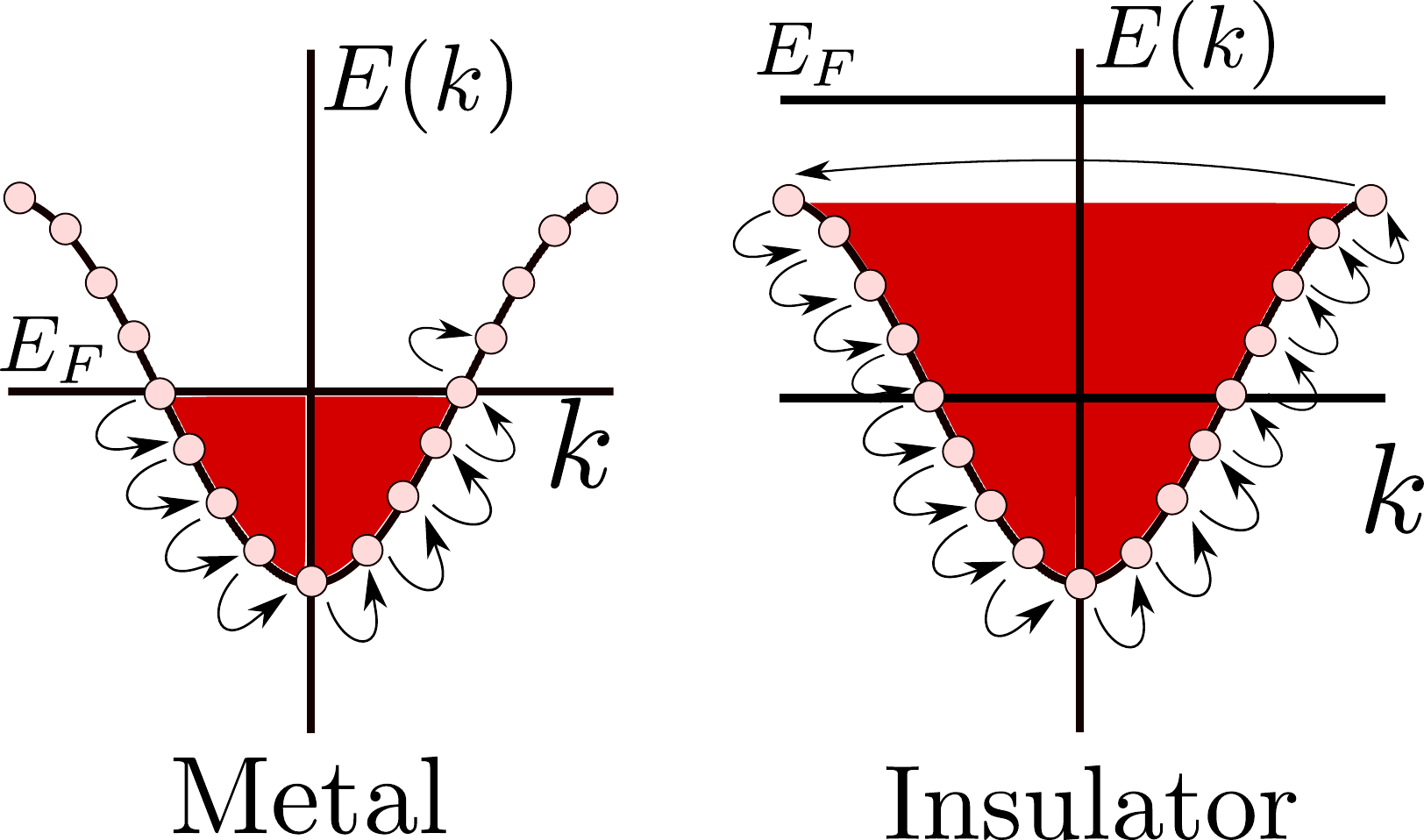}
    \caption{Action of operator $\hat{O}$ on a metal and an insulator. While on the metal it takes it to a new state, on an insulator it brings it to itself albeit with a phase.}
    \label{polarOp}
        \end{framed}
\end{figure}

So the action of operator $\hat{O}$ on a half-filled band, for instance where all the $k$ states aren't filled -- shifts each of the $k$ points by $2\pi/L$ thereby changing the many body state leading to $\langle \hat{O} \rangle=0$, while for a filled band where all the $k$ states are already filled -- it can't change it since $k=\pi+\frac{2\pi}{L} \rightarrow k=-\pi+\frac{2\pi}{L}$ leading to the fact that $ \langle \hat{O} \rangle \neq 0 $ (see \Fig{polarOp}). 

In fact for a multi-band system (such as SSH, where we have two bands for every $k$ point) one can fill one of the bands -- and as the state comes back to itself it can carry a non-trivial phase. The general value of the expectation on the operator is given by 
\beq
\langle \Psi| \hat{O}|\Psi\rangle = r e^{i\gamma}
\eeq
where $\gamma$, the phase turns out to play a crucial role, something we discuss in the next section.
    
\subsection{Geometric phase}

In the above discussion we found that for an insulating state, action of operator $\hat{O}$ on an insulator just shifts each $k$ point to the next $k$ point. Therefore the value of $\gamma$ could in principle be calculated from just the single particle wavefunctions when they are calculated at each value of $k$
\beq
\gamma=-\arg \left[\prod_{k=-\pi}^{\pi}\left\langle\psi_{k} \mid \psi_{k+2 \pi / L}\right\rangle\right] 
\eeq
For a multiband system where we have a band index one can calculate this for each band

\beq
\gamma_n=-\arg \left[\prod_{k=-\pi}^{\pi}\left\langle\psi^n_{k} \mid \psi^n_{k+2 \pi / L}\right\rangle\right] 
\eeq

Therefore if $|\psi^n_{k}\rangle$ is thought of as a vector, and given $k$ goes from $-\pi$ to $\pi$ (i.e a ring) the phase we are worrying about is in fact a geometric phase which is captured when a vector twists over a ring.

    \subsection{Geometric phase in discrete SSH chain}
    
    We now go back to the SSH model we were discussing before, and see how $\hat{O}$ acts. Diagonalizing the Hamiltonian (see \eqn{ssh_2mat}) gives us the following eigenstates
    \beq
    |\pm k\rangle=\frac{1}{\sqrt{2}}
    \begin{bmatrix}
    \pm e^{-i\phi(k)} \\
    1 \\
    \end{bmatrix}
    \ \ , \ \ \phi(k)=\tan ^{-1}\left(\frac{w \sin (k)}{v+w \cos (k)}\right)
    \eeq
    
    The creation operator for the two bands can be written as:
    \beq
    \alpha_{k \pm}^{\dagger} =\frac{1}{\sqrt{2}}\left[e^{-i \phi(k)} c_{k+}^{\dagger} \pm c_{k-}^{\dagger}\right]
    \eeq
    where for the lower band
    \beq
    \alpha_{k-}^{\dagger}  \equiv \frac{1}{\sqrt{2}} \frac{1}{r_{(k)}}\left[\left(v+w e^{-i k}\right) c_{k+}^{\dagger}-r_{(k)} c_{k-}^{\dagger}\right]
    \eeq
    
    Here, $r_{(k)}$ is the absolute value of the eigenvalue.  The ground state wavefunction is given by $|G S\rangle=\prod_{k\in B Z} \alpha_{k-}^{\dagger}|\Omega\rangle$. Therefore,
    
    \beq
    \hat{O}|G S\rangle=\prod_{k\in B Z} \hat{O} \alpha_{k-}^{\dagger} \hat{O}^{-1} \hat{O}|\Omega\rangle
    \eeq
    This simplifies to
    \beq
    \hat{O}|G S\rangle=\prod_{k\in B Z} \frac{1}{\sqrt{2}} \frac{1}{r_{(k+2 \pi / L)}}\left[\left(v+w e^{-i(k+2 \pi / L) a}\right) c_{k+}^{\dagger}-r_{(k+2 \pi / L)} c_{k-}^{\dagger}\right]
    \eeq
    This leads to
   
    \beq
    \gamma=-\sum_{k\in B Z} \arg \left[\frac{\left(v+w e^{i k}\right)\left(v+w e^{-i(k+2 \pi / L)}\right)}{r_{(k)} r_{(k+2 \pi / L)}}+1\right]
    \label{bphase_eqn}
    \eeq
    
    To do a sanity check on the above equation, we can put $v$ = 1 and $w$ = 0 to get $\gamma$ = 0 and putting $v$ = 0 and $w$ = 1 gives us $\gamma$ = $\pi$.  {\it Interestingly}, remember that at $v=0$ or $w=0$ we have a set of decoupled dimers. So these two insulators are {\it different} -- they polarize differently under a electric field -- something we had anticipated before.  Even more surprisingly, this feature just doesn't happen at extreme limits. For the entire regime when $v<w$, we have $\gamma=0$ and then it suddenly changes to $\pi$ at $v=w$ (see \Fig{fig:Opexp}). These transitions are sensitive to finite size -- as the system size increases the transition happens exactly at $v=w$.

    \begin{figure}[!b]
    \centering
    \begin{subfigure}{\textwidth}
        \centering
        \includegraphics[width=0.8\textwidth]{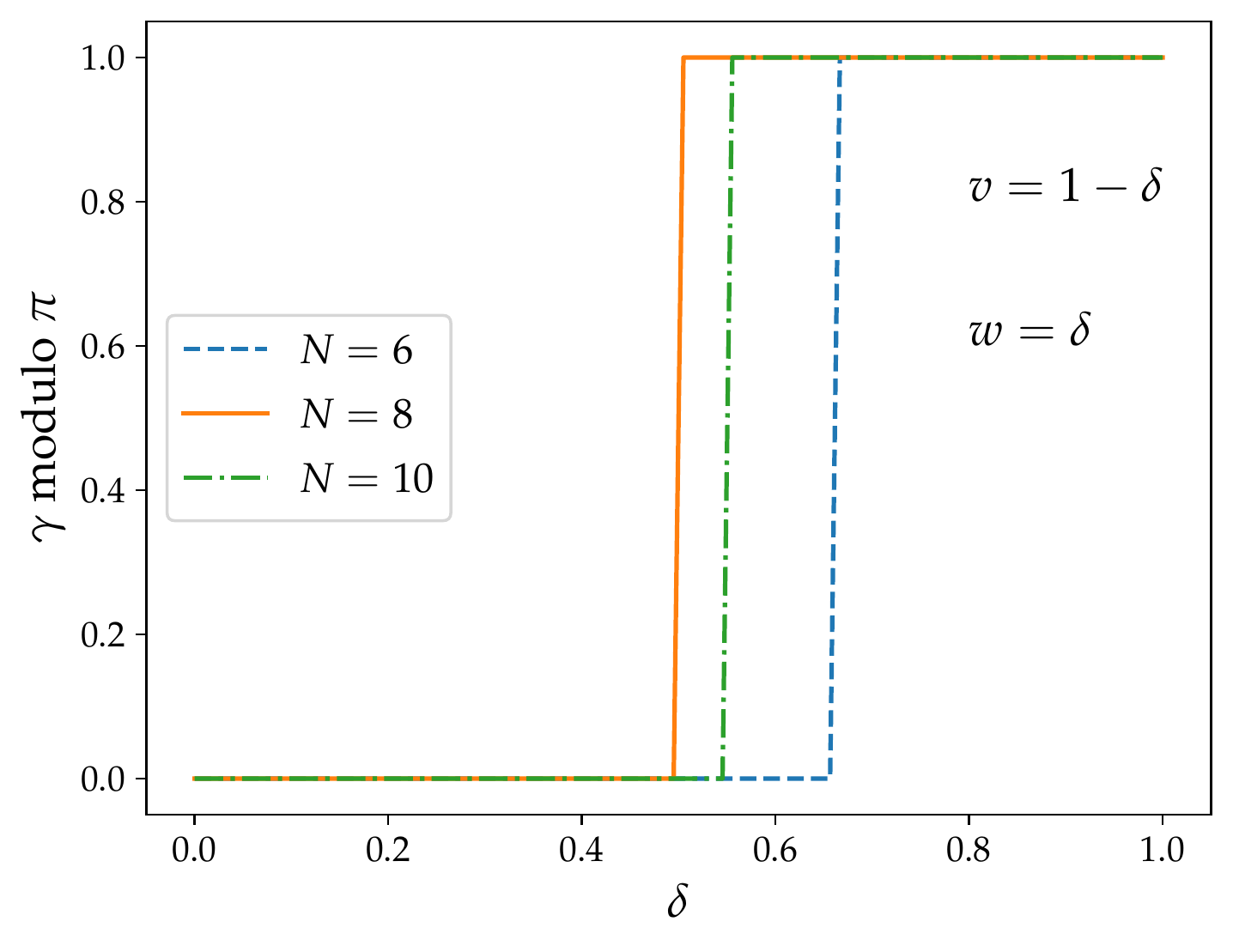}
        \caption*{(a)}
    \end{subfigure}%

    \begin{subfigure}{\textwidth}
        \centering
        \includegraphics[width=0.8\linewidth]{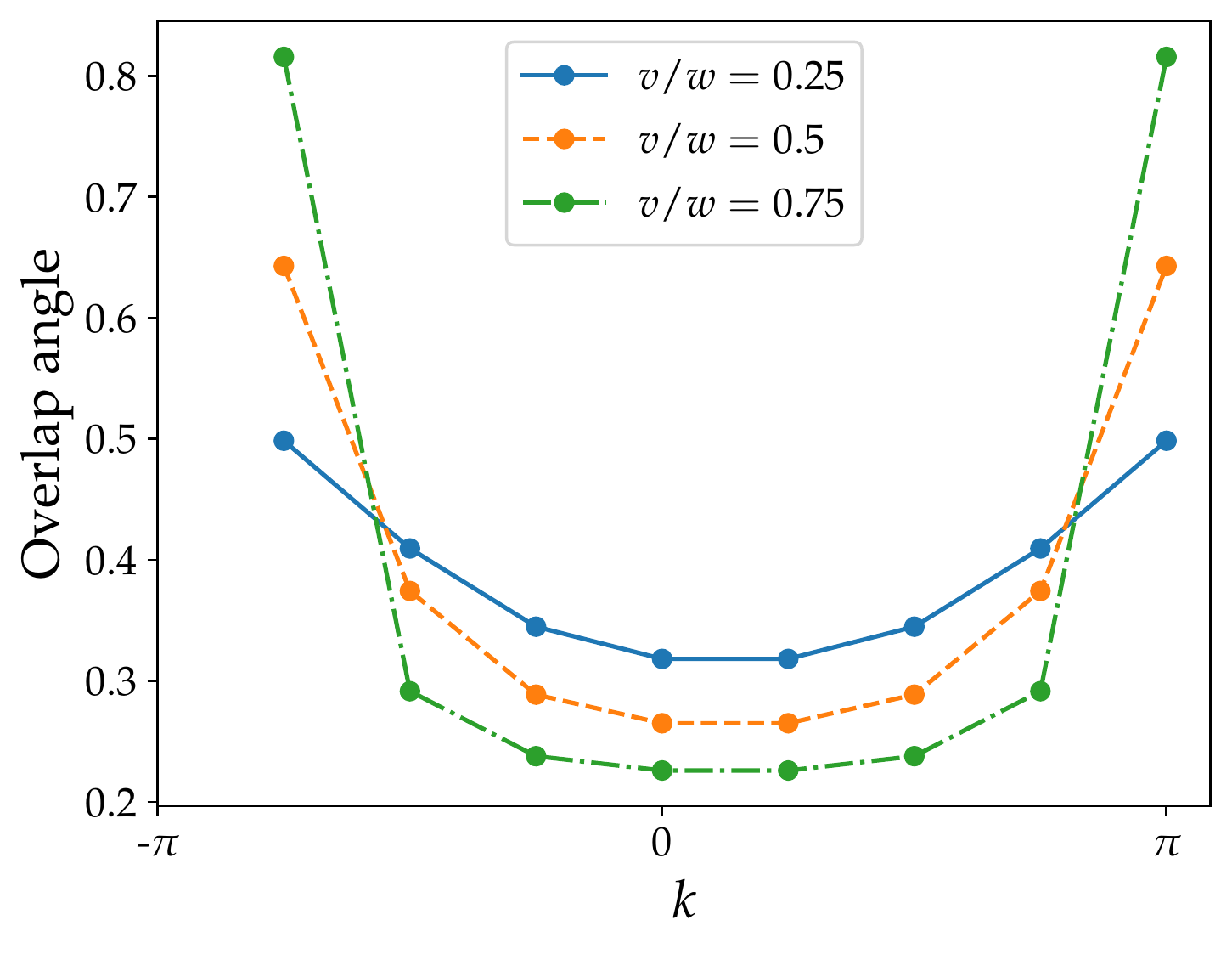}
        \caption*{(b)}
    \end{subfigure}%
    \end{figure}
    \begin{figure}\ContinuedFloat
    \centering
    \begin{subfigure}{\textwidth}
        \centering
        \includegraphics[width=0.8\linewidth]{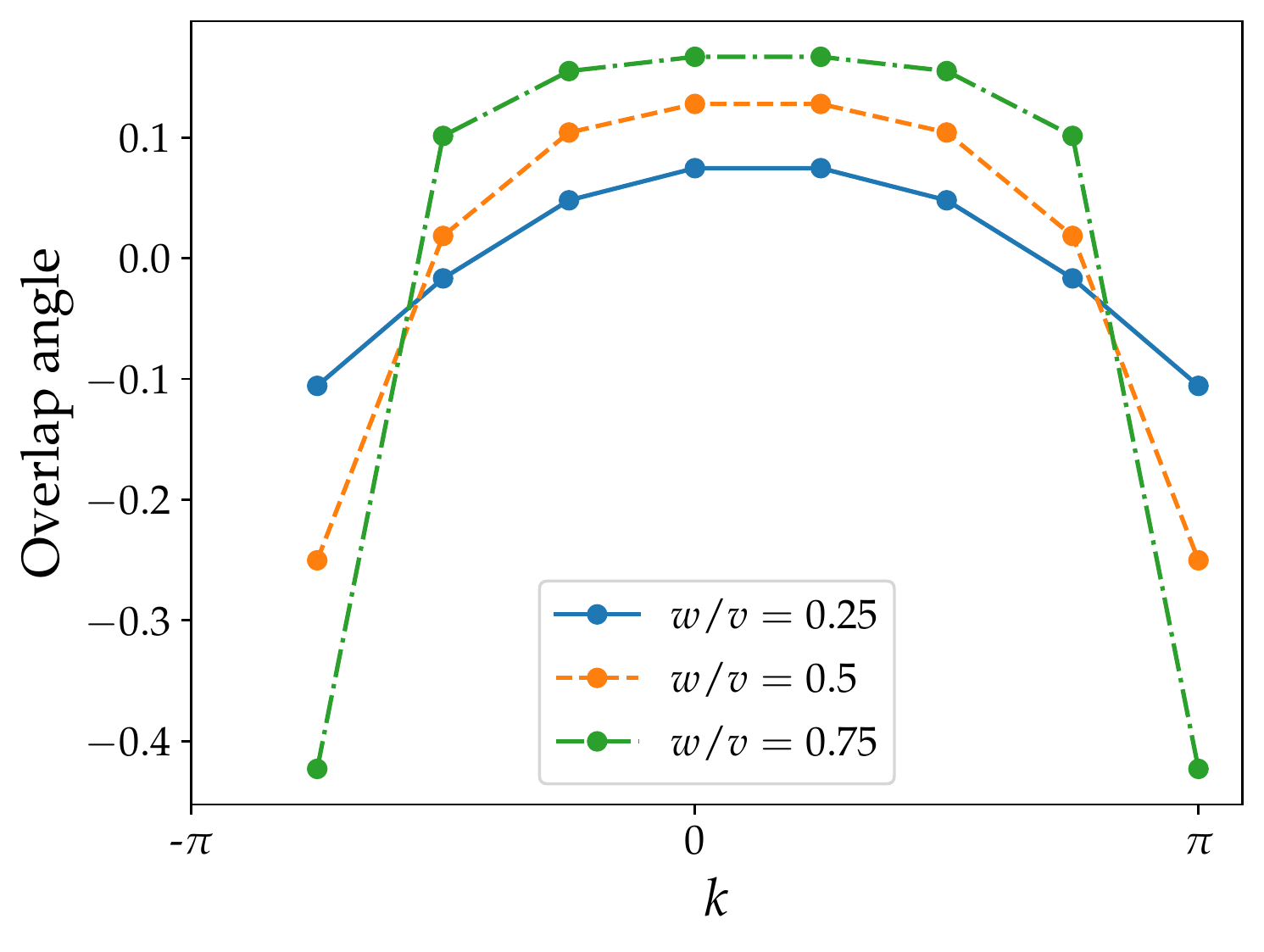}
        \caption*{(c)}
    \end{subfigure}%
    
    \begin{subfigure}{\textwidth}
        \centering
        \includegraphics[width=0.8\linewidth]{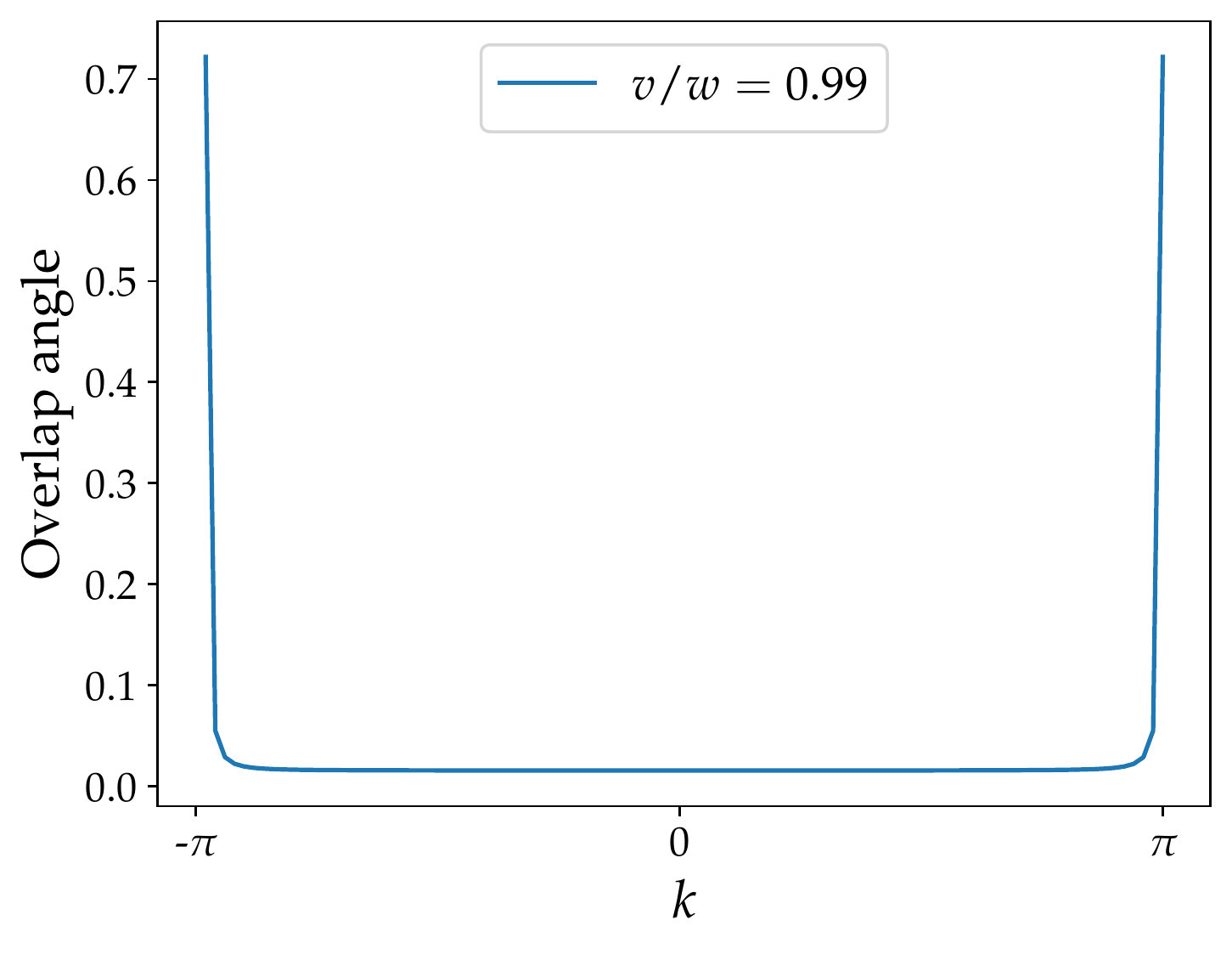}
        \caption*{(d)}
    \end{subfigure}%
    \caption{(a) Geometric phase ($\gamma$) switches from $0$ to $\pi$ at $\delta=0.5$ albeit with finite size effects. (b) Behavior of the overlap angle (\eqn{bphase_eqn}) as a function of $k$ for $N=16$ for $v/w<1$, i.e. the topological regime and for $v/w>1$ in (c). The behavior of overlap angle when very close to gap closing $(v\sim w)$ is shown in (d) for $N=200$ system.}
        \label{fig:Opexp} 
    \end{figure}

    The geometric phase which gets captured between one point of $k$ and the next $k$ point can tell us where in the band is the non-trivial geometric phase located. We calculate this for each individual $k$ and show as a function of $k$ in \Fig{fig:Opexp}. An interesting thing to note is while the qualitative  shape of the curve does not change for $v>w$ and $v<w$ (see (b) and (c) of \Fig{fig:Opexp}) they integrate to different quantities: zero or $\pi$ depending on whether we are in a trivial or a topological phase. Another observation is that when we are close to the phase transition point ($v=w$) the entire contribution to the geometric phase comes from the points where the band gap closes in the Brillouin zone! (see \Fig{fig:Opexp} (d)). 
    
    \leftHighlight{For a fun introduction to topology, do watch \href{https://www.youtube.com/playlist?list=PLTBqohhFNBE_09L0i-lf3fYXF5woAbrzJ}{YouTube lectures} by Tadashi Tokieda.}
    
    The reason we call the $v<w$ regime as a {\it topological} phase, is because it is characterized by a nontrivial integer $\gamma/\pi$ which by construction cannot change smoothly but only suddenly by an integer. Moreover, this was found by integrating a geometric phase over the {\it complete} Brillouin zone. This is remarkably different the way we think about conventional condensed matter phases where we can identify the nature of the phase by quantities which change smoothly and can be observed or measured locally. For example, a conventional magnet has magnetization everywhere in the sample which can smoothly go to zero as temperature is increased. 
    
    Another feature of a topological phase such as $v<w$ regime, is that the boundary physics is stable to disorder and perturbations -- something we won't delve into here. This property that there are physical systems which are characterized by quantities such as integers which remain stable to smooth changes shares a deep similarity with the subject of topology in mathematics, for instance such as the number of punctures in any material can only be an integer and cannot change smoothly. These connections can be made very concrete and goes under the name of topological quantum field theory -- something we won't discuss at all.

\begin{mdframed}[style = mystyle]
    
\rightHighlight{Example 6}
\hypertarget{example6}{{\it \textbf{Example:}}} Dipoles everywhere
    
Does the energy spectrum tell us all we need to understand about any system? Consider the two chains given below. The first one is just our usual SSH chain. The second one seems a little odd, the $w$ hopping is now connected to the next nearest unit cell instead of the nearest unit cell. Let's take a case where $v=0$ for both the chains. Can you write the real space Hamiltonian for both the chains?
    \begin{figure}[H]
        \centering
        \includegraphics[width = 0.7\linewidth]{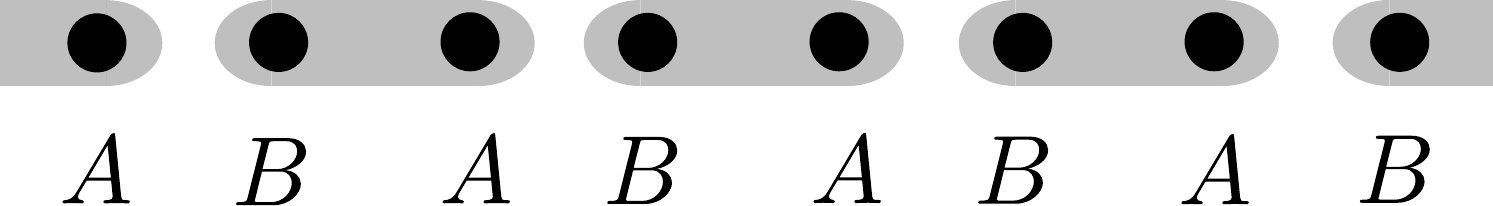}
      \caption{Chain 1}
      \label{SSH_schematicw1}
    \end{figure}
    
    \begin{figure}[H]
        \centering
        \includegraphics[width = 0.7\linewidth]{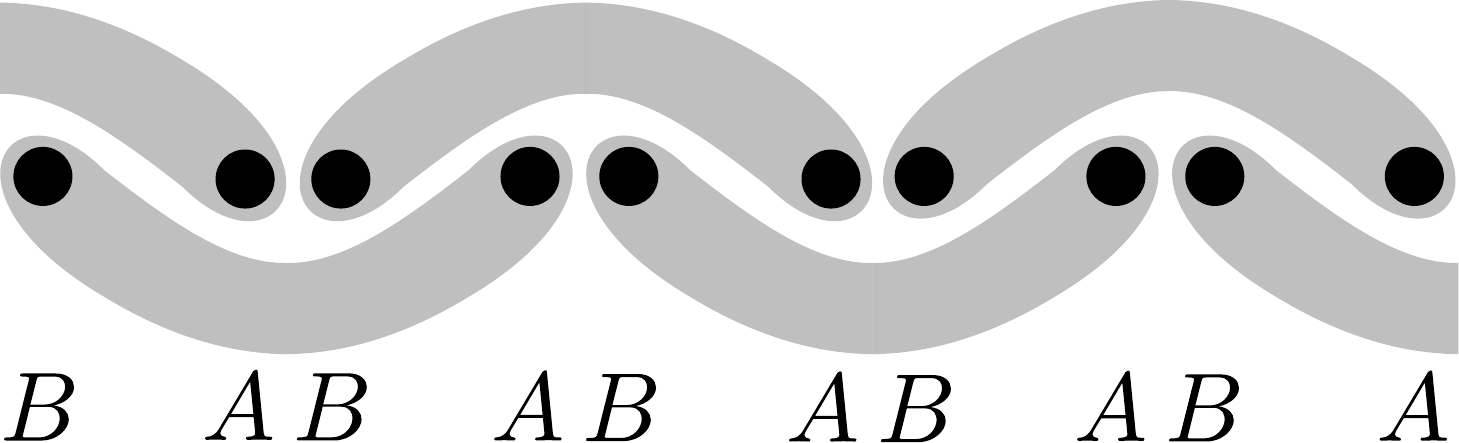}
      \caption{Chain 2}
      \label{SSH_schematicw2}
    \end{figure}

    It may seem that the second system is just a rearrangement of the first one, given both of them are a bunch of disconnected dimers with the same strength of hopping. In fact elongating the dimers in the first chain can smoothly take us to the second chain.
    
    The surprise is that these two systems are in fact {\it not} smoothly deformable, and one needs to necessarily go via a metallic phase. To see this one has to invoke the polarization framework we had discussed before. The two Hamiltonians in the $k$ space are
    \beq
    H_1 = \begin{bmatrix} 
    	 0 & e^{-ika} \\ 
    	 e^{ika} & 0 \\\end{bmatrix}
    \eeq
    \beq
    	 H_2 = \begin{bmatrix} 
    	 0 & e^{-i2ka} \\ 
    	 e^{i2ka} & 0 \\\end{bmatrix}
    \eeq	 
    where $a$ is the inter-unit cell distance. As is expected for disconnected dimers, the spectrum for both $H_1$ and $H_2$ are identical and have flat bands. To interpolate between them let us construct a new Hamiltonian, 
        \beq
        H_{\lambda} = \lambda H_1 \  + \ (1-\lambda)H_2
        \label{adiabatic_ham}
        \eeq
    Does our system remain an insulator while we vary $\lambda$ from 0 to 1? 
    
    \begin{figure}[H]
        \centering
        \begin{subfigure}{.5\linewidth}
        \centering
      \includegraphics[width=.9\linewidth]{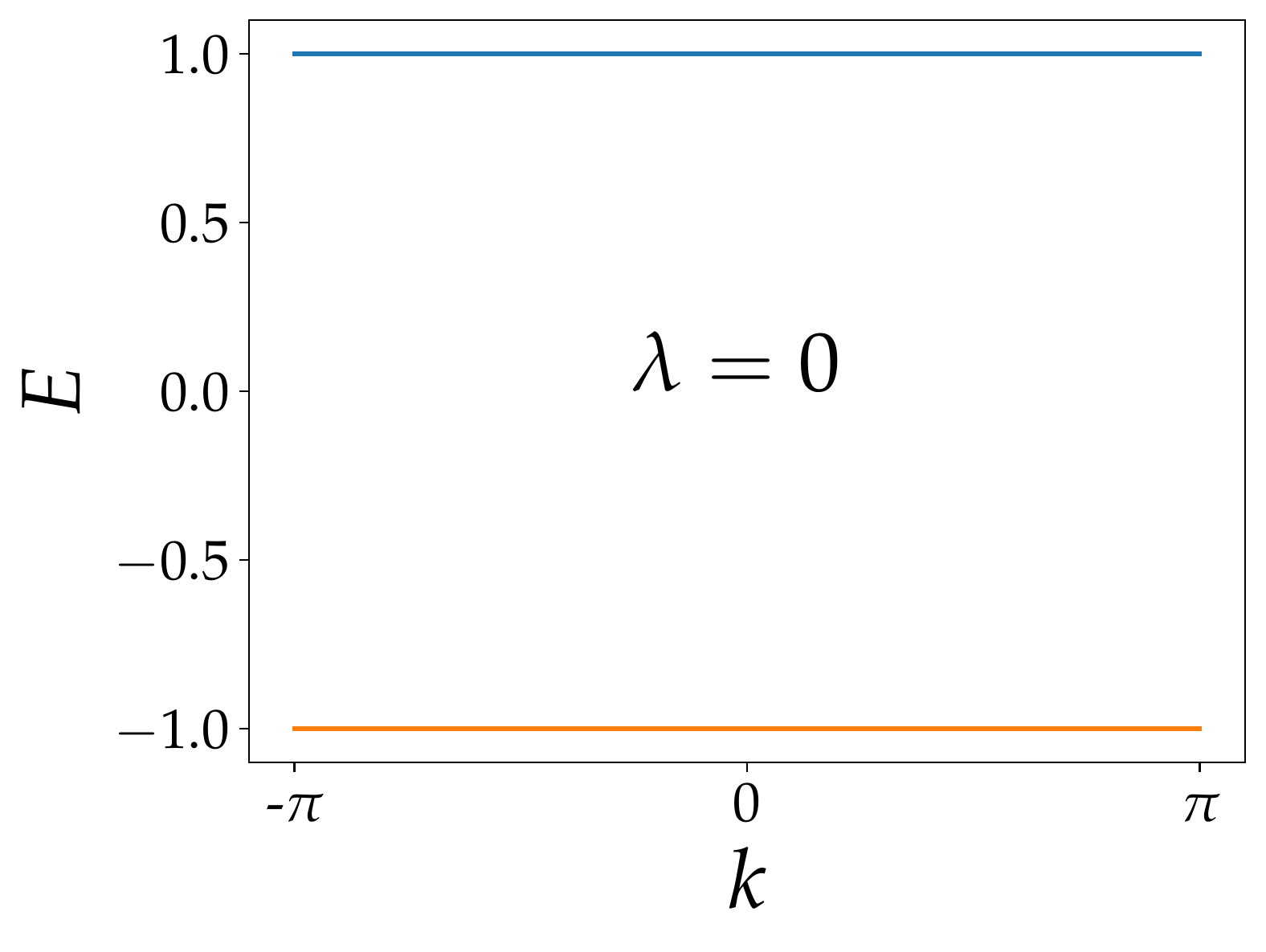}
      \caption{$\lambda$ = 0, \  $\lambda$ = 1}
      \label{adia0}
        \end{subfigure}%
        \begin{subfigure}{.5\textwidth}
        \centering
      \includegraphics[width=.9\linewidth]{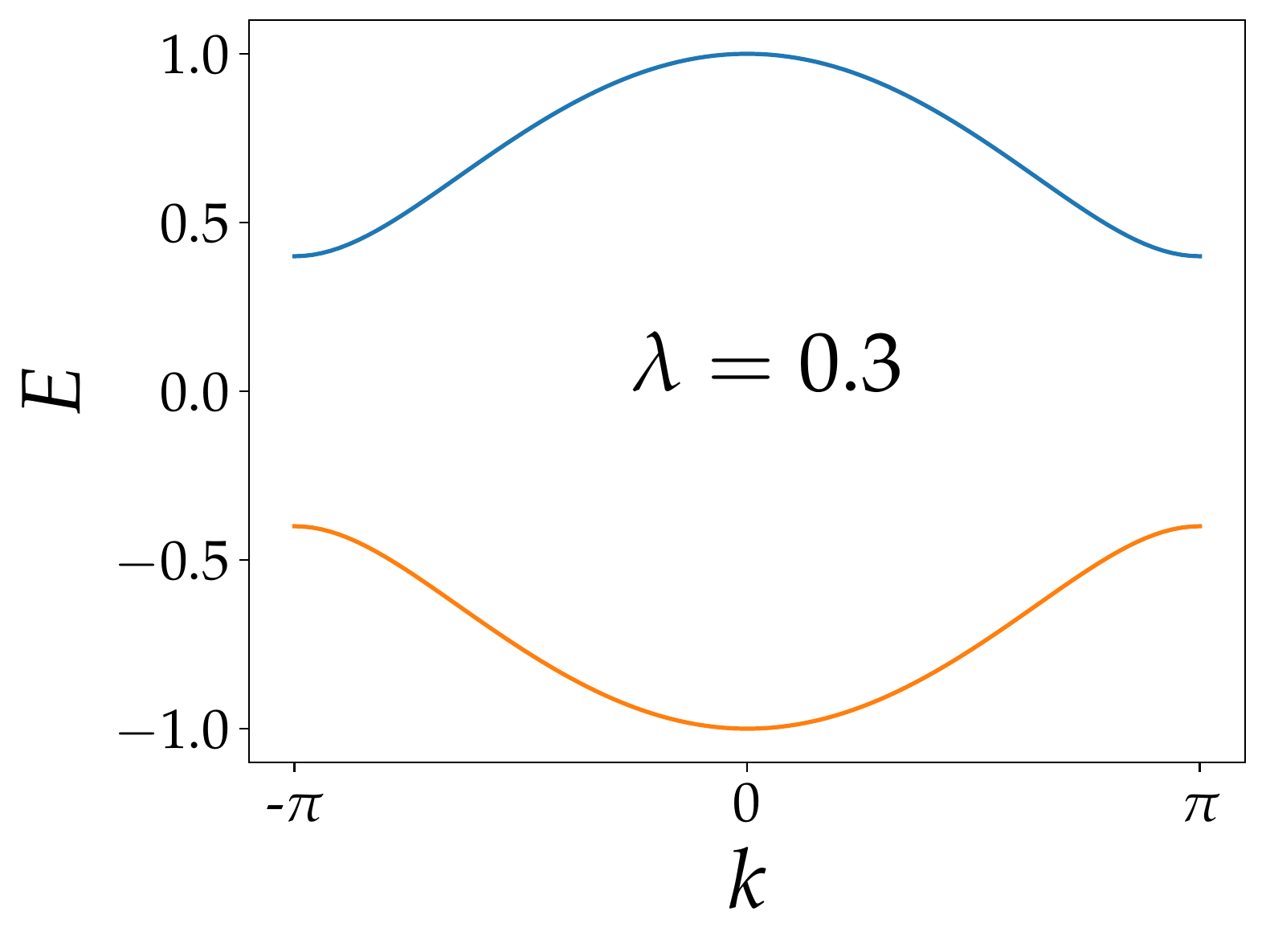}
      \caption{$\lambda$ = 0.3}
      \label{adia0p3}
        \end{subfigure}
          \begin{subfigure}{.5\linewidth}
        \centering
      \includegraphics[width=.9\linewidth]{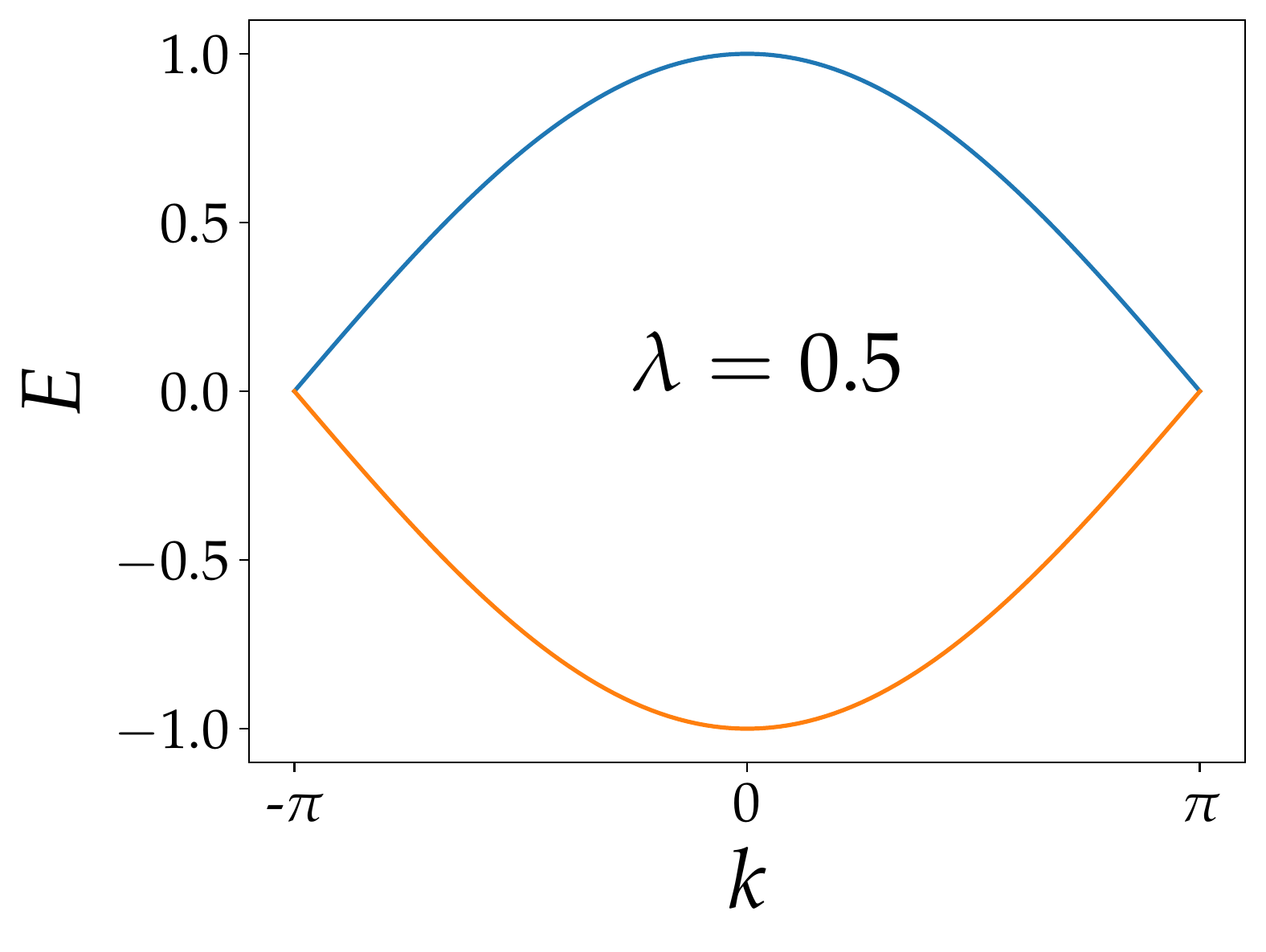}
      \caption{$\lambda$ = 0.5}
      \label{adia0p5}
        \end{subfigure}%
        \begin{subfigure}{.5\textwidth}
        \centering
      \includegraphics[width=.9\linewidth]{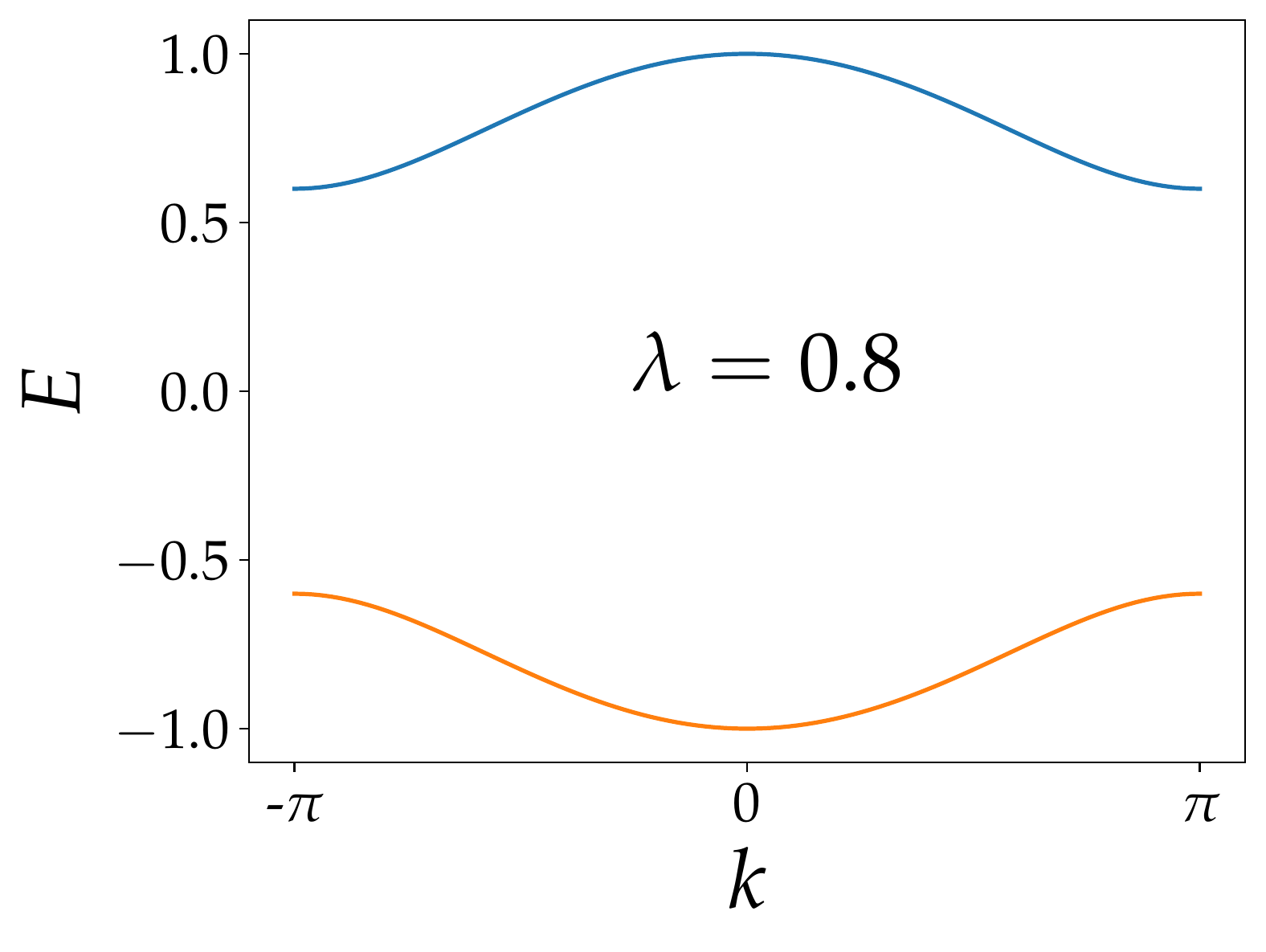}
      \caption{$\lambda$ = 0.8}
      \label{adia0p8}
        \end{subfigure}
        \caption{Energy spectrum for the Hamiltonian in \eqn{adiabatic_ham}, for different values of $\lambda$. The gap closes for $\lambda$ = 0.5}
        \label{adiabatic}
    \end{figure}
    
    Turns out, one can't do that because we encounter a metallic phase at $\lambda=0.5$ (see \Fig{adiabatic}). This suggests that $\lambda<0.5$ and $\lambda>0.5$ are two different phases. Are they topologically different? Can you calculate the geometric phase for $H_1$ and $H_2$? 
        
    \end{mdframed}

\section{Entanglement entropy}

Up until last section we found polarization is one way that can distinguish the two phases of the SSH system. We now discuss, quantum entanglement, an idea which has no classical analogue - and how it also plays a crucial role in these topological phases.

\subsection{What is entanglement entropy?}
Given any system we can partition the complete Hilbert space as the tensor product of two Hilbert sub-spaces.
\[H_{AB} = H_A \otimes H_B\]
where $A$ and $B$ represent the sub-spaces. In this context one can pose: for a wavefunction defined on the complete Hilbert space, can it also be written in a separable form between these two sub-spaces? One way to answer this is to Schmidt decompose a wave function
 \beq
 |\Psi\rangle\ = \sum_{i=1}^n \alpha_i |u_i\rangle_A \otimes |v_i\rangle_B
 \label{schmidt_decomp}
 \eeq
 where $|u_i\rangle_A$ and $|v_i\rangle_B$ are orthonormal states in the subsystem A and B respectively. The state is called a product (un-entangled) state if only one of the coefficients $\alpha_i$ is non-zero where such a separation is indeed possible; more generally such a separation is not possible and the state can carry a finite entanglement.
 
\rightHighlight{
  Consider a system with two spinless electrons on two sites. Are these two electrons entangled?
}
 Given the density matrix of the system
 \beq
 \rho^{AB} = |\Psi\rangle\langle \Psi|
 \eeq
 the partial trace of this density matrix over the subsystem A is called $\rho^A$ while over the subsystem B is called $\rho^B$.

 The entanglement entropy $S$ is defined as
 \beq
 S = -\text{Tr}[\rho^A\log(\rho^A)] = - \text{Tr}[\rho^B\log(\rho^B)]
 \label{Strace}
 \eeq
 This equals
 \beq
 S = -\sum_i e_i \log(e_i)
 \eeq
 where the $e_i$  are the eigenvalues of the partial traced density matrix $\rho^A$ or $\rho^B$.
 
It is fairly straightforward to see that in terms of the Schmidt decomposition coefficients (see \eqn{schmidt_decomp}) and definitions of the partially traced density matrix, $S$ can be represented in terms of the Schmidt coefficients $\alpha_i$ as
 \[S = - \sum_{i=1}^n |\alpha_i|^2\log(|\alpha_i|^2)\] 
The reader may heard that two spin 1/2 particles, when in a singlet state, has an entanglement entropy of $\log(2) \sim 0.693$. At this point it may be tempting to think that in order to discuss entanglement entropy, we need at least two or more particles. But rather than the number of particles -- it is the wavefunction and its separability in {\it real} space that determines the value of $S$ (see \hyperlink{example7}{Example 7}). 
 
 \begin{mdframed}[style = mystyle]
 \leftHighlight{Example 7}
\hypertarget{example7}{{\it \textbf{Example:}}}  Let's consider two sites $A$ and $B$ with a hopping term between them and with one electron. The ground state wavefunction is given by  
 \[ |\Psi\rangle = \frac{1}{\sqrt{2}} \begin{bmatrix} 
	 1 \\ 
	 -1 \\\end{bmatrix} \]
	 Or in second quantization notation  $|\Psi \rangle = \frac{1}{\sqrt{2}}( c_A^{\dagger} - c_B^{\dagger}) |\Omega\rangle$ where $|\Omega \rangle$ is the vacuum state. With two sites the total Hilbert space dimension is $2^2 = 4$. The basis states being, the $zero$ particle state $|\Omega \rangle$, the 1 particle states $c_A^{\dagger}|\Omega\rangle$ and $c_B^{\dagger}|\Omega\rangle$ and the two particle state $c_A^{\dagger}c_B^{\dagger}|\Omega\rangle$. The state $|\Psi\rangle$ in the complete Hilbert space is then given by  \[|\Psi\rangle = \frac{1}{\sqrt{2}} \begin{bmatrix} 
	  0\\
	  1\\
	 -1 \\ 
	  0 \\\end{bmatrix} \]
	  The subsystems we are considering are the site $A$ with states $|\Omega\rangle$ and $|c_A^{\dagger}|\Omega\rangle$ and similarly the single site $B$ with the states $|\Omega \rangle$ and $c_B^{\dagger}|\Omega \rangle$. The density matrix is given by \[ \rho^{AB} = \frac{1}{2} \begin{bmatrix}
	  0 & 0 & 0 & 0\\
	  0 & 1 & -1 & 0\\
	  0 & -1 & 1 & 0 \\ 
	  0 & 0 & 0 & 0 \\\end{bmatrix} \] 
	  The partial trace over subsystem A is \[ \rho^{A} = \frac{1}{2} \begin{bmatrix} 
	  1 & 0\\
	  0 & 1\\\end{bmatrix} \]
	  with the entanglement entropy given by \[S = - \Big( \frac{1}{2}\log(\frac{1}{2}) + \frac{1}{2}\log(\frac{1}{2})  \Big)\]
	  \[ S = \log(2) = 0.693\]
	  Therefore, even for a simple system with just one electron hopping between two sites, there is a  non-zero entanglement entropy associated with the partitioning the system into two halves. This toy problem also gives an intuition that there is an entropy of $\log(2)$ attached to breaking of a bond, something we will later find useful. 
	  \end{mdframed}

\begin{figure}[!htb]
		\centering
		\begin{framed}
	\includegraphics[width = 0.9\textwidth]{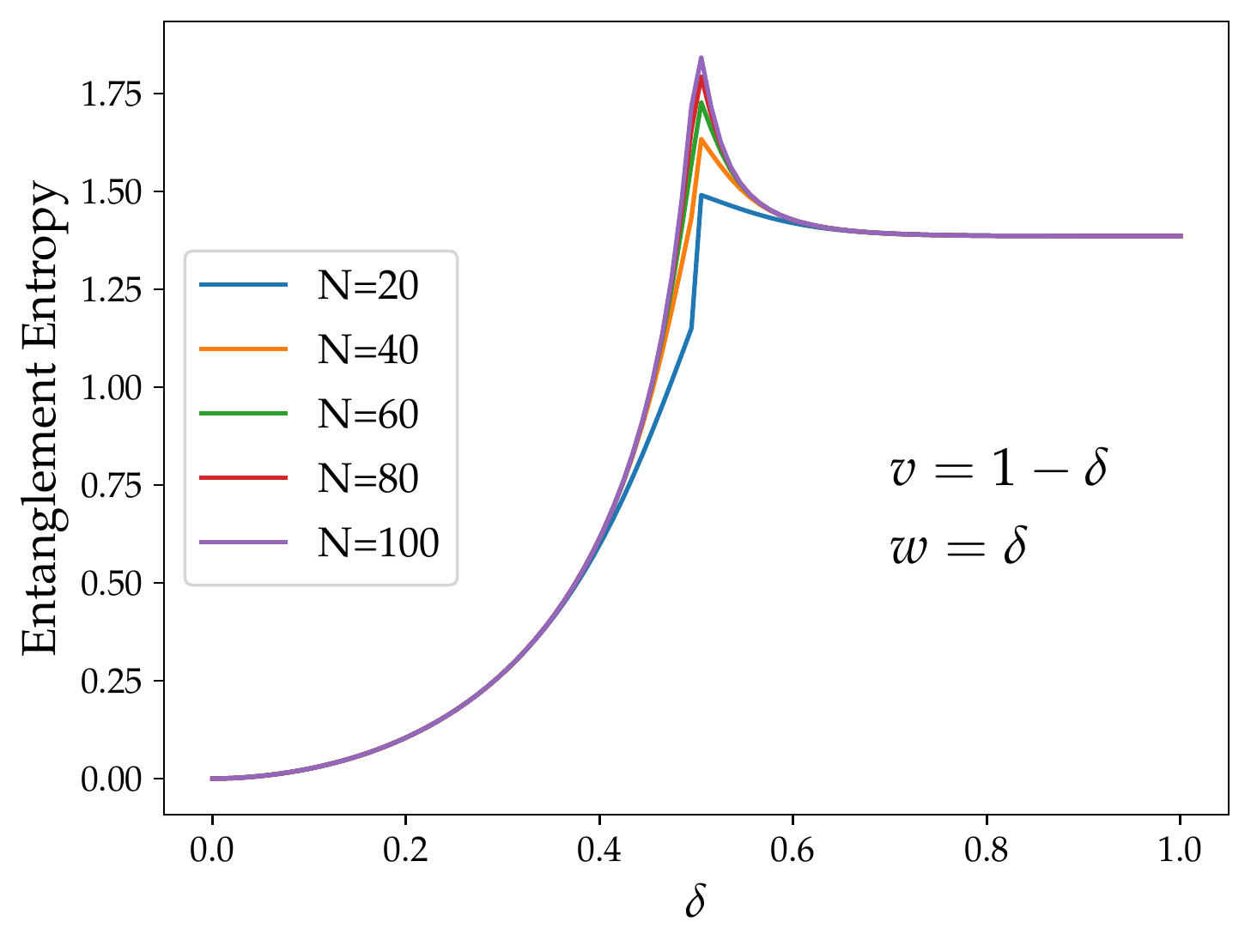}
	
	\caption{Entanglement entropy as a function of $\delta$ for halfway partition of a SSH chain with periodic boundary conditions for different lengths $N$.}
	\label{fig:my_label_pEE}
	\end{framed}
		\end{figure}

      Therefore given a wavefunction, one can evaluate a density matrix and trace it over partially to obtain a reduced density matrix. This when done for the ground state wavefunction in the SSH model leads us to \Fig{fig:my_label_pEE}. While for $v<w$ the entanglement entropy is small and close to zero, it saturates to $\log(2)$ when $v>w$! 
	  Therefore even entanglement entropy can provide us indications of having a topological phase!

	  While a density matrix calculation is easily done for small system sizes, as the system size increases, the dimensions required for our state and the density matrices would blow up exponentially (for a system of size $N$, the density matrix size is $2^N \times 2^N$). Is there a way then to extract the same information in another, but in a simpler way? Turns out that for a non-interacting fermion model, as is our case, a two-point correlator matrix could tell us all that we need to know \cite{Calabrese}.
	  
\begin{mdframed}[style = mystyle]
\rightHighlight{The number of bonds being cut is representative of the entanglement entropy between the two halves}
We had a look at the following dipoles (see below) earlier in \hyperlink{example6}{Example 6}. Would the entanglement entropy of the left half with the right half be different for the cases considered here? 

\begin{figure}[H]
    \centering
    \includegraphics[width = 0.7\linewidth]{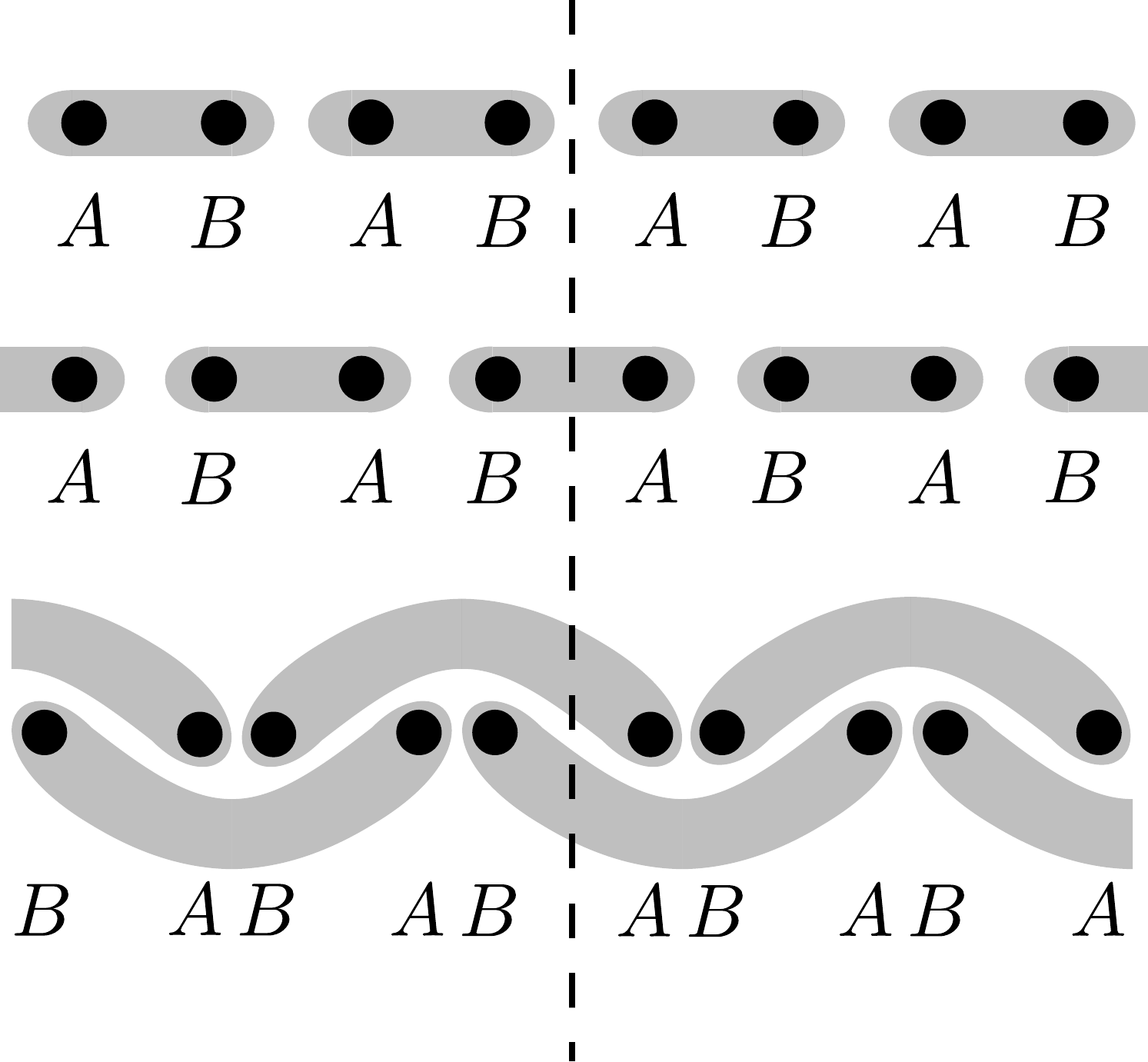}
\end{figure}

\end{mdframed}

\subsection{Correlators}
In condensed matter physics, the ``correlator'' plays a very important role in the understanding of ground state properties as well as the dynamics of the system. The single particle correlator, which will be invoked a lot in this work is defined as 
$C_{\alpha \beta}= \langle c_{\alpha}^{\dagger} c_{\beta} \rangle$ describes the correlations between indices $\alpha$ and  $\beta$ in any given state, where $\alpha$ and $\beta$ denote any arbitrary basis labels. For instance, consider an insulator where the ground state wave function is given by $|\Phi_G\rangle = \prod_{k}c_{k}^{\dagger} |\Omega \rangle$ (i.e., all the momentum states are filled).  Here, we simply get $C_{k,k'}=\langle\Phi_G|c_{k}^{\dagger}c_{k'} |\Phi_G\rangle=\delta_{kk'}$, which indicates that we do not have long ``distance'' correlation. Of course, here distance simply points to the difference in the label number in an arbitrary basis. 

To put the concept of distance to a real context, we can now calculate the averaged correlation amplitude of processes that connect two sites $i$ and $j$- namely the simultaneous annihilation of an electron at site $j$ and the creation of an electron at site $i$. One may immediately recognize that if $i=j$, we do have that the correlator represents the electron density at the site $i=j$. Let us investigate this a little further. Here we suggest the use of the word ``simultaneous'', literally meaning equal time, a concept that is meaningful in the unperturbed system. With the introduction of a time dependent perturbation, as we will shortly see, the correlators have to be extended to two different times. For now, let us focus on the equal time static correlators, in connection with unperturbed ground states. 

Let us see what happens to the long distance correlation between two sites in an eigenstate, i.e., $C_{ij}=\langle \Phi_G|c_{i}^{\dagger}c_{j}|\Phi_G \rangle$. In order to understand this, we ask the question: what happens when we remove an electron from one site and add to another simultaneously in (a) metallic state and b) insulating state? Before we get to the math, let us summarise the essentials. In the insulating state to be completely filled such that $|\Phi_G\rangle=\prod_{k\in BZ} c_{k}^{\dagger} |\Omega\rangle$, whereas in a metallic state we have
$|\Phi_G\rangle=\prod_{|k|<k_F}c_{k}^{\dagger}|\Omega \rangle$. In the former case, $BZ$ represents the Brillouin zone and hence a fully filled system. The latter case represents a partially filled system, such that at zero temperature, the states $k<k_F$, the Fermi wave vector, are filled. A simple pictorial representation of the two cases can be gathered from Fig.~\ref{polarOp}.

We can also quite well surmise what will happen in the former case. In a filled band, with the states being filled in the eigenspace, quite obviously the sites comprising the real space are also occupied. Keeping the Pauli exclusion principle in mind, it will not be possible to ``simultaneously'' remove an electron in site $i$ and place in site $j$, since site $j$ will also be occupied, of course unless $i=j$. Thus it is easy to make the right guess that $C_{ij}= \delta_{ij}$. In the latter case i.e.~for the metal, this might not be the case since it is a partially filled system in the eigenbasis and hence in real space also! let us now formalize this calculation. 

Using 
\beq
c_{i}^{\dagger}=\frac{1}{\sqrt{N}} \sum_{k \in BZ} e^{ik x_i} c_k^{\dagger}
\eeq
for a fully filled state one finds,
\beq
C_{ij} = \frac{1}{N} \sum_{k,k' \in BZ} \delta_{kk'} e^{i(kx_i -k'x_j)}= \frac{1}{N} \sum_{k\in BZ} e^{i\left (k\left(x_i -x_j\right)\right)} =\delta_{ij}.
\eeq
However, for the partially filled band where
\beq
|\Phi_G\rangle = \prod_{|k|<k_F} c_{k}^{\dagger} |\Omega\rangle
\eeq
one obtains 
\beq
C_{ij} = \sum_{\mid k \mid \mid k' \mid <k_F} \delta_{k,k'} e^{i(kx_i -k'x_j)}= \frac{1}{N} \sum_{\mid k \mid < k_F} e^{i\left (k\left(x_i -x_j\right)\right)} 
\eeq
%
where one uses the fact that $\langle\Phi_G| c_k^{\dagger} c_{k'}|\Phi_G \rangle = \delta_{k,k'}$ when $|k|<k_F$, denoting filled states and $\langle \Phi_G|c_k^{\dagger} c_{k'}|\Phi_G\rangle = 0$ when $|k|> k_F$, denoting empty states. When $i=j$, the correlator $C_{ii} = \frac{N_f}{N}$, denoting that the probability of occupying a given site $i$ is smaller than unity and equal to the density!

Interestingly for $i\neq j$, assuming $x_i=0$ and a thermodynamically large system

\beq
C_{j} \sim  \frac{2}{2\pi} \int_0^{k_F} \cos(k x_j) dk =  \frac{\sin(k_F x_j)}{ \pi x_j}
\eeq

This shows that for a metal not only the correlator has characteristic oscillations that depend on the Fermi vector but also a power law decay as a function of distance $x_j$ from $x_i=0$. This is remarkably different from an insulating state, and is an alternate way of thinking about metals and insulators!

Now that we have seen one interesting example of the use of the correlator, we will now see some more applications of the correlator, specifically when it comes to determining the topological aspect of a given state. Here, the correlator- most importantly the one-particle two point correlator will be used in lieu of the full density matrix that we encountered earlier. 
\subsection{Entanglement Entropy using the correlator matrix}  
\label{ee_correlator}

Here we point out that the two-point correlators and entanglement entropy are in fact intricately related and one can calculate the latter using the former, in the non-interacting case. We now define a two point correlator matrix. Given a $N$ site chain, it is a $N \times N$ matrix CM, such that 
 \[ \text{CM}_{ij} = \langle\Psi|c_i^{\dagger}c_j|\Psi\rangle \]
 where $c_i^{\dagger}$ is the creation operator for site $i$ and $c_j$ is the annihilation operator for site $j$ and $\text{CM}_{ij}$ is the expectation of this operator taken over the many body ground state we had encountered before: $|\Psi\rangle$.  But wait, does this really help? Even though CM is $N\times N$ , to calculate it we still need to calculate the $2^N \times $1 $|\Psi\rangle$.  Fortunately one can also write CM, in terms of the one particle first quantized $N\times$1 wavefunctions. 
 \beq
 \text{CM} = \sum_{i} |\Phi_i\rangle\langle\Phi_i|
 \label{CMt_phi}
 \eeq
 where the summation is over all the occupied states having energy less than the Fermi energy. The way to see this is given a second quantized Hamiltonian $\hat{H}$, we can write this in terms of a matrix $\mathcal{H}$ as
 \beq
 \hat{H} = \mathbf{C}^{\dagger}\mathcal{H} \mathbf{C}
 \label{12quant_ham}
 \eeq
 where \textbf{C} is a column vector comprising of the annihilation operators $c_i$ at site $i$. We can diagonalize $\mathcal{H}$ to rewrite it as 
 \beq
 \hat{H} = \mathbf{C}^{\dagger} R^T D R \mathbf{C}
 \label{12quant_ham_2}
 \eeq
 where $D$ is a diagonal matrix of the energy eigenvalues $\epsilon_n$ and \textbf{C}=$R\mathbf{C_f}$, where $C_f$ is column vector comprising of annihilition operators $f_n$ where $f^\dagger_n$ creates an eigenstate at the $n^{th}$ eigenstate. For a ground state
 \beq
 |\Psi \rangle = \prod_{\epsilon_n<E_f} f^\dagger_n |\Omega\rangle
 \eeq
 such that number of filled states are $N/2$ we get
 \beq
 \langle\Psi|c_i^{\dagger}c_j|\Psi\rangle = \sum_{qp}^{N} \langle0|f_{q}^{\dagger}R^T_{qn}R_{mp}f_{p}|0\rangle = \sum_{p}^{N/2} R^T_{pn}R_{mp}
 \label{Cmt_phi}
 \eeq
 
 Given that the columns of $R$ are the first quantized wave functions, we arrive at \eqn{CMt_phi}!
 
 So now, how do we get the entanglement entropy from the correlation matrix \cite{hatsugai,Peschel} and how do we partition it?\\
 For most of the examples we'd consider partitioning the chain into halves, but in general to partition over sites 1..$j$ and $j$+1...$N$, we can simply consider the truncated matrix CMt which is a $j\times j$ sub matrix of CM. Alternatively, we could have also considered the other sub-system which corresponds to a matrix size $(N-j)\times(N-j)$, but the entanglement entropy would be the same in both cases. Given a truncated CMt matrix, the entanglement entropy $S$ is given by
 \beq
 S = - \sum_{i=1}^{j} \Big( e_i\log(e_i) + (1-e_i)\log(1-e_i) \Big)
 \label{S_entropy_values}
 \eeq
 where $e_i$ is an eigenvalue of CMt. For a not so quick derivation on how to arrive at this see Appendix.\ref{EntanCorr}.

\begin{mdframed}[style = mystyle]
 \leftHighlight{Example 8}
\hypertarget{example8}{{\it \textbf{Example:}}}
 Let us try to obtain the entanglement entropy for a 4 atom SSH chain with periodic boundary conditions using the correlator matrix method we discussed. Using the density matrix method we would require a $16 \times 16$ size matrices. Let's see if we can do better this time.
 
 \begin{figure}[H]
    \centering
    \includegraphics[width = 0.5\linewidth]{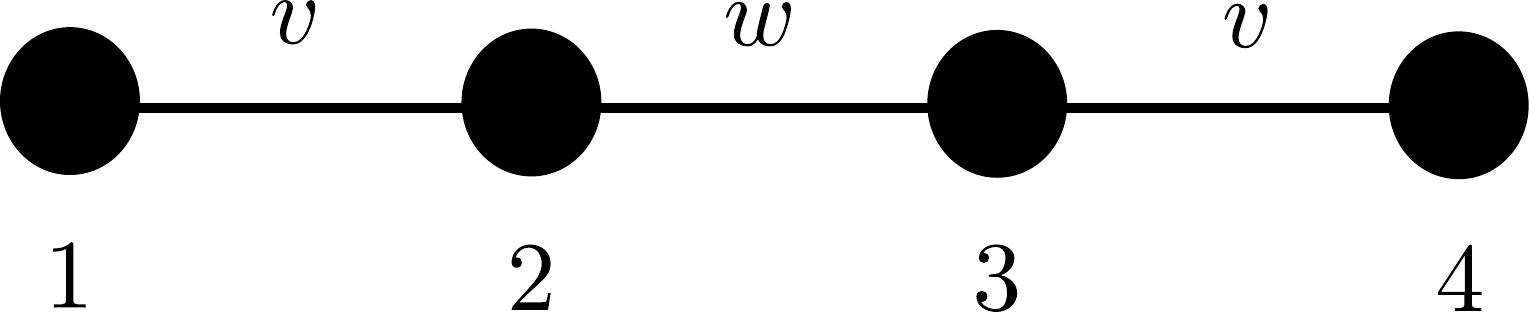}
  \label{SSH_correlator}
\end{figure}

Let's take the topological case with $v=0$ and $w=1$. The four single-particle eigenvalues of this setup are -1,-1,1,1. For the half filled system we look at the two lowest energy levels and their wavefunctions. 
\[|\Phi_1\rangle = \frac{1}{\sqrt{2}} \begin{bmatrix} 
	  0\\
	  -1\\
	  1 \\ 
	  0 \\\end{bmatrix} \ \ \ \ \ \ |\Phi_2\rangle = \frac{1}{\sqrt{2}} \begin{bmatrix} 
	  1\\
	  0\\
	  0 \\ 
	  -1 \\\end{bmatrix} \]
	  
Using \eqn{CMt_phi}, we have the correlator matrix $CM$ to be 

\[CM = \frac{1}{2} \begin{bmatrix} 
	  0 & 0 & 0 & 0\\
	  0 & 1 & -1 & 0\\
	  0 & -1 & 1 & 0 \\ 
	  0 & 0 & 0 & 0 \\\end{bmatrix} +
	  \frac{1}{2} \begin{bmatrix} 
	  1 & 0 & 0 & -1\\
	  0 & 0 & 0 & 0\\
	  0 & 0 & 0 & 0 \\ 
	  -1 & 0 & 0 & 1 \\\end{bmatrix} \]
	  
\[CM = \frac{1}{2} \begin{bmatrix} 
	  1 & 0 & 0 & -1\\
	  0 & 1 & -1 & 0\\
	  0 & -1 & 1 & 0 \\ 
	  -1 & 0 & 0 & 1 \\\end{bmatrix}  \]	 
	  
	 Now, since we want the entropy of entanglement of the first half with the second half, we truncate the top left $2\times2$ matrix of CM. Referring to this as CMt, we have
	 
\[CMt = \frac{1}{2} \begin{bmatrix} 
	  1 & 0\\
	  0 & 1\\\end{bmatrix}  \]
	  
Both the eigenvalues of $CMt$ are 0.5. Using \eqn{S_entropy_values}, we have the entanglement entropy S as,
\[S = - 2\Big( \frac{1}{2}\log(\frac{1}{2}) + (1-\frac{1}{2})\log(1-\frac{1}{2})  \Big)\]
\[S = 2\log(2) = 1.3863 \]
 We can see this value of entropy in \Fig{fig:my_label_pEE} and we only had to work with matrices of size $4 \times 4$!
	  
 \end{mdframed}

\subsection{A relook at the SSH Chain}
Great, equipped with correlator matrices, we can look at entanglement entropy in SSH chains. 

First we look at a SSH chain with periodic boundary conditions. In the plot \Fig{fig:my_label_pEE} we show the behavior of entanglement entropy as a function of $\delta$ where $v=1-\delta$ and $w=\delta$, for chains with different lengths $N$.  Let us just look at $N$=20 for the rest of this discussion. For $\delta=0$ we have $v=1$ and $w=0$ -- the trivial region. We choose a partition at $j=N/2$ to ask how entangled the first half is with the second half. 

What would one expect the entanglement entropy to be? Site 10 and 11 are connected by hopping parameter $w$ which is 0 in this case. Intuitively one would expect the entropy to be 0 and from the plot, it indeed is!
Now let's look at the other end of the plot where $\delta = 1$  and we have $v=0$ and $w=1$ i.e. the extreme topological region. Now, site 10 and 11 are connected and from our earlier toy example, we expect the entropy for one \q{broken} bond to be $\log(2)$. But the actual value from the plot is twice of it -- $2\log(2)$! Does this mean that two bonds are being broken? The answer is in affirmative, since what we have is a periodic chain or a ring, and hence by partitioning at $j=N/2=10$, we cut the bond between site 10 and 11 as well as the bond between site 20 and 1! 

At $\delta=0.5$ there appears to be a peak whose value increases with $N$. At this point we have $v=w=0.5$ and therefore the chain is a metal. Metals have higher sub-system entanglement and it grows with system size albeit slowly. Next, let us look at the SSH chain with open boundary conditions.

\begin{figure}[!htb]
    \centering
    \begin{framed}
    \begin{subfigure}{\linewidth}
    \centering
  \includegraphics[width=.9\linewidth]{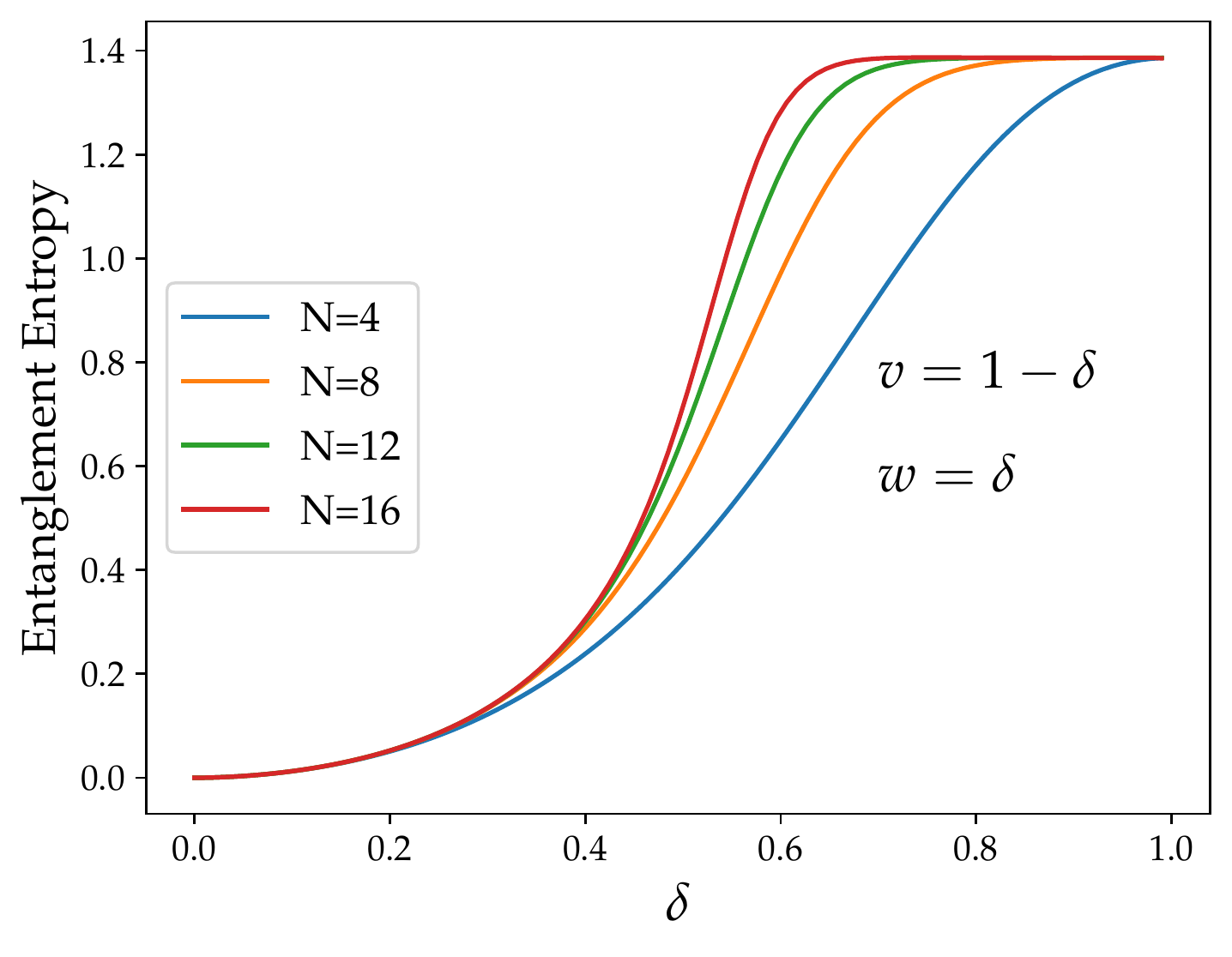}
  \caption{}
  \label{open_EE}
    \end{subfigure}\\
    \begin{subfigure}{\linewidth}
    \centering
  \includegraphics[width=.9\linewidth]{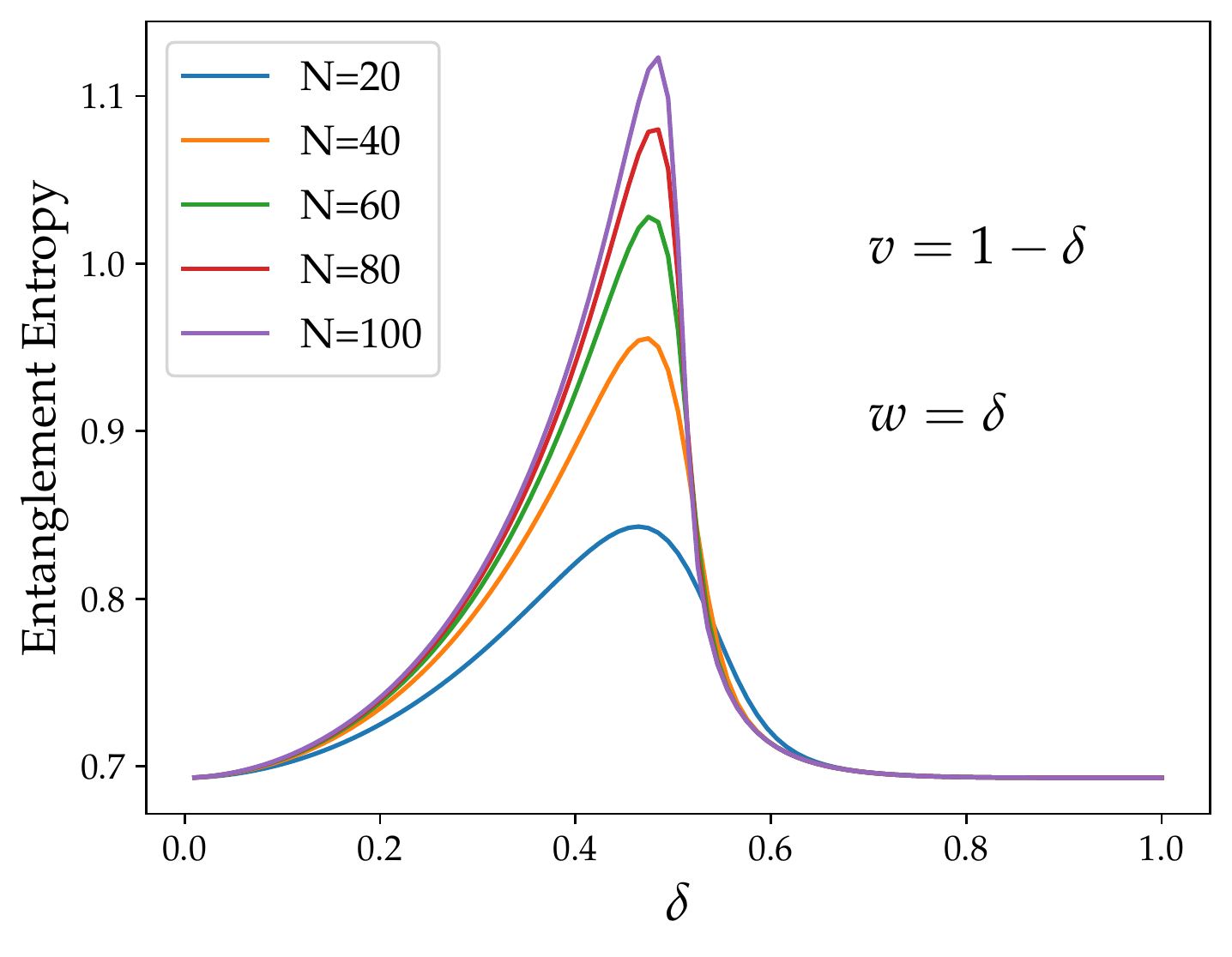}
  \caption{}
  \label{open_EE_1less}
    \end{subfigure}
    \caption{(a) Entanglement entropy for halfway partition in a SSH chain with open boundary conditions for different lengths $N$. (b) Entanglement entropy for halfway partition in a SSH chain with open boundary conditions, after removing the highest energy electron (so as to remove the edge state electron in the topological regime) for different lengths $N$.}
    \label{fig:my_label}
    \end{framed}
\end{figure}

In \Fig{open_EE} for $\delta = 0$ we have the entropy to be zero as earlier. However for $\delta = 0.99$ we have the entropy to be $2\log(2)$. This is counter intuitive, since we no longer have the second bond due to periodicity between the first and the $N^{th}$ site. So where does the  other $\log(2)$ come from? 
\leftHighlight{At $\delta\sim 0$ why does the system with $N/2-1$ electrons has $\log(2)$ entropy?}
In the topological regime, we also have two edge states in an open chain. For $v=0.01$ and $w=0.99$ these states hybridize leading to a bonding orbital between site 1 and site $N$. It is this bonding orbital which leads to an additional $\log(2)$ contribution.

To check this, let us keep the number of electrons in the system to be one less than $N$/2. This would lead to the edge state remaining unoccupied. The result is shown in \Fig{open_EE_1less} and as expected, the entanglement entropy now remains $\log(2)$ when $\delta>0.5$.

\subsection{\textit{U} can make all the difference}

So far, we have been looking at models where there is no electron electron interaction. Let's look at the Hubbard chain where such an interaction is considered and electrons are now assumed to be spinful. Therefore, each site has two states in the singly occupied sector\cite{Hubbard_BM}, a spin up state and a spin down state. 

\begin{figure}[t]
		\centering
		\begin{framed}
	\includegraphics[width=1.0\columnwidth]{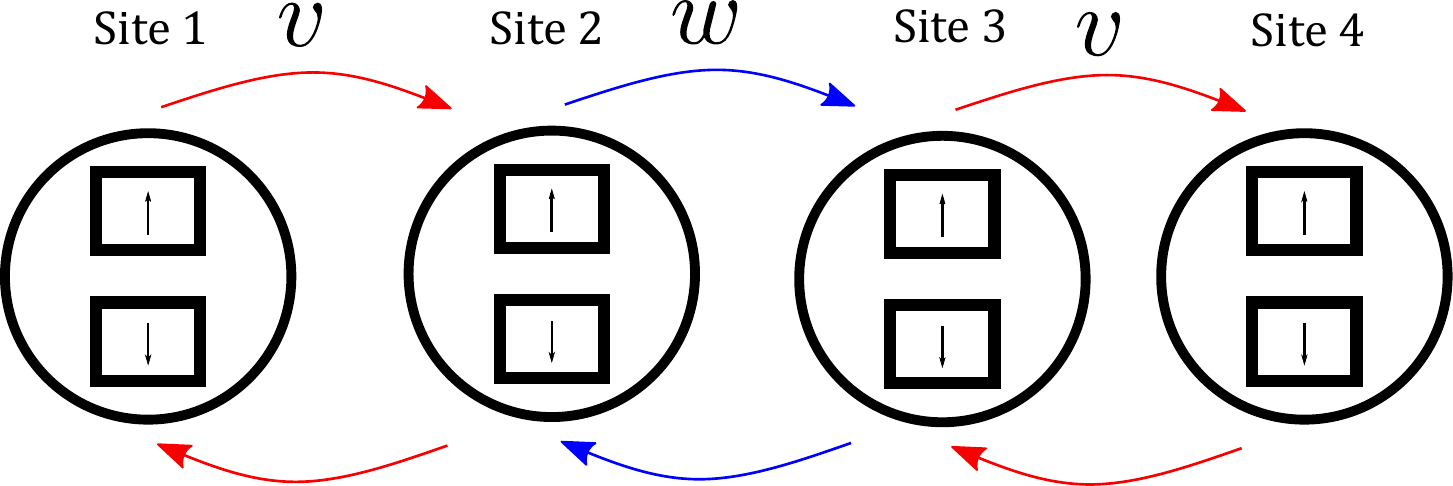}
	\caption{A diagram giving a pictorial depiction of the hopping terms given in \eqn{hubbard_ham}.}
	\label{hubbard_fig}
	\end{framed}
		\end{figure}

The Hamiltonian is given by, 
\beq
 H = -t \sum_{i,\sigma} \  c_{i,\sigma}^{\dagger}c_{i+1,\sigma} + c_{i+1,\sigma}^{\dagger} + U \sum_i \ n_{i\uparrow}n_{i\downarrow}
 \label{hubbard_ham}
\eeq
where $n_{i,\sigma}$ is the number operator on that site for a particular spin $\sigma$. Just like the SSH chain we now have $v$  and $w$ hoppings for every spin sector separately (see \Fig{hubbard_fig}).

\begin{figure}[!htb]
    \centering
    \begin{framed}
    \begin{subfigure}{\linewidth}
    \centering
  \includegraphics[width=.9\linewidth]{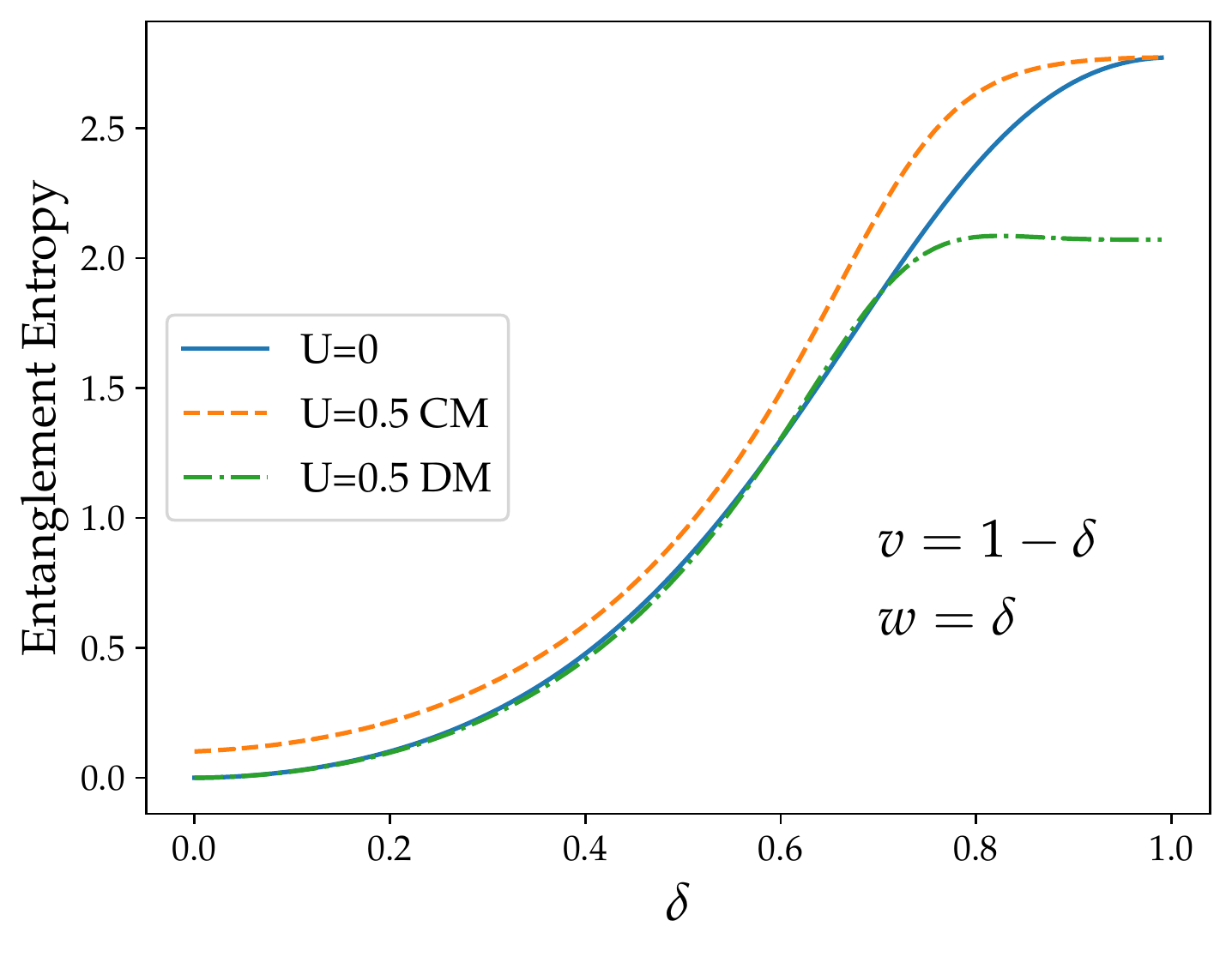}
  \caption{}
  \label{hubbard_u0p5}
    \end{subfigure}\\
    \begin{subfigure}{\linewidth}
    \centering
  \includegraphics[width=.9\linewidth]{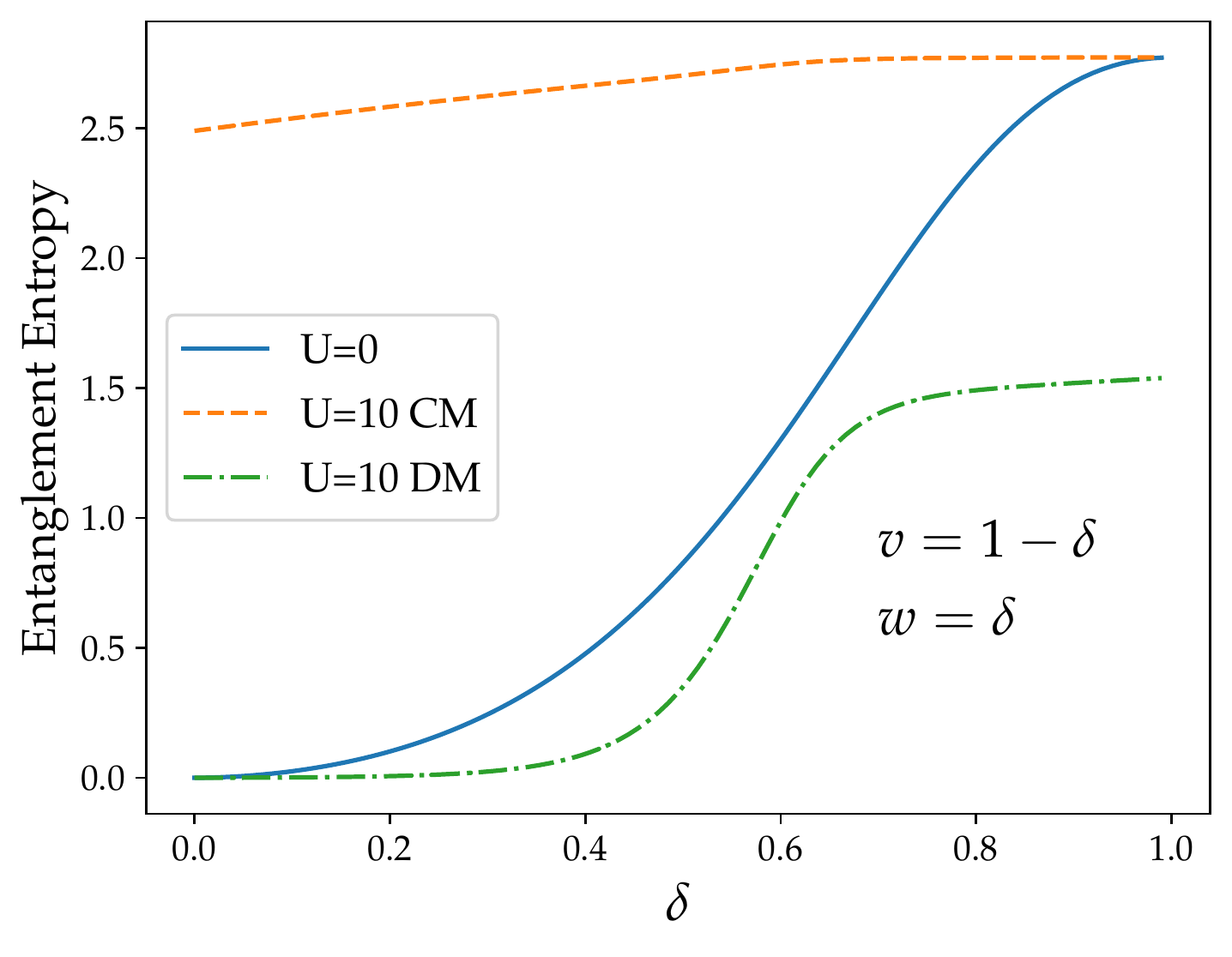}
  \caption{}
  \label{hubbard_u10}
    \end{subfigure}
    \caption{Entanglement entropy for halfway partition in a Hubbard chain(open boundary conditions) with 4 sites for two values of $U$, (a) $U$ = 0.5 and (b) $U$ = 10
    CM denotes that the entropy has been calculated using the correlator matrix and DM denotes that the entropy has been calculated using the density matrix.}
    \label{hubbardee}
        \end{framed}
\end{figure}

The parameter $U$ models the interaction where two electrons of opposite spins, at the same site, have an energy strength of $U$. If $U$ is negative, then there is an added incentive for electrons to occupy the same site and vice-versa. If $U=0$, then the chain in \Fig{hubbard_fig} is simply 2 parallel SSH chains, a spin up chain and a spin down chain that are not interacting with each other. Now, since we've introduced the model, we can try to calculate the entanglement entropy of the left half with the right half in presence of interactions. Here's the tricky part though, since the product of two number operators is no longer quadratic, the correlation matrix method to calculate the entropy would miss something compared to the density matrix method. But how bad could it be?

\Fig{hubbardee} shows this calculation using both the density matrix method and the correlator matrix method for a 4 site Hubbard chain with open boundary conditions. For the $U=0$ (blue curve), the entanglement entropy is just twice that of a single SSH chain and both methods lead to the same result. For $U=0.5$, the two methods are still quite similar only deviating from each other in the topological regime of $\delta>0.5$. But for $U=10$, the two methods do not agree with each other at all! Do we understand these numbers? Just like the number of bond cuts or other ideas we could use to interpret our earlier results?

Well, the upshot of this section is to make you appreciate that there is a lot we {\it do not} understand about topological phases of matter, in particular what role do the interactions play.

\section{Taking lead}

\subsection{Looking at electronic transport}

Up until now, we have looked at periodic systems and open chains. But real systems are also affected by externalities of the environment to which it is connected. Let us now try to model such systems. We will consider that the 1D open chains that we have been analyzing up until now are connected to the environment using \textit{leads}. Such systems are called \textit{open-quantum systems} and one way to analyze them is through the Keldysh non-equilibrium Green's function (NEGF) technique \cite{LNE_datta} \cite{datta_2005_chapter_8}\cite{datta_2005_chapter_9}. The Green's functions are themselves multi-point correlators taken at different times, and for the purposes of steady state calculations, we Fourier transform the time differences into the energy domain. We will leave the reader to these advanced considerations from other excellent pedagogical references \cite{Jauho}. Our pedagogy here begins with the consideration of steady state situations via the energy domain Green's functions.   

The leads are also referred to as \textit{contacts}. We will assume that these contacts are \textit{large sources or sinks} that are maintained at a constant chemical potential. These open systems are important to understand as they can help us simulate actual experiments and thus give us an idea as to what to expect. Employing NEGF, would help us simulate their local density of states, conductance and many other properties as we will see later. We will try to explain the basics of NEGF through a simple \hyperlink{example_NEGF}{example}.

\begin{mdframed}[style = mystyle, align=center,userdefinedwidth=1.0\columnwidth]
\leftHighlight{Example 9}
\hypertarget{example_NEGF}{{\it \textbf{Example:}}}
Consider a 1D chain of 8 atoms as shown in \Fig{fig:NEGF_example_schematic1}. Consider a potential $V(x)$ over this 1D lattice chain and let us assume that the chain is attached to contacts on both the ends. These contacts are themselves semi-infinite 1D tight-binding chains.

Take, $\psi_{i}$ to be the discretized value of the wavefunction $\psi(x)$ at the $i^{th}$ lattice position. 

\begin{figure}[H]
\centering
\includegraphics[width = \linewidth]{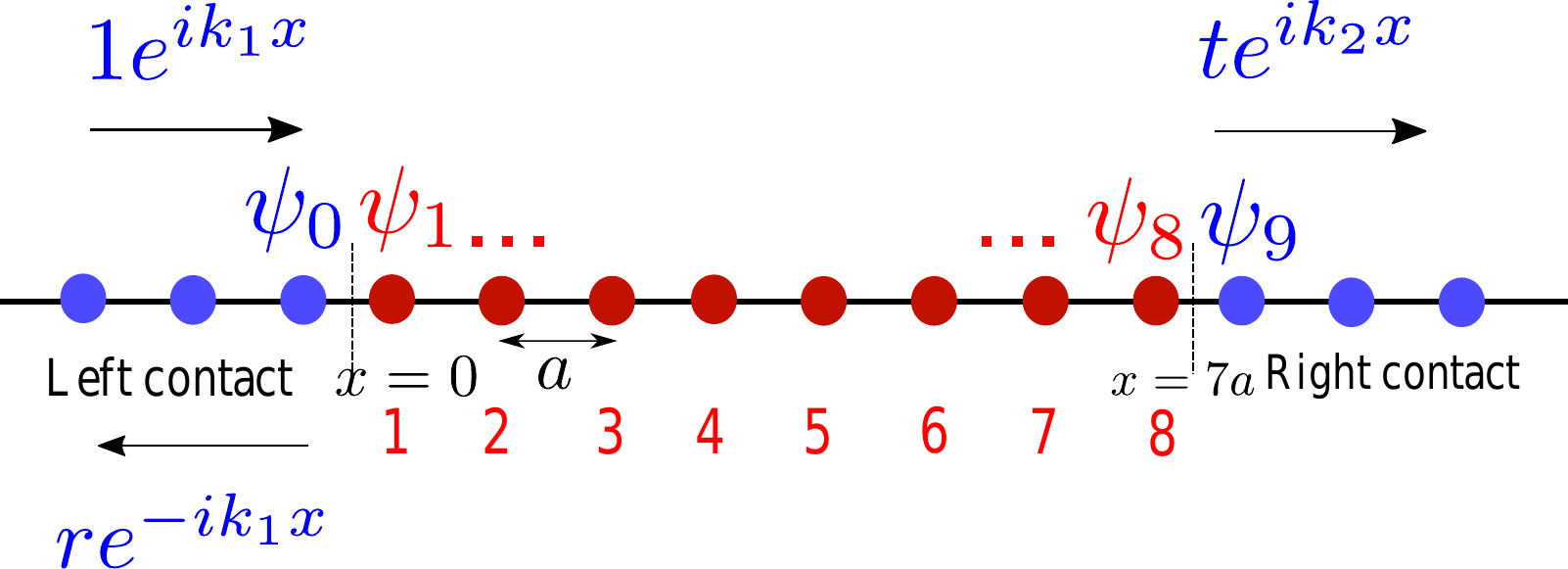}
    \caption{An eight atom 1D chain is considered. The setup is an open quantum system, i.e. boundaries of the lattice are connected to contacts (semi-infinite 1D chains)}
    \label{fig:NEGF_example_schematic1}
\end{figure}
 The lattice is connected to two contacts -- \textit{left contact} and \textit{right contact}. The left and the right contacts have a constant potential $V_{LC}$ and $V_{RC}$ respectively. We commonly refer the left contact as the \textit{source} and the right contact as the \textit{drain}. Why do we take contacts to be semi-infinite 1D chains? The semi-infinite chains help them act as \textit{large sinks} which maintain a constant chemical potential. Therefore, we can write wavefunctions in that contact regions as plane waves. 
 
 The wave function in the left contact is $e^{ik_{1}x}$ + $re^{-ik_{1}x}$. Similarly, wave function in the right contact is $te^{ik_{2}x}$. Here, $k_{1}$ is $\sqrt{\frac{2mV_{LC}}{\hbar^{2}}}$ and $k_{2}$ is $\sqrt{\frac{2mV_{RC}}{\hbar^{2}}}$. The lattice sites 1 and 8 are connected to the contacts. Note that, $V_{1}$ is equal to $V_{LC}$ and $V_{8}$ is equal to $V_{RC}$. Why?  This is because the first and the eighth lattice sites are a part of the contact and hence the potential there should be equal to the potential of the connected contact. Using \eqn{discrete_hamiltonian}, we can write the discretized Schrodinger's equation for the lattice region as:
\beq
E \psi_{i}-\left[-t_{0} \psi_{i-1}+\left(2 t_{0}+V_{i}\right) \psi_{i}-t_{0} \psi_{i+1}\right]=0
\label{discretized_Hamiltonian_NEGF}
\eeq
where, $t_{0}$ = $\hbar^{2}/2ma^{2}$. \eqn{discretized_Hamiltonian_NEGF} holds for all $i$ from 1 to 8. More specifically for $i=1$ we have
\beq
   E \psi_{1}-\left[-t_{0} \psi_{0}+\left(2 t_{0}+V_{1}\right) \psi_{1}-t_{0} \psi_{2}\right]=0
   \label{eqn:discretized_Hamiltonian_NEGF_site1}
\eeq
Now, we employ the boundary conditions to evaluate $\psi_0$ which is the wavefunction amplitude on the first site of the left contact. Given the first atom of the device (site 1) is located at $x=0$, $\psi_{1} = 1 + r$, and $\psi_{0} = \psi_{x=-a} =e^{-ik_{1}a} + re^{ik_{1}a}$. Placing these values, in \eqn{eqn:discretized_Hamiltonian_NEGF_site1} gives us:
\beq
E \psi_{1}-\left[\highlight{-t_{0} e^{ik_{1}a}\psi_{1}}+\left(2 t_{0}+V_{1}\right) \psi_{1}-t_{0} \psi_{2}\right]= \highlight{t_{0}(e^{ik_{1}a} - e^{-ik_{1}a})} 
\label{eqn:NEGFformulate_1} 
\eeq
where we have substituted $\psi_{0}=\psi_{1} e^{i k_{1} a}-\left(e^{i k_{1} a}-e^{-i k_{1} a}\right)$. Similarly, if we apply boundary conditions at lattice site 8, we would get $\psi_{8}$ = $te^{ik_{2}(x=7a)}$ = $te^{i7k_{2}a}$ and $\psi_{9}$ = $te^{ik_{2}(x=8a)}$ = $te^{i8k_{2}a}$.  Using, these substitutions in \eqn{discretized_Hamiltonian_NEGF} written for lattice site 8 would give us:
\beq
    E \psi_{8}-\left[\highlight{-t_{0}e^{ik_{2}a}\psi_{8}}+\left(2 t_{0}+V_{8}\right) \psi_{8}-t_{0} \psi_{7}\right]= 0 
    \label{eqn:NEGFformulate_2} 
\eeq

The highlighted terms in both \eqn{eqn:NEGFformulate_1} and\eqn{eqn:NEGFformulate_2} are different from the bulk sites ($2-7$) shown in \eqn{discretized_Hamiltonian_NEGF}. Notice that the complete system (with the two semi-infinite leads) and the device, is essentially an infinite system which cannot be solved. However by using the boundary conditions, and the effect of the leads (see \eqn{eqn:NEGFformulate_1} and \eqn{eqn:NEGFformulate_2}) we are able to capture the effective physics just within the the sites belonging to the device. 

If $H$ is the Hamitlonian for just the device with open boundary conditions, then the total effective Hamiltoian (including the effect of leads) is given by 
\beq
[EI - H - \Sigma_{1} - \Sigma_{2}]\psi = S
\label{eqn:NEGF1}
\eeq
where
\begin{figure}[H]
    \centering
    \includegraphics[width=\linewidth]{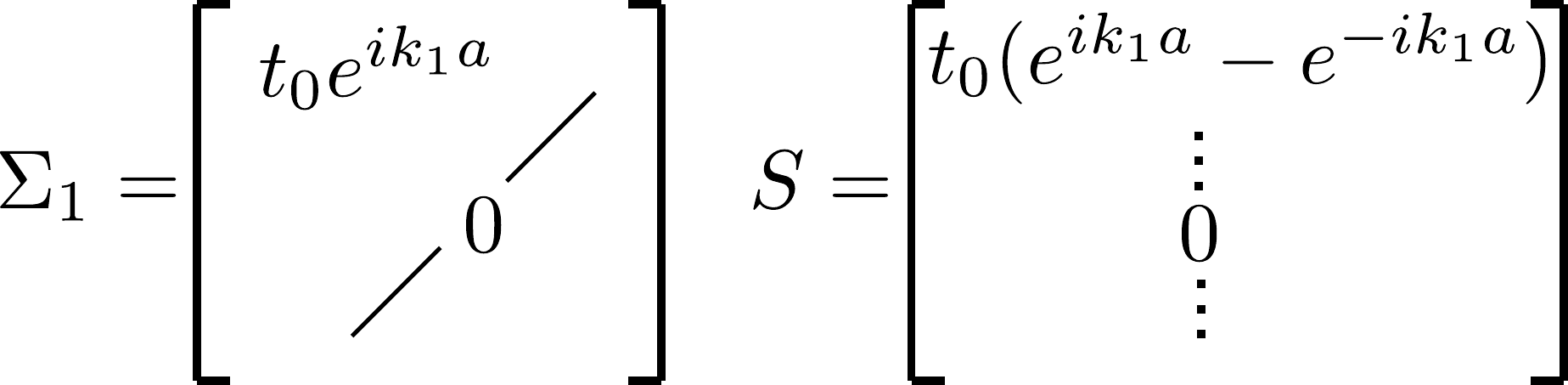}
    \caption{Self-energy matrices for the considered setup are non-Hermitian square matrices of size 8 with one element non-zero. Source matrix is a column matrix of size 8 with the top element non-zero. $\Sigma_{2}$ as similar to $\Sigma_{1}$ in its form except that it has the non-zero element at the right-bottom}
    \label{fig:NEGF_matrix_elements}
\end{figure}
$\Sigma_{1}$ and $\Sigma_{2}$ are called \textit{self-energy matrices} and $S$ is called the \textit{source} matrix. In the above case, $\Sigma_{1}$ is square matrix with all terms 0 except the term at (1,1) which is $t_{0}e^{ik_{1}a}$, $\Sigma_{2}$ has all terms 0 except the term at (8,8) which is $t_{0}e^{ik_{2}a}$ and $S$ is a $8 \times 1$ column matrix with all terms 0 except the first term which is $t_{0}(e^{ik_{1}a} - e^{-ik_{1}a})$. The size of the self energy matrices is the same as the system size but unlike the Hamiltonian they are not necessarily Hermitian. 

We can rearrange \eqn{eqn:NEGF1} and write 
\beq
\psi = {\cal G}(E)S
\label{Green}
\eeq
where $\cal{G}$ is called the Green's function and is equal to $[EI - H - \Sigma_{1} - \Sigma_{2}]^{-1}$. 

Briefly, if you remember, Green's functions provide us an alternate of solving differential equations where instead of finding a solution of a differential equation by either a qualified guess or explicit expansions (such as Frobenius method), one uses an integral form where a charge put in the system and one thinks about what response it creates \cite{ARFKEN2013447}. For example in electromagnetic theory, one can consider putting a single electric charge in real space and asking what electric field configuration it creates in its surrounding space. \Eqn{Green} is now of the same form where $S$ can be considered like a charge or a {\it source}, $\Psi$, the wavefunction in the device is the effect it creates. 

Is easy to see if $H+\Sigma_1+\Sigma_2$ has a set of eigenvalues $\epsilon_n = \epsilon^R_n + i \epsilon^I_n$ and eigenvectors $|n\rangle$, then

\beq
{\cal G}(E) = \sum_n \frac{|n \rangle  \langle n |}{E-\epsilon^R_n - i \epsilon^I_n }
\label{eqn:green_function_expansion}
\eeq

\end{mdframed}

Note that $H + \Sigma$ has complex eigenvalues in general and thus open systems gives a finite lifetime to eigenstates. While for a hermitian Hamiltonian, that has just a real eigenvalue $\epsilon_n$ for a stationary eigenstate $\Psi_n$, the time evolution of this state is given by
\beq
\Psi_n(t) = \exp(-i \epsilon_n t) \Psi_n
\eeq
which just adds a phase dependent on time. In the case of a complex eigenvalue which has a real and an imaginary part
($\epsilon_n = \epsilon^R_n + i \epsilon^I_n$)
\beq
\Psi_n(t) = \exp(-i \epsilon^R_n t) \exp(\epsilon^I_n t) \Psi_n
\eeq
which leads to a decay of the wavefunction if $\epsilon^I <0$ within a time $\tau \sim \frac{1}{\epsilon^I}$. So $\tau \sim \frac{1}{\epsilon^I} $ is called the lifetime of the electron. This essentially means that unlike a periodic Hamiltonian where an electron cannot {\it vanish}, in an open system due to the presence of leads, an electron can decay into the contacts. 

\rightHighlight{Show that the DOS for a one-dimensional tight-binding model goes as $D(E) \sim \frac{1}{\sqrt{E}}$ at low energies. How can DOS diverge? }

These decay physics also leads to a broadening of density of states (DOS) in the system. DOS essentially quantifies the number of eigenstates at energy $\epsilon$. For a set of real eigenvalues at energies $\epsilon_n$ which are fairly separated from each other
\beq
D(E) = \sum_n \delta(E-\epsilon_n)
\eeq
which is a $\delta$ function at every real energy $\epsilon_n$.
However when these energies have an imaginary part each of these $\delta$ functions gets modified to a  lorentzian where
\bea
D(E) &=& \sum_n \frac{1}{2\pi} \frac{|\epsilon^I_n|}{(E-\epsilon^R_n)^2 + \frac{1}{2}(\epsilon^I_n)^2} \\ &=& \frac{i}{2\pi} \Big[ \Big( \frac{1}{E-\epsilon^R_n - i \epsilon^I_n}  \Big) - \Big( \frac{1}{E-\epsilon^R_n + i \epsilon^I_n}  \Big) \Big]
\label{eqn:LDOS_description}
\eea
This is called spectral broadening which is characteristic to systems in which particles have a finite lifetime.

While such a broadening for a single eigenvalue is easy to capture within lifetimes of every eigenstate, for a lattice system it is more natural to define a matrix called $\Gamma$, which is the imaginary part of the self-energy matrix 
\beq
\Gamma = i(\Sigma - \Sigma^{\dagger})
\eeq
that characterises the effect of leads on the device, in terms of creating finite broadening and corresponding decay of the electrons from within the device. Therefore corresponding to each contact there is a $\Gamma$ matrix for e.g.,~$\Gamma_1= i (\Sigma_1-\Sigma_1)$ and $\Gamma_2= i (\Sigma_2-\Sigma_2)$.

The NEGF formalism can be used to calculate important properties of a \textit{device} like I-V characteristics, conductance (often called transmission here etc.). Here we would would not go into the details and refer to an excellent book \cite{datta_2005_chapter_8} \cite{datta_2005_chapter_9}, but present the essential quantities which one calculates such as
\beq
 T(E) = \text{Tr}[\Gamma_{1}{\cal G}(E)\Gamma_{2}{\cal G}(E)^{\dagger}] \label{eqn:transmission}
\eeq
$T(E)$ is the transmission at an energy $E$. Calculating transmission tells us `conductance' through the device at energy $E$. Naturally it depends on how the electrons decay from both the contacts ($\Gamma_1, \Gamma_2$) and the way electrons propagate within the device ${\cal G}$.

Given the transmission it is easy to calculate the current ($I$) in the system which is
\beq
 I=(q / h) \int_{-\infty}^{+\infty} \mathrm{d} E T(E)\left(f_{1}(E-\mu_{LC})-f_{2}(E-\mu_{RC})\right) \label{eqn:current}
\eeq
which depends on the chemical potentials of the right and left contacts, their corresponding occupancies ($f(\epsilon)$) and their transmission. 
The DOS ($D(E)$) of the complete system is given by
\beq
D(E) = \text{Tr}[i({\cal G}(E) - {\cal G}(E)^{\dagger})/2\pi] \label{eqn:LDOS}
\eeq
which can be seen by combining \eqn{eqn:green_function_expansion} and \eqn{eqn:LDOS_description}. 

\rightHighlight{What should be the transmission through a clean one dimensional wire? Can you evaluate this using NEGF?}

Now, for a contact which is semi-infinite and has a clean tight-binding description we found specific forms for $\Sigma$ and  $\Gamma$ matrices as discussed in the \hyperlink{example_NEGF}{previous example}. This also allows us to evaluate the local DOS (LDOS) which refers to the contribution of the DOS at a real space position. Given a set of complex eigenvalues, the LDOS at site $j$ should be

\beq
D(E, x=j) = \sum_n \frac{1}{2\pi} \frac{|\epsilon^I_n| |\langle n| x=j\rangle|^2 }{(E-\epsilon^R_n)^2 + \frac{1}{2}(\epsilon^I_n)^2}
\eeq
Clearly when summed over all sites $j$, the LDOS at energy $E$ equals $D(E)$. This can be calculated quite simply by evaluating

\beq
D(E, x=j) = \langle j |i({\cal G}(E) - {\cal G}(E)^{\dagger})/2\pi   | j\rangle 
\label{eqn:LDOS_1}
\eeq
which nothing but the $j^{th}$ diagonal element of 
$i({\cal G} - {\cal G}^{\dagger})/2\pi$.

However, more realistically, often we are unaware of the exact nature of the leads. Even one can use the NEGF equations (see \eqn{eqn:NEGF1}) to model the junction by choosing $\Gamma, \Sigma$ as some effective parameters. $\Sigma$ matrices in most simplest assumption can have an imaginary value on the first and the last site of the device such that both the ($1,1$) element of $\Sigma_1$ and ($N,N$) element of $\Sigma_2$ is again $-\frac{i\gamma}{2}$. This leads to the corresponding $\Gamma$ matrices to have the corresponding (1,1) and ($N,N$) term to be $\gamma$. This simplifies the calculations remarkably, where even in absence of any information regarding the leads, and just the device Hamiltonian ($H$) we can calculate the current and the conductance using NEGF formalism. We will now apply these tools to a SSH chain in order to investigate its transport signatures, if any.

\begin{figure}[!b]
\begin{framed}
\begin{minipage}{\textwidth}
    \centering
    \includegraphics[width=0.9\linewidth]{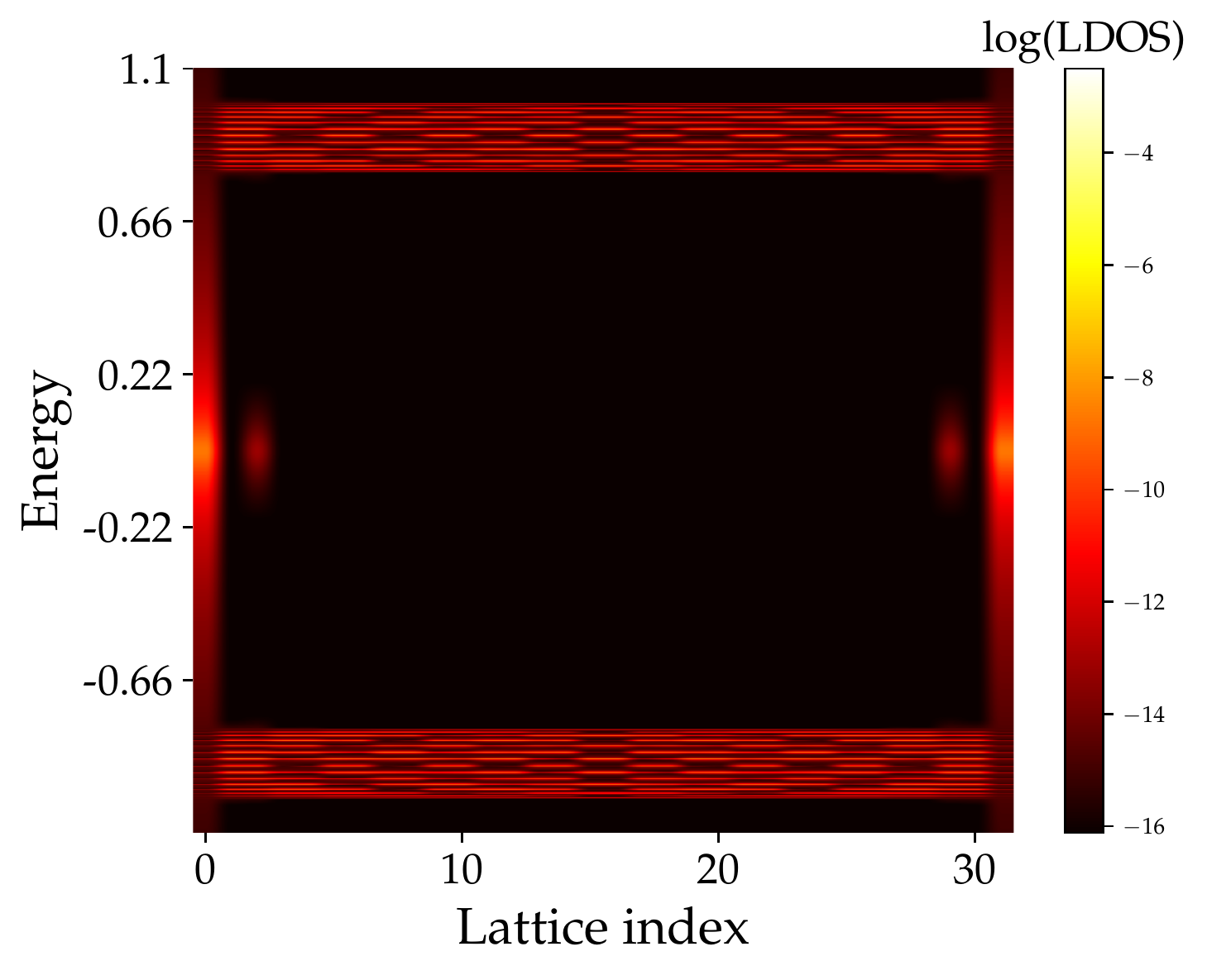}
    \caption*{(a)}
     \label{fig:topological_trivial_LDOS_A}
\end{minipage}
\begin{minipage}{\textwidth}
    \centering
    \includegraphics[width=0.9\linewidth]{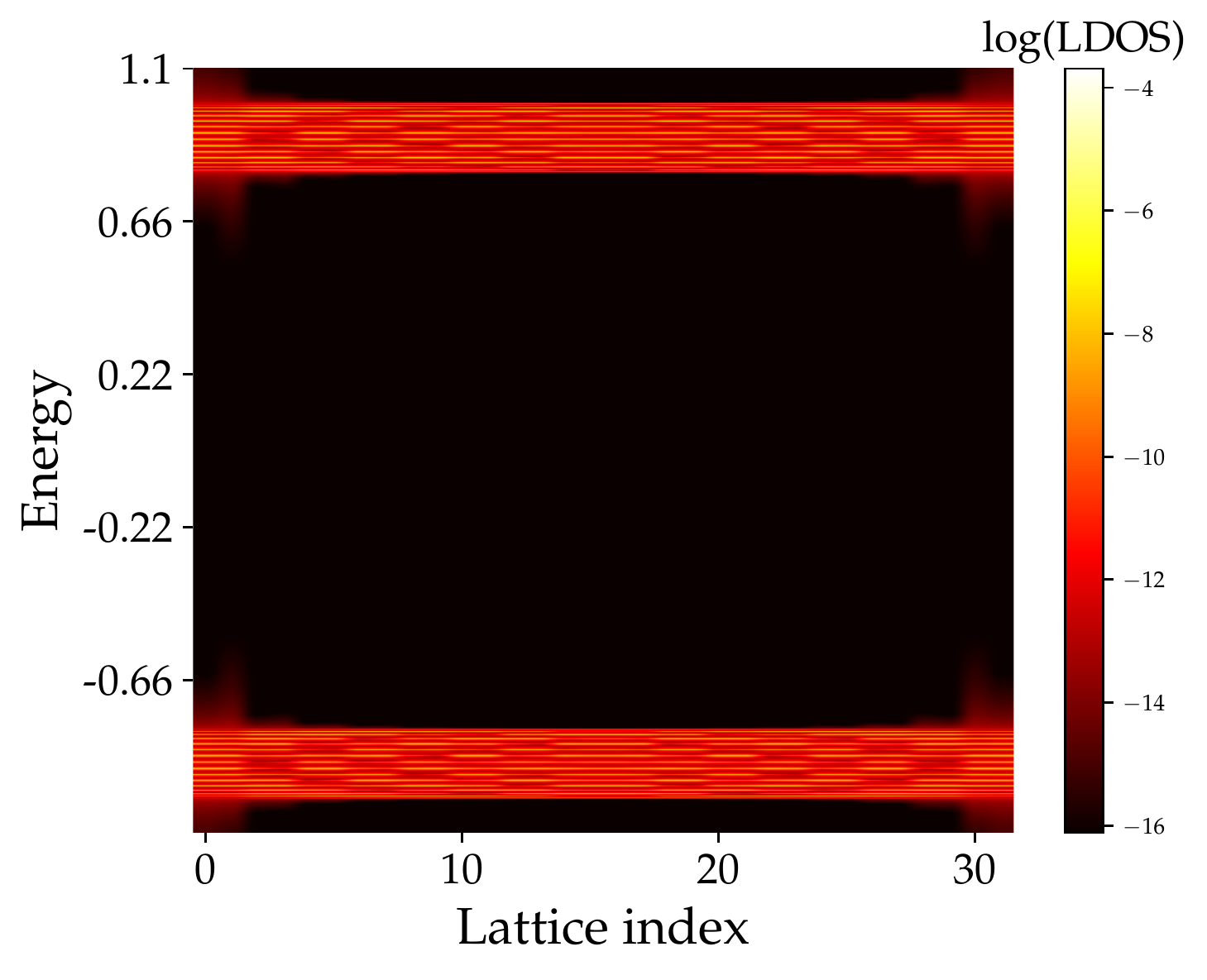}
    \caption*{(b)}
   \label{fig:topological_trivial_LDOS_B}
\end{minipage}
\end{framed}
\end{figure}%
\begin{figure}[ht]\ContinuedFloat
\begin{framed}

\begin{minipage}{\textwidth}
    \centering
    \includegraphics[width=0.9\linewidth]{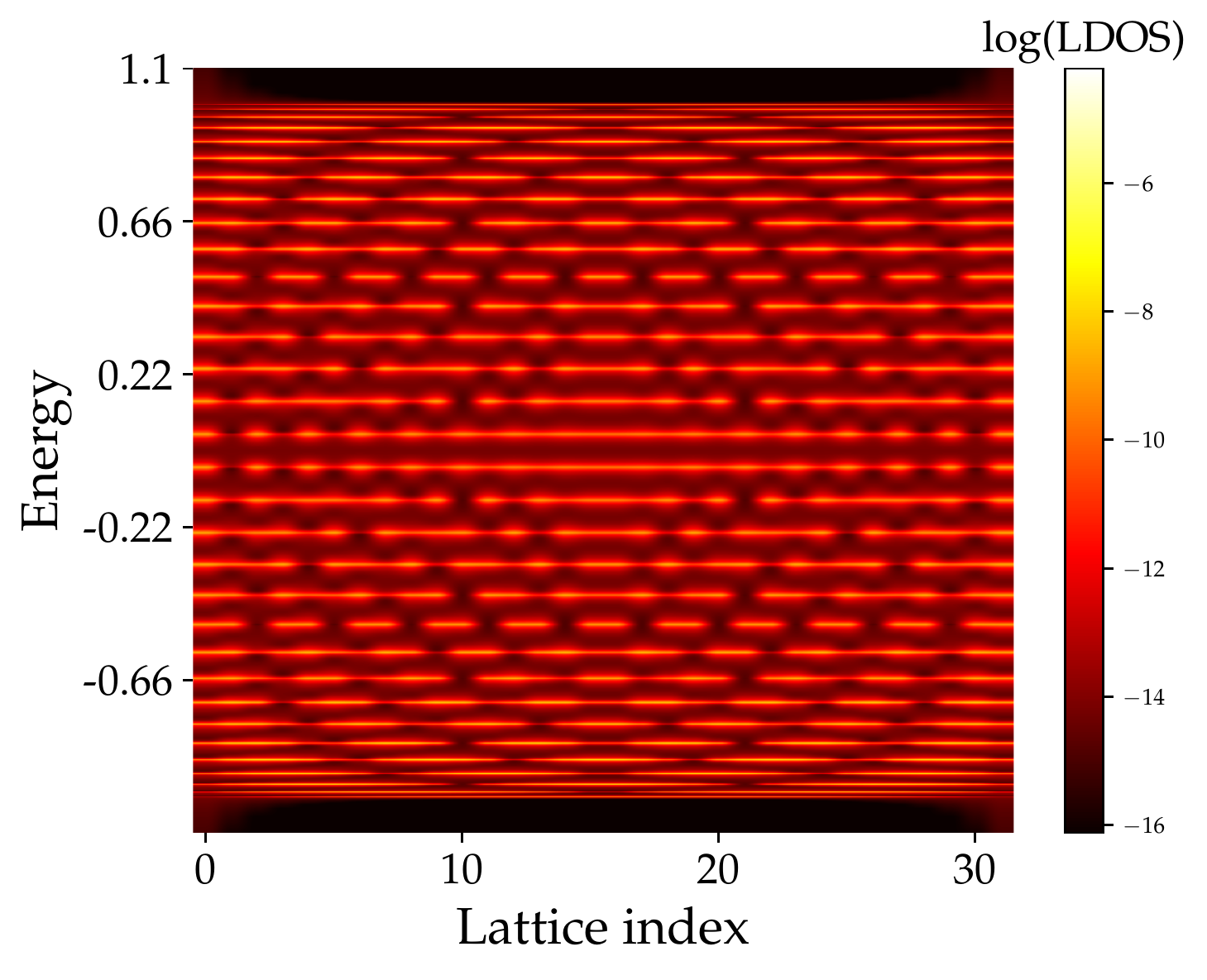}
    \caption*{(c)}
     \label{fig:topological_trivial_LDOS_C}
\end{minipage}
\caption{Local density of states (LDOS) in an SSH chain as a function of lattice index and energy $E$ for three regimes (a) topological ($w = 0.9, v = 0.1$), (b) trivial  ($w = 0.1, v = 0.9$)  (c) metal ($w = 0.5, v = 0.5$) for a 32 atom system. A finite coupling to the leads is considered where $\gamma=0.1w$.}
\label{fig:topological_trivial_LDOS}
\end{framed}
\end{figure}

\subsection{Transport properties of the SSH chain}

We have already seen that the SSH chain can exist in two phases -- (i) a topological phase when $v<w$ and (ii) a trivial phase when $v>w$. The transition point from one phase to another is when $v=w$ and the system is a metal. Just to remind ourselves (see \fig{fig:energy_spectrum}), in the topological regime the system has $\sim zero$ energy eigenstates at the edges. We will now use the NEGF formalism to see if the topological regime has any transport signatures.

Intuitively, in a thermodynamically large system, even if the edge states exist -- they {\it cannot} lead to a finite conductance because these edge states decay exponentially (see discussion near \eqn{open_ssh_exp}) and therefore with increasing system size, the overlap between the wavefunctions on the right and the left edge will continue to fall exponentially, leading to no transmission at close to $\sim zero$ energies. However in a finite size system, these edge states 
can hybridize and lead to some weak signatures in the LDOS.

\Fig{fig:topological_trivial_LDOS} plots the local density of states (LDOS) (see \eqn{eqn:LDOS_1}) of a finite SSH chain in the three different regimes -- topological, trivial and a metal as a function of energy $E$ when their is finite coupling to the leads. The first figure shows the SSH chain in the topological limit and we can clearly notice LDOS due to the edge sites hosting $\sim zero$ energy states. The second figure shows the SSH chain in the trivial limit where there are no edge states. The third figure shows the SSH chain in the conductance limit. As expected, LDOS are spread uniformly across the entire spectrum of energy.

\leftHighlight{What do you think would be the transmission signature (transmission as a function of energy) for these three phases of the SSH chain? (for a hint, see the \href{https://github.com/hnoend/SSH_codes/blob/main/Figure24.m}{code provided}.)}

\begin{figure}[!b]
\centering
\begin{framed}

\begin{subfigure}{\linewidth}
 \centering
 \includegraphics[width=.8\linewidth]{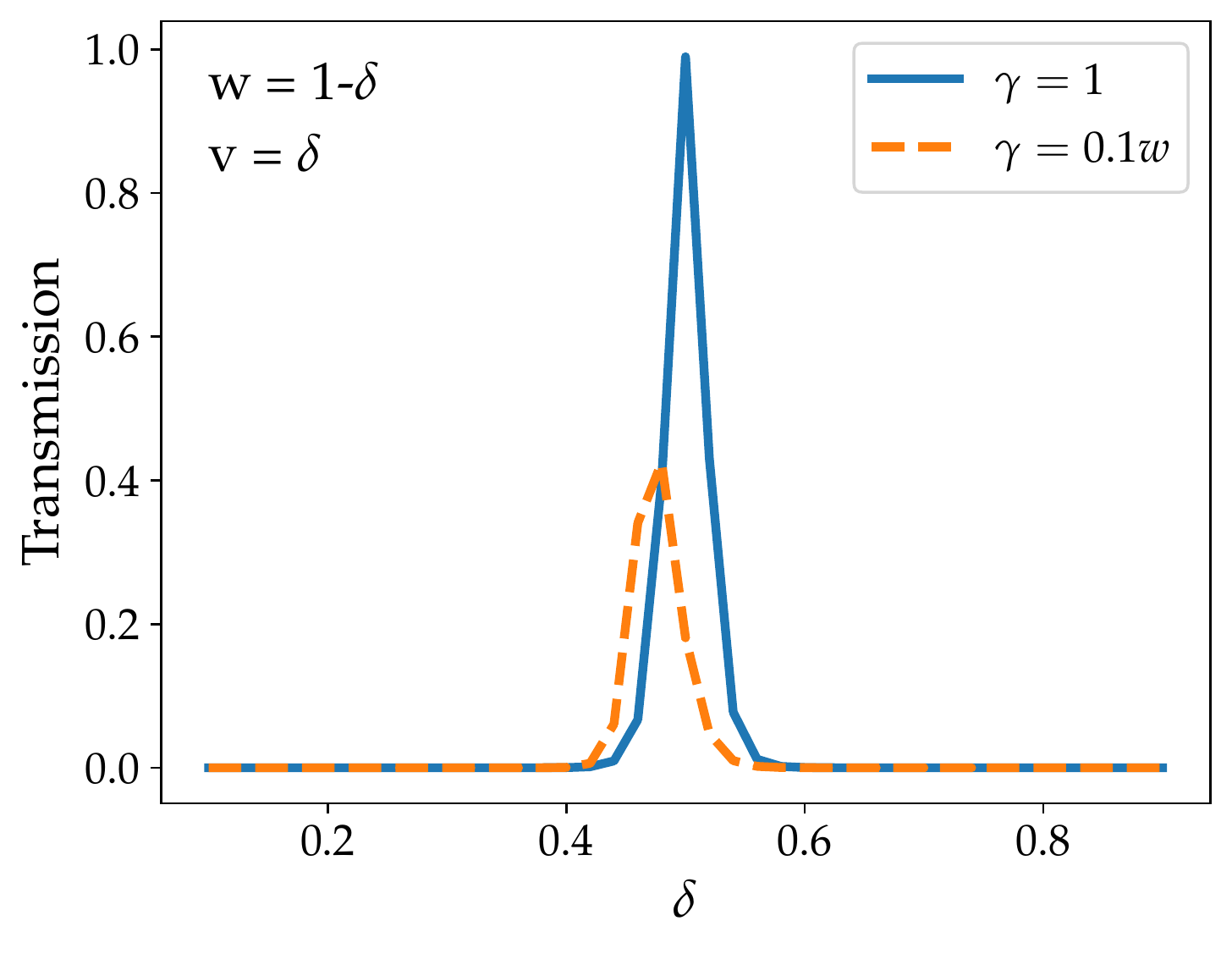}
   \caption*{(a)}
\end{subfigure}
 \begin{subfigure}{\linewidth}
 \centering
 \includegraphics[width=.8\linewidth]{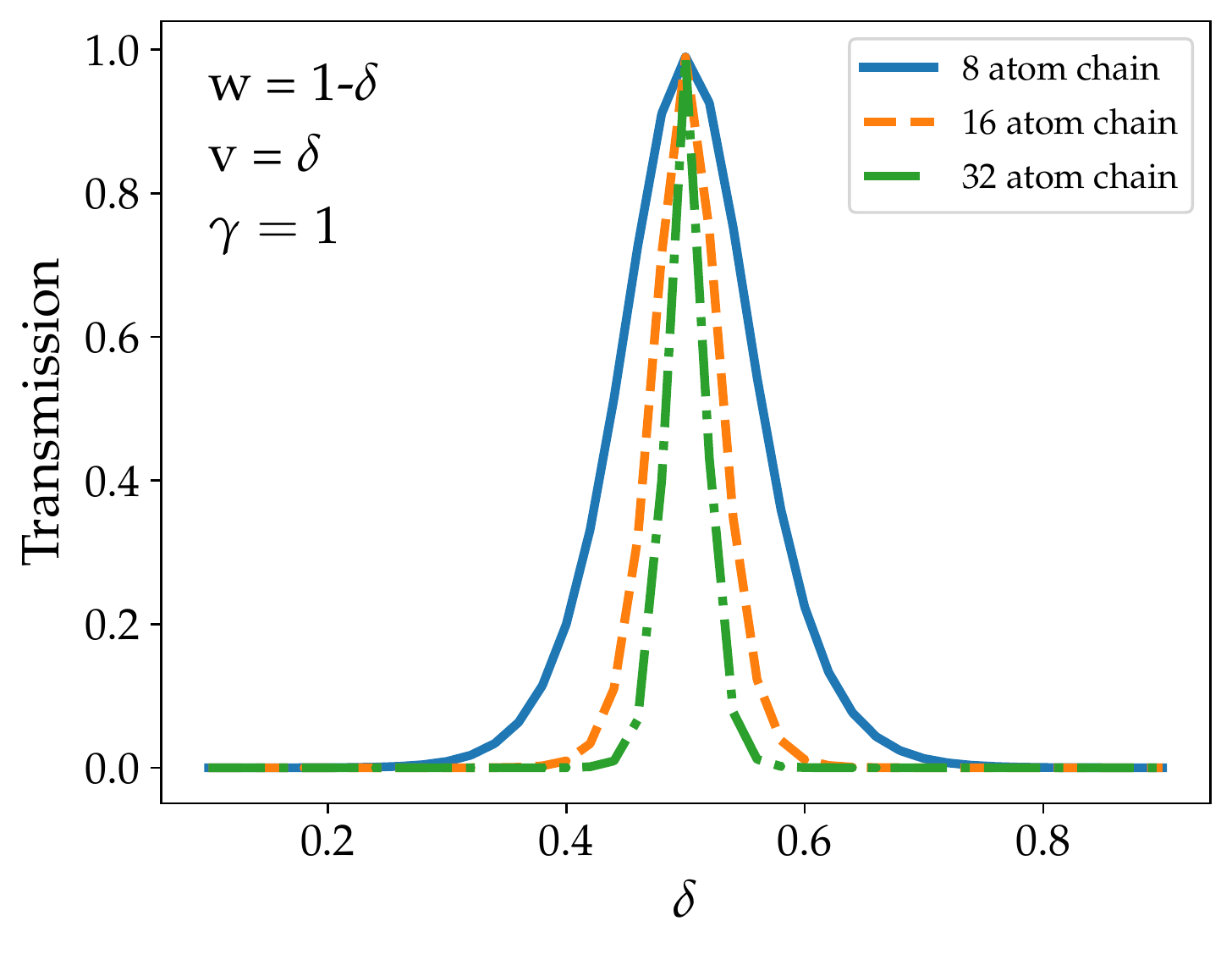}
   \caption*{(b)}
\end{subfigure}
\caption{ The behavior of transmission ($T(E)$) as a function of $\delta$ at $E=0$ for (a) different values $\gamma$ given a system size $N=32$ and (b) different values of $N$ for $\gamma=1$.}
\label{linear_conductance_plots}

\end{framed}
\end{figure}

We will now observe what happens if we try and apply voltages on the two ends of the chain and pass current through it. We would try and observe the difference, if any, in the two regimes. The behavior of transmission (conductance) as a function of $\delta $ where $v=\delta$ and $w=1 - \delta$ is shown in \fig{linear_conductance_plots} for $E=0$. The conductance, as is expected shows a peak when $\delta=0.5$, i.e.,~we are in a metallic system. In both $v<w$ or in $v>w$ the transmission is close to $zero$ as is expected for insulating systems. Interestingly if the coupling to the leads (approximated by a parameter $\gamma$) is varied one can tune the transmission. For instance at $\gamma=0$ where the system is not coupled then the transmission is zero while with increasing $\gamma$ the transmission increases when $\delta=0.5$. Some representative plots are shown in \fig{linear_conductance_plots} for $\gamma = 0.1 w, 1.0$. In \fig{linear_conductance_plots} (b) we look at the conductance as a function of increasing system size and find that except at $\delta=0.5$ the conductance falls in the $\delta<0.5$ and $\delta>0.5$ regime, consistent with the expectations that the insulators even in the {\it topological limit} show no conductance signature.

We must now remark that the transport discussion presented here is for the $U=0$ limit (absence of interactions). The NEGF technique in the strict sense works well when the contact coupling $\gamma \gg U $. In the opposite regime, we are in a situation of weakly coupled quantum dots and the transport is typically treated within more complicated methods such as the density matrix master equation approach in the Fock space. A detailed description of that technique is carried out in various works including \cite{Muralidharan_2008,LNE_datta} which we will just refer to here. 

You may wonder what surprise did we find for the transport signature of the SSH chain -- and the answer {\it none}! In fact the topological regime cannot be detected with a conventional transport experiment for the SSH model. However a LDOS measurement (via using a scanning tunneling microcope (STM)) can still lead to some signatures (see \Fig{fig:topological_trivial_LDOS}) but we have not been able to successfully do this in experiments yet. This reflects that even while topological phases such as SSH model can very well be understood in periodic systems etc., it is hard to detect them experimentally. 

However the above programme of simulating such toy systems under a theoretical framework has allowed us to visit a range of fantastic physics ideas which span from ideas of polarization, entanglement, correlators, Green's functions, and many more. In fact it is studies like this which guides both the experiments and the theoretical work which can lead to discoveries of new and exciting phases of quantum condensed matter.

\section{The way ahead}
In the last sections we visited many-a-ideas in quantum condensed matter physics to investigate a particular model Su-Schriefer-Heeger chain. SSH chain being a prototypical example has been addressed in beautiful pedagogical articles, we particularly recommend for the uninitiated
\cite{Anderson:1997vm,short_topo, Batra_2020}. Here we have addressed some of the ideas which are often not discussed in detail -- particularly the idea of polarization, the entanglement entropy, and electronic transport under the non-equilibrium Green's function. We hope this will encourage you, the reader, to take a more serious study of these ideas and explore more!

If this article made you feel a bit excited about topological quantum phenomena, then it is the right time to take the dive into the world of quantum materials and devices. With the concepts of geometric phase and winding numbers discussed here, one can move on to the exciting world of topological electronics.

Just as we noticed that the topological phase of the 1D chain gives us an ``end-state'', these ideas can be generalized to higher dimensions (called bulk-edge correspondence) where in two dimensions, we will have an edge state and in three dimensions we will have a surface state! This brings in the concept of topologically protected states along the boundary. One can immediately envision that such topologically protected states along a boundary, if they can propagate, can give rise to possibly dissipation less currents \cite{RevModPhys.82.3045}. This is because such states are ``unimpeded'' by impurities and defects to a large extent. In order to start exploring along these lines, we suggest a short road map. 

For two dimensional structures (2D), a few more ideas such as the calculation of Berry curvature and its relation to the Hall conductivity are needed \cite{shankar2018topological} to take the plunge in the 2D topological systems. A detailed study of the quantum Hall effect and the deep connection of the Hall conductance quantization with Berry curvature needs to be grasped. 
The extension of the 2D system to the hexagonal lattice finally opens up a world of possibilities including the quantum spin Hall effect and various related topological phases. With that, a whole range of limitless device possibilities \cite{Gilbert} featuring such topologically protected states open up. 

In 1972, Anderson told us ``more is different" (read this today \cite{Anderson393}, if you haven't), it is fair to say that quantum condensed matter {\it today} is finding that more and more things are {\it  topologically} different. Don't miss this exciting physics journey if you haven't already started!

\section{Acknowledgements}
A.A. would like to acknowledge Abhishodh Prakash for initiating the TQFT lecture series at ICTS, where part of this material was discussed as in a set of \href{http://live.icts.res.in/videos/video/4772/}{lectures}. A.A. gladly acknowledges his wonderful years of learning and research at IISc and ICTS, particularly with Vijay B. Shenoy and Subhro Bhattacharjee, which seeded many of the ideas. A.A. acknowledges partial funding from Max Plank Partner Grant at ICTS, Bangalore. A part of the material was also developed by B.M. during the 2020 iteration of the course titled ``Topological Electronics'' (EE-787) at IIT Bombay. 
B. M. would like to acknowledge useful discussions with Sven Rogge, Gilles Buchs, Alestin Mawrie and Ashwin Tulapurkar. B.M. also acknowledges funding  under the Visvesvaraya Ph.D Scheme of the Ministry of Electronics and Information Technology (MEITY), Government of India, implemented by Digital India Corporation (formerly Media Lab Asia). This work is also supported by the Science and Engineering Research Board (SERB), Government of India, Grant No. STR/2019/000030, the Ministry of Human Resource Development (MHRD), Government of India, Grant No. STARS/APR2019/NS/226/FS under the STARS scheme.

\appendix

\section{Entanglement entropy from correlator matrix}
\label{EntanCorr}

The density matrix when a system is in a state $|\Psi \rangle$
\beq
\rho = |\Psi\rangle \langle \Psi |
\label{eqn:mat}
\eeq
So any operator expectation on this state
is 
\beq
 \langle \hat{O} \rangle = \text{Tr}\Big( \hat{O} \rho  \Big)
 \label{expec}
\eeq
While for a pure state, density matrix is indeed given by \eqn{eqn:mat}, in general the density matrices can correspond to a mixed state or can be used to model a system with an effective temperature or an ensemble of systems. There while \eqn{eqn:mat} is not valid, however \eqn{expec} continues to hold.

For example, let's a consider a single spin in a magnetic field $B_0$ along the $z$ direction \cite{2020Density}. The Hamiltonian is given by $H=-\gamma \mathbf{S}.\mathbf{B}= -\gamma S_z B_0$. We have two possible eigenstates, the $|\uparrow\rangle$ state along the z direction with energy $ -\gamma \hbar B_0/2$ and the $|\downarrow\rangle$ state with energy $ \gamma \hbar B_0/2$. Let $E_0$ be $\gamma \hbar B_0/2$. The thermal density matrix is then given by 
\beq
 \rho = \frac{1}{e^{-\frac{E_0}{kT}}+e^{\frac{E_0}{kT}}} \Big( e^{\frac{E_0}{kT}} |\uparrow\rangle \langle \uparrow| + e^{-\frac{E_0}{kT}} |\downarrow\rangle \langle \downarrow| \Big) 
\eeq

In order to calculate the expectation of the $S_x$ operator using \eqn{expec} we get,
\beq
 \langle S_x \rangle = \text{Tr}\Big( \frac{1}{e^{-\frac{E_0}{kT}}+e^{\frac{E_0}{kT}}}\begin{bmatrix} 
	  0 & 1\\
	  1 & 0\\\end{bmatrix} \begin{bmatrix} 
	  e^{\frac{E_0}{kT}} & 0\\
	  0 & e^{-\frac{E_0}{kT}}\\\end{bmatrix}  \Big)
	  \eeq
which leads to $\langle S_x \rangle= 0$. The expectation for $S_y$ would also be 0. It would  be non zero for $S_z$ indicating that the spin would like to align itself along the z direction. This  example also illustrates that the thermal density matrices can be written as $\rho = e^{-\beta H}$ with appropriate normalization constants.

Let's go back to the density matrix of a pure state as shown in \eqn{eqn:mat}. Subsystem density matrices which are obtained after tracing over part of the system are mixed states and can often be written as \beq
\rho_A = Ke^{-\beta H}
\label{rhoH}
\eeq
where $H$ is now called the entanglement Hamiltonian. $K$ is a normalization constant. $\beta$ is not the temperature but just appears as a scaling factor here.

Can every density matrix be written in the form of \eqn{rhoH}? In principle, yes. But one intriguing thing happens when traced density matrix belongs to a free fermionic state -- here the entanglement Hamiltonian is also a ``free Hamiltonian" i.e. it just has quadratic operators. Therefore expectation of operators such as $c^\dagger_i c_j$ which really is the correlator we discussed before can now be written as
\beq
\langle c^\dagger_i c_j  \rangle = \text{Tr}\Big(   c^\dagger_i c_j e^{ -\beta H}\Big)
\eeq

Note given a ground state wavefunction, the truncated correlator matrix (CMt) is easily obtainable (see  \sect{ee_correlator}). However for density matrix, we were interested in its eigenvalues to obtain the entanglement entropy.  Now we interpret the above equation as a thermal expectation over a different Hamiltonian $H$ at a temperature $1/\beta$.

Therefore a set of wavefunctions which diagonalizes $H$ has a thermal occupancy of every single particle state as 
\beq
n_f(\epsilon) = \frac{1}{1+ \exp(\beta \epsilon)}
\eeq
So any observable, including a correlator, when measured on this thermal state is given by
\beq
\langle c^\dagger_i c_j \rangle = \sum_n \phi^*_{in} \phi_{jn} \frac{1}{1+ \exp(\beta \epsilon_n)}
\eeq

But the above equation also shows that $C_{ij}$ matrix is itself diagonalizable with the wavefunctions $\phi$. Hence
the eigenvalues of CMt matrix can tell us the eigenvalues of $H$! If CMt has eigenvalues $e_i$, 
then 
\beq
e_i= \frac{1}{ 1 + e^{\beta\epsilon_i}}
\eeq
or 
\beq
e^{\beta \epsilon_i} = \frac{1-e_i}{e_i}
\eeq

Given $\rho =e^{-\beta H}$, the trace of $\rho$

\beq
\text{Tr}(\rho) =  K\prod_i (1+ e^{-\beta \epsilon_i})
\eeq
where $K$ is a normalization constant. Here the states are summed over all possible occupancies of the free fermionic states.  Given properties of density matrix \cite{Sakurai}
\beq
 K = \prod_i \frac{1}{(1+ e^{-\beta \epsilon_i})} = \prod_i (1-e_i)
\eeq

Notice that entanglement entropy is given by
\begin{align}
S &=-\sum_i \rho_i \log \rho_i  \\
&= - K \log (K) -  K \sum_i e^{-\beta \epsilon_i} \log K e^{-\beta \epsilon_i} \notag \\ & \qquad - K \sum_{i,j, i\neq j} e^{-\beta (\epsilon_i+\epsilon_j)} \log K e^{-\beta (\epsilon_i+\epsilon_j)} + \ldots
 \end{align}

It is easy to see that in terms of eigenvalues of the Hamiltonian we have
\beq
S =  -\log K + \sum_i   \frac{e^{-\beta \epsilon_i}  \beta \epsilon_i}{1+\exp(-\beta \epsilon_i)}
\eeq

\beq
S =\sum_i -e_i \log e_i - (1-e_i) \log (1-e_i)  
\eeq

It is interesting to see that $e_i$ is same as a Fermi function which tells us the occupancy of any state $i$. In this sense the entropy gets contributions both of filled and empty phase space!

\bibliographystyle{unsrt}
\bibliography{ref}

\end{sloppypar}

\end{document}